\documentclass[11pt]{article}
\usepackage{verbatim,amsmath,amssymb}
\usepackage{epsfig,float,color}
\usepackage{geometry}
\usepackage{setspace}
\usepackage{natbib}
\usepackage{amsthm,mathrsfs,amsfonts,dsfont} 
\usepackage{hyperref}
\usepackage{subfigure}

\definecolor{darkblue}{rgb}{0,0,1}
\definecolor{darkgreen}{rgb}{0.1,.8,0}
\definecolor{magenta}{rgb}{1,0,1}
\definecolor{darkgr}{rgb}{0.1,.8,0}
\hypersetup{pdftex=true, colorlinks=true, breaklinks=true, linkcolor=darkblue, menucolor=darkblue, pagecolor=darkblue, citecolor=darkblue, urlcolor=darkblue}

\newcommand {\eqb}[1]{\begin{equation}\begin{array}{#1}}
\newcommand {\eqe}{\end{array}\end{equation}}

\newcommand {\esb}[1]{\begin{equation*}\begin{array}{#1}}
\newcommand {\ese}{\end{array}\end{equation*}}
\newcommand {\ds}{\displaystyle}

\newcommand {\pa}[2]{\frac{\partial{#1}}{\partial{#2}}}

\newcommand {\paqq}[3]{\frac{\partial^2{#1}}{\partial{#2}\,\partial{#3}}}
\newcommand {\back}{\! \! \!}
\newcommand {\is}{\back &=& \back}
\newcommand {\dis}{\back &:=& \back}

\newcommand {\ais}{\back &\approx& \back}

\newcommand {\plus}{\back &+& \back}

\newcommand {\norm}[1]{\|#1\|}
\newcommand {\tr}{\mathrm{tr}\,}

\newcommand {\divz}{\mathrm{div}\,}

\newcommand {\dif}{\mathrm{d}}

\newcommand {\II}{{I\kern-.3em I}}
\newcommand {\III}{{I\kern-.3em I\kern-.3em I}}

\newcommand {\mrc}{\mathrm{c}}

\newcommand {\mrf}{\mathrm{f}}
\newcommand {\mrg}{\mathrm{g}}

\newcommand {\mrm}{\mathrm{m}}

\newcommand {\mrp}{\mathrm{p}}

\newcommand {\mrr}{\mathrm{r}}
\newcommand {\mrs}{\mathrm{s}}
\newcommand {\mrt}{\mathrm{t}}

\newcommand {\mrv}{\mathrm{v}}
\newcommand {\mrw}{\mathrm{w}}

\newcommand {\ma}{\mathbf{a}}
\newcommand {\mcc}{\mathbf{c}}
\newcommand {\md}{\mathbf{d}}

\newcommand {\mf}{\mathbf{f}}
\newcommand {\mg}{\mathbf{g}}

\newcommand {\mk}{\mathbf{k}}

\newcommand {\mm}{\mathbf{m}}

\newcommand {\mpp}{\mathbf{p}}
\newcommand {\mq}{\mathbf{q}}
\newcommand {\mr}{\mathbf{r}}

\newcommand {\muu}{\mathbf{u}}
\newcommand {\mv}{\mathbf{v}}
\newcommand {\mw}{\mathbf{w}}
\newcommand {\mx}{\mathbf{x}}

\newcommand {\ba}{\boldsymbol{a}}

\newcommand {\be}{\boldsymbol{e}}
\newcommand {\bff}{\boldsymbol{f}}
\newcommand {\bg}{\boldsymbol{g}}

\newcommand {\bi}{\boldsymbol{i}}
\newcommand {\bj}{\boldsymbol{j}}

\newcommand {\bn}{\boldsymbol{n}}

\newcommand {\bt}{\boldsymbol{t}}

\newcommand {\bv}{\boldsymbol{v}}
\newcommand {\bw}{\boldsymbol{w}}
\newcommand {\bx}{\boldsymbol{x}}

\newcommand {\bnu}{\mbox{\boldmath$\nu$}}

\newcommand {\bvphi}{\mbox{\boldmath$\varphi$}}

\newcommand {\bxi}{\mbox{\boldmath$\xi$}}

\newcommand {\bome}{\mbox{\boldmath$\omega$}}

\newcommand {\mB}{\mathbf{B}}

\newcommand {\mD}{\mathbf{D}}

\newcommand {\mF}{\mathbf{F}}
\newcommand {\mG}{\mathbf{G}}
\newcommand {\mH}{\mathbf{H}}

\newcommand {\mL}{\mathbf{L}}

\newcommand {\mN}{\mathbf{N}}

\newcommand {\mX}{\mathbf{X}}

\newcommand {\bA}{\boldsymbol{A}}

\newcommand {\bD}{\boldsymbol{D}}

\newcommand {\bF}{\boldsymbol{F}}

\newcommand {\bL}{\boldsymbol{L}}

\newcommand {\bN}{\boldsymbol{N}}

\newcommand {\bW}{\boldsymbol{W}}
\newcommand {\bX}{\boldsymbol{X}}

\newcommand {\sig}{\sigma}

\newcommand {\bsig}{\mbox{\boldmath$\sigma$}}

\newcommand {\bone}{\mathbf{1}}

\newcommand {\bbC}{\mathbb{C}}

\newcommand {\bbR}{\mathbb{R}}

\newcommand {\IR}{{\rm\kern.24em
   \vrule width.02em height1.53ex depth-.05ex
   \kern-.3em R}}
\newcommand {\ic}{{\rm\kern.20em
   \vrule width.02em height1.0ex depth-.05ex
   \kern-.22em c}}
\newcommand {\ia}{{\rm\kern.20em
   \vrule width.02em height1.05ex depth-.0ex
   \kern-.25em a}}
\newcommand {\IC}{{\rm\kern.24em
   \vrule width.02em height1.4ex depth-.05ex
   \kern-.26em C}}
\newcommand {\ID}{{\rm\kern.34em
   \vrule width.02em height1.5ex depth-.05ex
   \kern-.36em D}}
\newcommand {\IS}{{\rm\kern.24em
   \vrule width.02em height1.6ex depth.05ex
   \kern-.26em S}}
\newcommand {\IT}{{\rm\kern.50em
   \vrule width.02em height1.55ex depth-.05ex
   \kern-.52em T}}

\newcommand {\IE}{{\rm\kern.24em
   \vrule width.02em height1.55ex depth-.05ex
   \kern-.33em E}}
\newcommand {\IEa}{{\rm\kern.24em
   \vrule width.02em height1.55ex depth-.05ex
   \kern-.33em E}^{1}_{ijkl}}
\newcommand {\IEb}{{\rm\kern.24em
   \vrule width.02em height1.55ex depth-.05ex
   \kern-.33em E}^{2}_{ijkl}}

\newcommand {\sC}{\mathcal{C}}
\newcommand {\sD}{\mathcal{D}}

\newcommand {\sF}{\mathcal{F}}
\newcommand {\sG}{\mathcal{G}}

\newcommand {\sQ}{\mathcal{Q}}

\newcommand {\sS}{\mathcal{S}}

\newcommand {\sW}{\mathcal{W}}

\newcommand {\Ass}[2]{\kern 0.9ex \vrule width0.45em height0.2ex depth0ex \kern -2.1ex \bigwedge_{#1}^{#2}}
\newcommand {\ASS}[2]{\kern 1.45ex \vrule width0.5em height0.2ex depth0ex \kern -2.65ex \bigwedge_{#1}^{#2}}

\newcommand{\mrT}{\mathrm{T}}

\newcommand{\bsigs}{\bsig_{\!\mrs}}
\newcommand{\sigs}{\sig}

\begin{document}

\begin{center}
\Large{\bf{A monolithic fluid-structure interaction formulation for solid and liquid membranes including free-surface contact}}\\

\end{center}

\begin{center}

\large{Roger A. Sauer\footnote{corresponding author, email: sauer@aices.rwth-aachen.de} and Tobias Luginsland\footnote{current affiliation: Daimler AG, 71059 Sindelfingen, Germany}}\\
\vspace{4mm}

\small{\textit{
Aachen Institute for Advanced Study in Computational Engineering Science (AICES), \\ RWTH Aachen
University, Templergraben 55, 52056 Aachen, Germany}}

\vspace{4mm}

Published\footnote{This pdf is the personal version of an article whose final publication is available at \href{http://dx.doi.org/10.1016/j.cma.2018.06.024}{www.sciencedirect.com}} 
in \textit{Comput. Methods Appl. Mech. Engrg.}, 
\href{http://dx.doi.org/10.1016/j.cma.2018.06.024}{DOI: 10.1016/j.cma.2018.06.024} \\
Submitted on 28.~March 2017, Revised on 18.~June 2018, Accepted on 20.~June 2018 

\end{center}

\vspace{2mm}

%\doublespacing

\rule{\linewidth}{.15mm}
{\bf Abstract}

A unified fluid-structure interaction (FSI) formulation is presented for solid, liquid and mixed membranes.
Nonlinear finite elements (FE) and the generalized-$\alpha$ scheme are used for the spatial and temporal discretization.
The membrane discretization is based on curvilinear surface elements that can describe large deformations and rotations, and also provide a straightforward description for contact.  
The fluid is described by the incompressible Navier-Stokes equations, and its discretization is based on stabilized Petrov-Galerkin FE.
The coupling between fluid and structure uses a conforming sharp interface discretization, and the resulting non-linear FE equations are solved monolithically within the Newton-Raphson scheme.
An arbitrary Lagrangian-Eulerian formulation is used for the fluid in order to account for the mesh motion around the structure.
The formulation is very general and admits diverse applications that include contact at free surfaces.
This is demonstrated by two analytical and three numerical examples exhibiting strong coupling between fluid and structure. 
The examples include balloon inflation, droplet rolling and flapping flags. 
They span a Reynolds-number range from 0.001 to 2000.
One of the examples considers the extension to rotation-free shells using isogeometric FE.


\vspace{-4mm}
\rule{\linewidth}{.15mm}

\section{Introduction}\label{s:intro}

Fluid-structure interaction (FSI) problems are challenging problems due to various reasons.
They combine the computational challenges of (generally non-linear) fluid and structural mechanics, and
they introduce new challenges, both physical and numerical, due to the coupling.
If the structure is highly flexible, such as a thin membrane, large deformations can be expected.
Those, in turn, have a large influence on the fluid flow. 
A comprehensive overview of FSI and its challenges is given by the monographs of \citet{ohayon04}, \citet{Bazilevs:2013vi} and \citet{Bazilevs:2016dg}. 
The classical focus in FSI problems is on solid structures.
However, some structures are not solids but rather fluids or fluid-like objects.
Examples are liquid menisci, soap films and lipid bilayers.
Lipid bilayers surround biological cells. 
They are characterized by both solid-like (i.e.~elastic bending) and fluid-like behavior (i.e.~in-plane flow).
Further, liquid (and solid) membranes can come into contact with surrounding objects.
A classical example is a liquid droplet rolling on a substrate.
The problem is characterized by fluid flow, surface tension and contact.
\\
While there are various formulations available in the present literature that capture all these aspects, there is no formulation that unifies them all into a single framework.
This is the objective of the present work.
In doing so, we build on our recent computational work on contact, membranes, shells and fluid dynamics.

The presented formulation is based on finite elements (FE) using an interface tracking technique based on a sharp interface formulation.
There is a large literature body on FE-based work on membrane-FSI that is surveyed in the following.
The computational approaches on interactions between fluids and membrane-like structures can be sorted into two groups.
The first group deals with solid structures like elastic membranes and flexible shells, while the second group is concerned with liquid membranes and menisci.
The first group can be further sorted into approaches that use surface formulations (based on shell and membrane theories) and contributions that use bulk formulations.
The second group can be further sorted into approaches that only account for the shape equation in order to characterize the liquid membrane (like the Young-Laplace equation), and approaches that also account for in-plane equations (such as the surface Navier-Stokes equations).
The latter case is necessary for liquid membranes that are not surrounded by a fluid, and consequently the FSI problem is due to the interplay of membrane shape and surface flow.
If a surrounding medium is considered, and no-slip conditions are applied on the membrane surface, the flow within the membrane is already captured by the bulk flow, and so no further equations are needed.
The method presented here is based on a surface formulation that accounts for both shape and in-plane equations.

%- solid membranes \\
%-- Surface formulations:\\
The following references deal with solid membranes using surface formulations.
In \citet{1997IJNMF..24.1091L} the authors employ a deformable spatial domain space-time FEM to study the interaction of an incompressible fluid with an elastic membrane.
\citet{bletzinger06} compute the flow around a tent structure using a staggered coupling between a shell code and a CFD code. 
\citet{Tezduyar:2007eb} review their FSI formulation based on space-time FE and introduce advancements regarding accuracy, robustness and efficiency.
Benchmark examples include the inflation of a balloon, the flow through a flexible diaphragm in a tube as well as a descending parachute.
Parachutes are also analyzed in \citet{karagiozis11} and \citet{Takizawa:2012il} using thin-shell formulations.
\citet{Le:2009eo} developed an implicit immersed boundary method for the incompressible Navier-Stokes equations to simulate membrane-fluid interactions.
Their examples include an oscillating spherical ball immersed in a fluid and the stretching of a red blood cell in a pressure driven shear flow.
\citet{vanOpstal:2015ip} present a hybrid isogeometric finite-element/boundary element method for fluid-structure interaction problems of inflatable structures such as airbags and balloons.
Boundary elements are also used in a recent isogeometric FSI formulation for Stokes flow around thin shells \citep{heltai16}.
\\
%-- Bulk formulations:\\
The following references deal with solid membranes using bulk formulations.
\citet{kloeppel11} numerically investigate the flow inside red blood cells (RBC) by means of monolithically coupling an incompressible fluid to a lipid bilayer represented by incompressible solid shell elements.
In \citet{2016CMAME.298..520F} the authors develop a monolithic strategy for the description of purely Lagrangian FSI problems.
For the solid, the FEM is used, while the fluid is discretized using the so-called Particle FEM \citep{Idelsohn:2004py}.
\citet{Yang:2016gp} introduce a finite-discrete element method for bulk solids and combine the developed numerical model with a finite element multiphase flow model. 
Only 2D examples are considered, such as a rigid structure floating on a liquid-gaseous interface.
\\
Recent reviews on computational FSI methods for solids have been given by \citet{Dowell:2001vu}, \citet{vanLoon:2007kk} and \citet{Bazilevs:2013vi}.
For an introduction to immersed-boundary methods as an alternative to conforming FE discretizations we refer to \citet{Peskin:2003go}.

% -- Liquid membrane formulations:\\
The following references deal with liquid membranes governed only by a shape equation.
\citet{Walkley:2005ui} present an arbitrary Lagrangian-Eulerian (ALE) framework for the solution of free surface flow problems including a dynamic contact line model and show its capabilities for the case of a sliding droplet.
\citet{saksono06b} propose a 2D finite element formulation for  surface tension and apply it to oscillating droplets and stretched liquid bridges.
\citet{Montefuscolo:2014hg} introduce high-order ALE FEM schemes for capillary flows.
The schemes are demonstrated on oscillating and sliding droplets accounting for varying contact angles.
\\
The following references deal with liquid membranes governed by shape and in-plane equations.
\citet{Barett:2015wr} present a numerical study of the dynamics of lipid bilayer vesicles.
A parametric finite element formulation is introduced to discretize the surface Navier-Stokes equations.
\citet{rangarajan15} introduce a spline-based finite-element formulation to compute equilibrium configurations of liquid membranes. 
\citet{liquidshell} present a 3D isogeometric finite element formulation for liquid membranes that accounts for the in-plane viscosity and incompressibility of the liquid.
\\
A general introduction to fluid membranes and vesicles and their configurations observed in nature is given by \citet{Seifert:1997cq}.
For a review on the droplet dynamics within flows, see \citet{Cristini:2004ac}.

There is also earlier work on combining contact and FSI.
It can be grouped into two categories:
Either contact is considered between solids submerged within the fluid (e.g.~see \citet{tezduyar06,mayer10}),
or contact is considered at free liquid surfaces.  
For liquid surfaces the same classical contact algorithms as for solid surfaces can be used \citep{droplet}.
An alternative treatment of free surface contact appears naturally in the Particle FEM \citep{Idelsohn06}. 
Additionally, the contact behavior between liquids and solids is also governed by a contact angle and its hysteresis during sliding contact.
A general computational algorithm for contact angle hysteresis is given in \citet{dropslide}.

Existing work is motivated by specific examples that either focus on solid or liquid membranes. 
The aim of this paper therefore is to provide a new unified FSI formulation that is suitable to describe solid membranes -- such as sheets, fabrics and tissues -- liquid membranes -- such as menisci and soap films -- and membranes with both solid- and liquid-like character, like lipid bilayers.
The formulation is based on a new membrane model that has been recently proposed to unify solid and liquid membranes \citep{membrane}.
The membrane model readily admits general constitutive laws \citep{shelltheo}, it extends to Kirchhoff-Love shells \citep{solidshell} and it is suitable to describe the coupling with other field equations \citep{sahu17}.
Further, the explicit surface formulation of the membrane provides a natural framework for free-surface contact such that any existing contact algorithm can be used.
The present work considers a monolithic coupling scheme between fluid and structure, and solves the resulting non-linear system of equations with the Newton-Raphson method.
Finite elements and the generalized-$\alpha$ scheme are used for the spatial and temporal discretization.
The formulation uses a conforming interface discretization and an ALE formulation for the mesh motion.

Compared to partitioned solvers, monolithic solvers are more complicating to implement (as they require the full tangent matrix and thus need a single code environment). 
But in terms of robustness, monolithic solvers are superior since the coupling between fluid and structure is fully accounted for without further approximation (beyond the usual FE discretization error).  
Also in terms of computational efficiency, recent works have shown that pre-conditioned monolithic solvers are competitive to partitioned ones \citep{Heil08,Kuettler10,Ha17}.
For these reasons the present work uses a monolithic FSI solver.

The following aspects are new in this work:
\begin{itemize}
\item A unified monolithic FSI formulation for liquid and solid membranes is presented.
\item It includes contact on free liquid surfaces, and 
\item it easily extends to rotation-free shells with general constitutive behavior.
\item Two simple analytical FSI examples are presented. 
\item The formulation is suitable for a wide range of applications, including free-surface flows, liquid menisci, flags and flexible wings.
\item The examples include a flow and contact analysis of a rolling 3D droplet.
\end{itemize}

The remainder of this paper is structured as follows. 
Sec.~\ref{s:theo} presents the governing theory of incompressible fluid flow, nonlinear membranes and their coupling.
The theory is used to solve two simple analytical FSI examples in Sec.~\ref{s:ana}.
The computational treatment is then presented in Sec.~\ref{s:FE} using finite elements for the spatial discretization of fluid and membrane,
and the generalized-$\alpha$ scheme for the temporal discretization of the coupled system.
Sec.~\ref{s:ex} presents three numerical examples ranging from very low to quite large Reynolds numbers.
The paper concludes with Sec.~\ref{s:concl}.

\section{Governing equations}\label{s:theo}

This section summarizes the governing equations for fluid flow, membrane deformation, membrane contact and their coupling.
The symbols $\sF$ and $\sS$ are used to denote the fluid domain and the membrane surface, cf.~Fig.~\ref{f:infl_cyl} in Sec.~\ref{s:ana_infl} and Fig.~\ref{f:flag_ex} in Sec.~\ref{s:flag}.

\subsection{Fluid flow}\label{e:theo_f}

The fluid motion is described by an arbitrary Lagrangian-Eulerian (ALE) formulation. 
It is therefore necessary to distinguish between the material motion and the mesh motion. 
An ALE formulation contains the special cases of a purely Lagrangian description, for which the material and mesh motion coincide, and a purely Eulerian description, for which the mesh motion is zero. 

\subsubsection{Fluid kinematics}\label{s:Fkin}

The material motion of a fluid particle $\bX$ within domain $\sF$ is characterized by the deformation mapping
\eqb{l}
\bx = \bvphi(\bX,t)
\eqe
and the corresponding deformation gradient (or Jacobian)
\eqb{l}
\bF := \ds\pa{\bvphi}{\bX}\,.
\label{e:bF}\eqe
The volume change during deformation is captured by the Jacobian determinant $J:=\det\bF$.
The velocity of the material is given by the time derivative of $\bx$ for fixed $\bX$, written as
\eqb{l}
\bv := \ds\pa{\bx}{t}\Big|_{\bX}
\label{e:bv}\eqe
and commonly referred to as the \textit{material time derivative}. 
It is also often denoted by the dot notation $\bv=\dot\bx$.
An important object characterizing the fluid flow is the velocity gradient
\eqb{l}
\bL := \nabla\bv = \ds\pa{\bv}{\bx}
\label{e:bL}\eqe
that can also be written as $\bL = \dot\bF\bF^{-1}$, where $\dot\bF$ is the material time derivative of the deformation gradient.
The symmetric part of the velocity gradient is denoted by $\bD:=\big(\bL+\bL^\mrT\big)/2$.
\\
Likewise to Eq.~\eqref{e:bv}, the material acceleration is given by
\eqb{l}
\ba:=\dot\bv = \ds\pa{\bv}{t}\Big|_{\bX}\,.
\label{e:ba}\eqe
It is related to the acceleration for fixed $\bx$,
\eqb{l}
\bv' := \ds\pa{\bv}{t}\Big|_{\bx}\,,
\label{e:bvprime}\eqe
according to
\eqb{l}
\dot\bv = \bv' + \bL\,(\bv-\bv_\mrm)\,,
\label{e:bvdot}\eqe
where $\bv_\mrm$ is the mesh velocity \citep{donea}.
For a purely Lagangian description $\bv_\mrm = \bv$, while for a purely Eulerian description $\bv_\mrm = \mathbf{0}$.

\textbf{Remark 2.1}: The gradient operator appearing in Eq.~\eqref{e:bL} (and likewise in Eq.~\eqref{e:bF}), is defined here as $\nabla\bv:=v_{i,j}\,\be_i\otimes\be_j$.\footnote{Following index notation, summation is implied on repeated indices. 
Latin indices range from 1 to 3 and refer to Cartesian coordinates. 
Greek indices range from 1 to 2 and refer to curvilinear surface coordinates.}
In matrix notation this corresponds to the square $3\times3$ matrix $[v_{i,j}]$.

\subsubsection{Fluid equilibrium}

From the balance of linear momentum within $\sF$ follows the equilibrium equation
\eqb{l}
\divz\bsig + \bar\bff = \rho\,\dot\bv \quad $in $\sF\,,
\label{e:sf_f}\eqe
which governs the fluid flow together with the boundary conditions
\eqb{rlll}
\bv \is \bar\bv ~& $on $\partial_x\sF\,, \\[2mm]
\bsig\bn = \bt \is \bar\bt ~& $on $\partial_t\sF\,.
\label{e:bc_f}\eqe
Here, $\bsig$ denotes the stress tensor within $\sF$, $\bt$ denotes the traction vector on the surface characterized by normal vector $\bn$, and $\rho$ denotes the fluid density, while $\bar\bff$, $\bar\bv$ and $\bar\bt$ are prescribed body forces, surface velocities and surface tractions. 
$\partial_x\sF$ and $\partial_t\sF$ denote the corresponding Dirichlet and Neumann boundary regions of the fluid domain $\sF$.
Boundary $\partial_x\sF$ can be split into the two parts
\eqb{l}
\partial_x\sF = \sS \cup \partial_{\hat x}\sF\,,
\eqe
where $\sS$ is the surface of the membrane, which is considered to impose its velocity onto the fluid, and $\partial_{\hat x}\sF$ denotes the remaining Dirichlet boundary of the fluid domain.
In order to solve PDE \eqref{e:sf_f} for $\bv(\bx,t)$, the initial condition
\eqb{l}
\bv(\bx,0) = \bv_0(\bx)
\eqe
is needed.

\subsubsection{Fluid constitution}

We consider an incompressible Newtonian fluid with kinematic viscosity $\nu$ and dynamic viscosity $\eta=\nu\rho$.
In that case the stress tensor is given by
\eqb{l}
\bsig = -p\,\bone + 2\eta\,\bD\,,
\eqe
where $p$ is the Lagrange multiplier to the incompressibility constraint
\eqb{l}
g := J - 1 = 0\,,
\label{e:g1}\eqe
which is equivalent to the condition
\eqb{l}
\divz\bv = 0\,.
\label{e:g2}\eqe
A consequence of this condition is that the fluid pressure, defined as $-\tr\bsig/3$, is equal to the Lagrange multiplier $p$.
It is an additional unknown that needs to be solved for together with $\bv$.
In case of pure Dirichlet boundary conditions ($\partial_t\sF=\emptyset$), the value of $p$ needs to be specified at one point in the fluid domain in order for the pressure field to be uniquely determinable.

\subsubsection{Fluid weak form}\label{s:wfF}

In order to solve the problem with finite elements the strong form equations \eqref{e:sf_f}, (\ref{e:bc_f}.2) and \eqref{e:g2} are reformulated in weak form.
They are therefore multiplied by the test functions $\bw$ and $q$, and integrated over the domain $\sF$. 
Function $\bw$ is assumed to be zero on the Dirichlet boundary $\partial_{\hat x}\sF$, but non-zero on the surface $\sS$.
Functions $\bw$ and $q$ are further assumed to possess sufficient regularity for the following integrals to be well defined.
In the framework of SUPG\footnote{Streamline upwind/Petrov-Galerkin \citep{Brooks82}} and PSPG\footnote{Pressure stabilizing/Petrov-Galerkin \citep{Hughes86}} stabilization, the weak form takes the form
\eqb{rll}
G_\sF := G_{\sF\mathrm{in}} + G_{\sF\mathrm{int}} + G_\mathrm{supg} - G_{\sF\mrs} - G_{\sF\mathrm{ext}} \is 0 \quad \forall~\bw\in\sW\,, \\[1mm]
G_\sG := G_\mrg + G_\mathrm{pspg} \is 0 \quad \forall~q\in\sQ\,,
\label{e:wfF}\eqe
where
\eqb{l}
G_{\sF\mathrm{in}} := \ds\int_\sF \bw \cdot \rho\,\dot\bv\,\dif v
\eqe
is the virtual work associated with inertia,
\eqb{l}
G_{\sF\mathrm{int}} := \ds\int_\sF \nabla\bw : \bsig\,\dif v
\eqe
is internal virtual work,
\eqb{l}
G_{\sF\mrs} := \ds\int_{\sS} \bw \cdot \bt\,\dif a
\eqe
is the virtual work of the fluid traction $\bt=\bsig\bn$ on boundary $\sS$,
\eqb{l}
G_{\sF\mathrm{ext}} := \ds\int_\sF \bw \cdot \bar\bff\,\dif v + \int_{\partial_t\sF} \bw \cdot \bar\bt\,\dif a
\eqe
is the external virtual work\footnote{In the following examples we consider zero Neumann BC ($\bar\bt=\mathbf{0}$) and constant gravity loading with 
$\bar\bff=\rho\,\bg$.},
\eqb{l}
G_\mrg := \ds\int_\sF q\,\divz\bv\,\dif v
\eqe
is the virtual work associated with incompressibility constraint \eqref{e:g2},
\eqb{l}
G_\mathrm{supg} := \ds\int_\sF \tau_\mrv\,\bff_{\!\mathrm{res}} \cdot \nabla \bw\,(\bv-\bv_\mrm)\,\dif v
\eqe
is the SUPG term,
\eqb{l}
G_\mathrm{pspg} := \ds\int_\sF \tau_\mrp \nabla q \cdot \bff_{\!\mathrm{res}}\,\dif v 
\eqe
is the PSPG term, and
\eqb{l}
\bff_{\!\mathrm{res}} := \rho\,\dot\bv - \divz\bsig - \bar\bff
\eqe
is the residual of Eq.~\eqref{e:sf_f}. 
Dimensionally, the residual is a force per volume.
Since in theory $\bff_{\!\mathrm{res}} = \mathbf{0}$, stabilization terms $G_\mathrm{supg}$ and $G_\mathrm{pspg}$ do not affect the physical behavior of the system.
In Cartesian coordinates $\bff_{\!\mathrm{res}} \cdot \nabla \bw\,(\bv-\bv_\mrm) = f_i^\mathrm{res}\,w_{i,j}\,(v_j-v_{\mrm j})$.
The scalars $\tau_\mrv$ and $\tau_\mrp$ are stabilization parameters that are discussed in Sec.~\ref{s:FE}.

\subsection{Deforming membranes}\label{s:theo_s}

This work focuses on pure membranes that do not resist bending and out-of-plane shear. 
The description of those membranes is based on the formulation of \citet{membrane}, which admits both solid and liquid membranes. 
What follows is a brief summary.

\subsubsection{Membrane kinematics}

The motion of a membrane surface $\sS$ is fully described by the mapping
\eqb{l}
\bx = \bx(\xi^\alpha,t)\,,
\label{e:bx}\eqe
where $\xi^\alpha$, for $\alpha=1,2$, are curvilinear coordinates that can be associated with material points on the surface. 
They can be conveniently taken from the parameterization of the finite element shape functions.
Based on mapping \eqref{e:bx}, the tangent vectors $\ba_\alpha:=\partial\bx/\xi^\alpha$ to surface $\sS$, the metric tensor components $a_{\alpha\beta}:=\ba_\alpha\cdot\ba_\beta$,\footnote{following the notation where $g_{ij}$ is the metric in the bulk, and $a_{\alpha\beta}$ is the metric on the surface} and the surface normal $\bn = \ba_1\times\ba_2/\sqrt{\det[a_{\alpha\beta}]}$ can be determined.
From the matrix inverse $[a^{\alpha\beta}]=[a_{\alpha\beta}]^{-1}$, the dual tangent vectors $\ba^\alpha:=a^{\alpha\beta}\ba_\beta$ can be defined such that $\ba^\alpha\cdot\ba_\beta$ is equal to the Kronecker delta~$\delta^\alpha_\beta$.\\
In order to characterize deformation, a stress-free reference configuration $\sS_0$ is introduced.
It will be considered here as the initial membrane surface, i.e.~$\sS_0:=\sS|_{t=0}$.
In the reference configuration the tangent vectors, metric tensor components, inverse components and normal vector are denoted by capital letters, i.e.~$\bA_\alpha$, $A_{\alpha\beta}$,  $A^{\alpha\beta}$ and $\bN$.
The in-plane deformation of surface $\sS$ is fully characterized by the relation between $A^{\alpha\beta}$ and $a^{\alpha\beta}$.
The surface stretch for instance is given by $J_\mrs := \sqrt{\det[a_{\alpha\beta}]/\det[A_{\alpha\beta}]}$.
\\
Following definitions~\eqref{e:bv} and \eqref{e:ba}, the membrane velocity $\bv$ and acceleration $\ba$ are obtained from Eq.~\eqref{e:bx}.

\subsubsection{Membrane equilibrium}

From the balance of linear momentum within $\sS$ follows the equilibrium equation
\eqb{l}
(\bsigs\,\ba^\alpha)_{;\alpha} + \bff_{\!\mrs} = \rho_\mrs\,\dot\bv
\quad $in $\sS\,,
\label{e:sf_s}\eqe
which governs the membrane deformation together with the boundary conditions
\eqb{rlll}
\bx \is \bar\bx & $for $\bx\in\partial_x\sS\,, \\[1mm]
\bsigs\,\bnu = \bt_\mrs \is \bar\bt_\mrs & $for $\bx\in\partial_t\sS\,,
\label{e:bc_s}\eqe
e.g.~see \citet{shelltheo}.
Here, $\bsigs$ denotes the stress tensor within $\sS$, $(...)_{;\alpha}$ denotes the covariant derivative w.r.t.~$\xi^\alpha$,
$\bt_\mrs$ denotes the traction vector on the membrane boundary characterized by normal vector $\bnu$, and $\rho_\mrs$ denotes the membrane density, while $\bar\bx$ and $\bar\bt_\mrs$ are prescribed boundary velocities and boundary tractions. 
The body force $\bff_{\!\mrs}$ is considered here to have contributions coming from the flow field, contact and external sources, i.e.
\eqb{l}
\bff_{\!\mrs} = \bff_{\!\mrf} + \bff_{\!\mrc} + \bar\bff_{\!\mrs}\,.
\eqe
In order to solve PDE \eqref{e:sf_s} for $\bx(\xi^\alpha,t)$, the initial conditions
\eqb{lll}
\bx(\xi^\alpha,0) \is \bX(\xi^\alpha)\,,\\[1mm]
\bv(\xi^\alpha,0) \is \bv_0(\xi^\alpha)\,,
\eqe
are needed.

\subsubsection{Membrane constitution}

For pure membranes, the stress tensor only has in-plane components, i.e.~it has the format
$\bsig_{\!\mrs} = \sig^{\alpha\beta}\,\ba_\alpha\otimes\ba_\beta$.
Two material models are considered in this work.
The first,
\eqb{l}
\sigs^{\alpha\beta} = \ds\frac{\mu}{J_\mrs}\bigg(A^{\alpha\beta} - \frac{1}{J^2_\mrs}\,a^{\alpha\beta}\bigg)\,,
\label{e:sig_sol}\eqe
is suitable for solid membranes. 
It can be derived from the 3D incompressible Neo-Hookean material model \citep{membrane}.
The second,
\eqb{l}
\sigs^{\alpha\beta} = \gamma\,a^{\alpha\beta}\,,
\label{e:sig_liq}\eqe
models isotropic surface tension, and is suitable to describe liquid membranes, e.g.~see \citet{droplet}.
The parameters $\mu$ and $\gamma$ denote the shear stiffness and the surface tension, respectively. 
Both are considered constant here.

\subsubsection{Membrane contact}

This work also considers that sticking contact can occur on the membrane surface $\sS_\mrc\subset\sS$. 
During sticking contact no relative motion occurs between the membrane and a neighboring substrate surface $\sS_\mathrm{sub}$.
Mathematically this corresponds to the constraint
\eqb{l}
\bg = \mathbf{0}\quad\forall\,\bx\in\sS_\mrc\,,
\label{e:bgc}\eqe
where
\eqb{l}
\bg(\bx) = \bx-\bx^0_\mrp
\eqe
denotes the contact gap between the membrane point $\bx\in\sS_\mrc$ and its initial projection point on the substrate surface, $\bx_\mrp^0\in\sS_\mathrm{sub}$, i.e.~$\bx^0_\mrp$ is the location where $\bx$ initially touched $\sS_\mathrm{sub}$.
Here, constraint \eqref{e:bgc} will be enforced by a penalty regularization.
For this, the contact traction at $\bx\in\sS$ is given by
\eqb{l}
\bff_{\!\mrc} = \left\{\begin{array}{ll}
-\epsilon\,\bg & $if $\bg\cdot\bn_\mrc < 0\,, \\[1mm]
\mathbf{0} & $else$\,,
\end{array}\right.
\label{e:fc}\eqe
where $\bn_\mrc$ is the surface normal of $\sS_\mathrm{sub}$. 
Instead of the penalty formulation, also any other contact formulation can be used to enforce \eqref{e:bgc}.
Further details on large deformation contact theory can be found in the textbooks of \citet{laursen} and \citet{wriggers-contact}.

\subsubsection{Membrane weak form}

In order to employ finite elements, the strong form equations \eqref{e:sf_s} and (\ref{e:bc_s}.2) are reformulated in weak from.
As shown in \citet{shelltheo}, the weak form for the membrane can be written as
\eqb{l}
G_\sS := G_{\sS\mathrm{in}} + G_{\sS\mathrm{int}} + G_\mrc - G_{\sS\mrf} - G_{\sS\mathrm{ext}} = 0\quad\forall~\bw\in\sW\,,
\label{e:wfS}\eqe
with the virtual work contributions
\eqb{rll}
G_{\sS\mathrm{in}} \dis \ds\int_\sS\bw\cdot\rho_\mrs\,\dot\bv\,\dif a\,, \\[4mm]
G_{\sS\mathrm{int}} \dis \ds\int_{\sS}\sigs^{\alpha\beta}\,\bw_{;\alpha}\cdot\ba_\beta\,\dif a\,, \\[4mm]
G_\mrc \dis-\ds\int_\sS\bw\cdot\bff_{\!\mrc}\,\dif a \,, \\[4mm]
G_{\sS\mrf} \dis \ds\int_\sS\bw\cdot\bff_{\!\mrf}\,\dif a \,, \\[4mm]
G_{\sS\mathrm{ext}} \dis \ds\int_\sS\bw\cdot\bar\bff_{\!\mrs}\,\dif a
+ \int_{\partial_t\sS} \bw\cdot\bar\bt_\mrs\,\dif s\,,
\label{e:Gicfe}\eqe
due to inertia, internal forces, contact forces, fluid forces and external forces acting on $\sS$ and $\partial_t\sS$.
Test function $\bw$ is the same as in \eqref{e:wfF}. 
Therefore, space $\sW$ needs to additionally satisfy the requirement that all integrals appearing above are well defined. 
Further $\bw$ is assumed to be zero on $\partial_x\sS$. \\
Pure membranes are inherently unstable in the quasi-static case ($\bv=\dot\bv=\mathbf{0}$) and therefore need to be stabilized \citep{membrane,droplet}. 
Here, no stabilization is required as the fluid forces $\bff_\mrf$ stabilize the membrane, even when $\rho_\mrs=0$ (as is considered in some of the following examples).
In the numerical examples following later, $\bar\bff_{\!\mrs}$ and $\bar\bt_\mrs$, and consequently $G_{\sS\mathrm{ext}}$, are considered zero.

\textbf{Remark 2.2}: It is straight forward to extend weak form~\eqref{e:wfS} to Kirchhoff-Love shells: 
$G_{\sS\mathrm{int}}$ and $G_{\sS\mathrm{ext}}$ simply need to be extended by the bending moments acting within $\sS$ and on $\partial\sS$, e.g.~see \citet{solidshell}. 
Kirchhoff-Love shells are suitable for thin membrane-like surface structures.
Such a structure is considered in Sec.~\ref{s:flag} using isogeometric finite elements.

\subsection{Coupling conditions}

The membrane deformation $\bx$ moves the fluid such that
\eqb{l}
\bv = \dot\bx~~$on $\sS 
\label{e:coupx}\eqe
is a Dirichlet BC for the fluid.
This choice assumes no tangential slip between membrane and fluid.
In response, the flow exerts a traction on the membrane such that
\eqb{l}
\bff_{\!\mrf} = -\bt~~$on $\sS 
\label{e:coupt}\eqe
is a `body force' of the membrane.
Eq.~\eqref{e:coupx} is the kinematic coupling condition between the two domains, while Eq.~\eqref{e:coupt} is the kinetic coupling condition.
If the membrane is surrounded by fluid on both sides, $\bt$ in \eqref{e:coupt} is replaced by the traction jump $[\![\bt]\!]:=\bt^+-\bt^-$, where $\bt^+$ is the traction on the front side (with outward normal $\bn$) and $\bt^-$ is the traction on the back side (with outward normal $-\bn$) of the membrane.
The combined FSI problem is then characterized by the two governing equations
\eqb{lll}
G_\sF + G_\sS  \is 0\quad\forall~\bw\in\sW\,, \\[1mm]
G_\sG \is 0\quad\forall~q\in\sQ\,,
\label{e:wf}\eqe
which can be solved for the unknown velocity $\bv$ and pressure $p$ in $\sF$. 
The membrane deformation can then be obtained from integrating $\bv$. 
Coupling condition \eqref{e:coupt} simply leads to the cancelation of terms $G_{\sF\mrs}$ and $G_{\sS\mrf}$ in the combined weak form \eqref{e:wf}.
This cancelation will carry over to the discretized weak form, as long as surface $\sS$ is discretized conformingly on the fluid and membrane side.

\section{Analytical examples}\label{s:ana}

This section presents the analytical solution of two simple examples. They serve as verification examples for the computational implementation discussed later.

\subsection{Solid membrane example: Fluid-inflated cylinder}\label{s:ana_infl}

\begin{figure}[h]
\begin{center} \unitlength1cm
\begin{picture}(0,5)
\put(-4.8,-.2){\includegraphics[height=50mm]{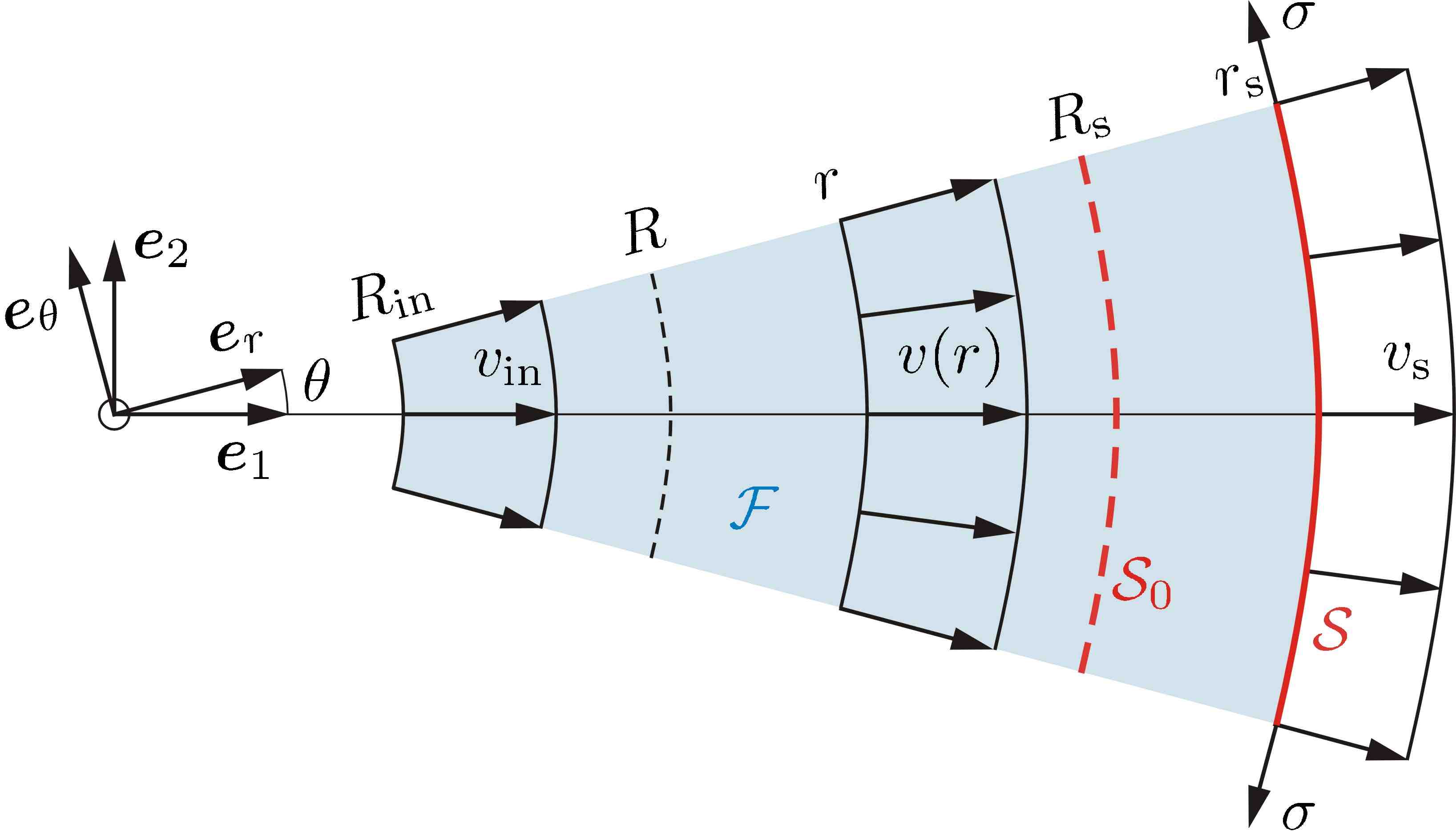}}
\end{picture}
\caption{Fluid-inflated cylinder: Membrane deformation $\sS_0\rightarrow\sS$ and fluid velocity $v(r)$ due to a radial inflow at $R_\mathrm{in}$.} 
\label{f:infl_cyl}
\end{center}
\end{figure}

As a first example we consider the radial inflation of a membrane cylinder due to a constant radial inflow as is illustrated in Fig.~\ref{f:infl_cyl}.
%-------------------------------------------------------------------------------------------------------------------------------
%-------------------------------------------------------------------------------------------------------------------------------
The example is chosen since it can be fully solved analytically and thus used for verification of the computational formulation, which is then considered in Sec.~\ref{s:ex1}.
Given the inflow velocity $v_\mathrm{in}$ at the inner boundary $R_\mathrm{in}$, the radial fluid velocity at location $r$ is given by
\eqb{l}
v(r) = \ds\frac{v_\mathrm{in}\, R_\mathrm{in}}{r}
\label{e:ana1_v}
\eqe
due to continuity.
Since $v=\dot r$, we obtain
\eqb{l} 
%r(R,t) = \sqrt{R^2+2v_\mathrm{in}(t)\,R_\mathrm{in}\,t}\,,
r(R,t) = \sqrt{R^2+2v_\mathrm{in}\,R_\mathrm{in}\,t}\,,
\label{e:ana1_r}
\eqe 
as the current position of the fluid particle initially at $R$.
The current membrane position is thus given by $r_\mrs=r(R_\mrs,t)$, where $R_\mrs$ is the initial position of the membrane.
In vectorial notation, the flow field can thus be characterized by the position, velocity and acceleration
\eqb{lll}
\bx(R,t) \is r\,\be_r\,, \\[1mm]
\bv(R,t) \is v\,\be_r\,, \\[1mm]
\ba(R,t) \is -\ds\frac{v^2}{r}\,\be_r\,,
\eqe
where $\be_r=\cos\theta\,\be_1+\sin\theta\,\be_2$ is the radial unit vector.
From this follows
\eqb{l}
\bD %=  -\sin^2\theta\, (\be_1\otimes\be_1) -\cos^2\theta\, (\be_2\otimes\be_2) -2\sin\theta\,\cos\theta\, (\be_1\otimes\be_2 + \be_2\otimes\be_1)\,,
= \displaystyle \frac{v}{r} \big( \bar\bone - 2\, \boldsymbol{e}_r \otimes\boldsymbol{e}_r \big)\,,\quad 
%$with$~~\bar\bone := \be_1\otimes\be_1 + \be_2\otimes\be_2 \,,
\eqe
with the 2D identity $\bar\bone := \be_1\otimes\be_1 + \be_2\otimes\be_2$,
such that $\divz\bD=\mathbf{0}$. 
The equation of motion thus reduces to $-\nabla p=\rho\,\ba$, which can be integrated to give the pressure field
\eqb{l}
p(R,t) = p_\mrs + \ds\frac{\rho}{2}\big(v_\mrs^2-v^2\big)\,,
\label{e:ana1_p}
\eqe
where $v_\mrs=v(r_\mrs)$ is the current membrane velocity, and $p_\mrs$ is the pressure acting on the membrane.
Neglecting membrane inertia, this pressure equilibrates the membrane stress
\eqb{l}
\sigma = \mu\,\bigg(\lambda - \ds\frac{1}{\lambda^3} \bigg)
\label{e:ana1_sig}
\eqe
caused by the membrane stretch $\lambda=r_\mrs/R_\mrs$ according to Eq.~(\ref{e:sig_sol}); see Appendix~\ref{s:ana_mem}.
From $p_\mrs = \sigma/r_\mrs$ follows
\eqb{l}
p_\mrs = \ds\frac{\mu}{R_\mrs} \left [1-\left(\frac{R_\mrs}{r_\mrs}\right)^4  \right]\,.
\eqe

\subsection{Liquid membrane example: Spinning droplet}\label{s:ana_spin}

As a second example we consider a spinning droplet.
This example is considered for comparison with the computational example of a rolling droplet in Sec.~\ref{s:ex2}. 
At very small length scales the influence of gravity is negligible, so that a rolling droplet remains approximately spherical.
Considering the axis of rotation to be $\be_2$, the motion of a spinning droplet can be expressed as
\eqb{l}
\bx(r,t) = r\,\be_r\,,
\eqe
where $\be_r=\cos\theta\,\be_1 - \sin\theta\,\be_3$, $\theta = \omega t$ and $\omega$ denotes the angular velocity around $\be_2$.
Consequently,
\eqb{lll}
\bv(r,t) \is \omega\,r\,\be_\theta\,, \\[1mm]
\ba(r,t) \is -\omega^2 r\,\be_r\,,
\eqe
where $\be_\theta=-\sin\theta\,\be_1 - \cos\theta\,\be_3$. 
Since we can write $x_1=r\cos\theta$ and $x_2=-r\sin\theta$, we find $\nabla\bv=\omega(\be_1\otimes\be_3-\be_3\otimes\be_1)$ such that $\bD=\mathbf{0}$ and
\eqb{l}
\bsig = -p\,\bone\,.
\eqe
The spin tensor, defined as $\bW:=\big(\bL-\bL^T\big)/2$, then becomes $\bW=\bL=\nabla\bv$. 
The axial vector of $\bW$, denoted by $\bome$, thus is $\bome =\omega\,\be_2$. 
It denotes the orientation and magnitude of the droplet's spin, and it is equal to half of the vorticity $\nabla\times\bv$.
Solving Eq.~\eqref{e:sf_f} (with $\bar\bff=\mathbf{0}$) for $p$ now gives
\eqb{l}
p(r) = \ds\omega^2\rho\frac{r^2}{2} + p_0\,.
\eqe
The constant $p_0$ follows from the boundary condition $p(r_0)=2\gamma/r_0$, where $\gamma$ is the surface tension of the droplet and $r_0$ is the droplet radius.
This condition enforces the Young-Laplace equation, which is contained inside Eq.~\eqref{e:sf_s}, see \citet{droplet}. 
Applying the boundary condition, we find
\eqb{l}
p(r) = \ds\frac{2\gamma}{r_0} - \frac{\rho\,\omega^2}{2}\big(r_0^2-r^2\big)\,.
\eqe
If desired, the constant velocity $\bv_0=\omega\,r_0\,\be_1$ can be added to $\bv(r,t)$, such that the resulting velocity is zero at the contact point (where $\theta=\pi/2$).

\section{Finite element formulation}\label{s:FE}

The coupled fluid-membrane problem of Sec.~\ref{s:theo} is solved with the finite element method using the generalized-$\alpha$ scheme. 
This section presents the required discretization steps and the resulting algebraic equations.

\subsection{Spatial discretization}

The computational domain is discretized into $n_\mathrm{el}$ finite elements, numbered $e=1,...,n_\mathrm{el}$. 
Some of these elements are 3D fluid elements, others are 2D surface elements or 1D line elements.
Element $e$ contains $n_e$ nodes and occupies the domain $\Omega_e$ in the current configuration. 
Each fluid element has four degrees-of-freedom (dofs) per node (three velocity components and a pressure), while the membrane elements each have three unknown displacements per node.
Each fluid element therefore contributes $4n_e$ force components, while each membrane element contributes $3n_e$ force components that need to be assembled into the global system. 
Those elemental forces are discussed in the following two sections.

\subsubsection{Fluid flow}

\textit{4.1.1.1 Basic flow variables}

Within a fluid element, the fluid velocity is approximated by the interpolation
\eqb{l}
\bv \approx \bv^h = \ds\sum_{I=1}^{n_e}N_I\,\bv_I \,,
\eqe
where $N_I$ and $\bv_I$ are the nodal shape function and nodal velocity, respectively.
In short, this can also be written as
\eqb{l}
\bv \approx \bv^h = \mN\,\mv_e \,,
\label{e:bvh}\eqe
where $\mN:=[N_1\bone,\,N_2\bone,\,...,\,N_{n_e}\bone]$ and $\mv_e := [\bv_1,\,\bv_2,\,...,\,\bv_{n_e}]^\mrT$.
The corresponding test function (or variation) is approximated in the same fashion, i.e.
\eqb{l}
\bw \approx \bw^h = \mN\,\mw_e \,.
\label{e:bwh}\eqe
The fluid pressure is approximated by the interpolation
\eqb{l}
p \approx p^h = \tilde\mN\,\mpp_e \,,
\eqe
where $\tilde\mN:=[N_1,\,N_2,\,...,\,N_{n_e}]$.
Likewise,
\eqb{l}
q \approx q^h =  \tilde\mN\,\mq_e \,.
\eqe
The structure of \eqref{e:bvh} is also used to interpolate the mesh motion, i.e.
\eqb{l}
\bv_\mrm \approx \bv^h_\mrm = \mN\,\mv_{\mrm e} \,.
\label{e:bvm}\eqe
In the present work, the $\mv_{\mrm e}$ are not treated as unknowns. Instead they will be defined through the membrane motion.

\textit{4.1.1.2 Derived flow variables}

As a consequence of the above expressions, we find the approximation of the acceleration (from Eq.~\eqref{e:bvdot})
\eqb{l}
\dot\bv \approx \dot\bv^h = \mN\,\mv'_e + \bL\mN\big(\mv_e-\mv_{\mrm e}\big)\,,
\eqe
the velocity gradient
\eqb{l}
\bL \approx \bL^h = \ds\sum_{I=1}^{n_e} \bv_I \otimes \nabla N_I\,, 
\eqe
the pressure gradient
\eqb{l}
\nabla p \approx \nabla p^h =  \mG\,\mpp_e\,, 
\eqe
and the velocity divergence
\eqb{l}
\divz\bv \approx \divz\bv^h = \mD\,\mv_e\,, 
\eqe
where
\eqb{l}
\nabla N_I = \left[\begin{matrix}
  N_{I,1} \\
  N_{I,2} \\
  N_{I,3}
\end{matrix}\right],
\eqe
$\mG := [\nabla N_1,\,\nabla N_2,\,...,\,\nabla N_{n_e}]$ and $\mD := [(\nabla N_1)^\mrT,\,(\nabla N_2)^\mrT,\,...,\,(\nabla N_{n_e})^\mrT]$.
Further, we introduce the classical B-matrix $\mB := [\mB_1,\,\mB_2,\,...,\,\mB_{n_e}]$, with
\eqb{l}
\mB_I := \left[\begin{matrix}
  N_{I,1} & 0 & 0\\
  0 & N_{I,2} & 0 \\
  0 & 0 & N_{I,3} \\
  0 & N_{I,3} & N_{I,2}\\
  N_{I,3} & 0 & N_{I,1} \\
  N_{I,2} & N_{I,1} & 0
\end{matrix}\right],
\eqe
in order to express the symmetric velocity gradient and its corresponding variation in Voigt notation (indicated by index `v') as
\eqb{rllll}
\nabla^s\bv_\mrv \ais \nabla^s\bv_\mrv^h \is \mB\,\mv_e\,, \\[1mm]
\nabla^s \bw_\mrv \ais \nabla^s\bv_\mrw^h \is \mB\,\mw_e\,,
\eqe
i.e.~arranged as $\nabla^s\bv_\mrv := [v_{1,1},\,v_{2,2},\,v_{3,3},\,v_{2,3}+v_{3,2},\,v_{1,3}+v_{3,1},\,v_{1,2}+v_{2,1}]^\mrT$.
The stress tensor, arranged as $\sig_\mrv:=[\sig_{11},\,\sig_{22},\,\sig_{33},\,\sig_{23},\,\sig_{13},\,\sig_{12}]$, can thus be written as
\eqb{l}
\bsig_\mrv \approx \bsig_\mrv^h = \bbC\,\mB\,\mv_e- \bone_\mrv\,\tilde\mN\,\mpp_e\,,
\eqe
with $\bbC:=\mathrm{diag}(2\eta\bone,\,\eta\bone)$ and $\bone_\mrv = [1,\,1,\,1,\,0,\,0,\,0]^\mrT$. 
Here, $\bone$ is the usual identity tensor in $\bbR^3$.
Due to the symmetry of the stress and since $\mB^\mrT\bone_\mrv = \mD^\mrT$, the integrand of $G_{\sF\mathrm{int}}$ becomes
\eqb{l}
\nabla\bw^h:\bsig^h = \mw_e^\mrT\,\mB^\mrT\,\bbC\,\mB\,\mv_e - \mw_e^\mrT\,\mD^\mrT\,\tilde\mN\,\mpp_e
\eqe
within element $\Omega^e$. 
\\
In order to represent the SUPG term, we introduce the arrays $\mB_\mrf := [\mB_{\mrf1},\,\mB_{\mrf2},\,...,\,\mB_{\mrf n_e}]$, with the $3\times 3$ blocks
\eqb{l}
\mB_{\mrf I} := \nabla N_I\otimes\bff_{\!\mathrm{res}}\,,
\eqe
and 
$\mB_\mrv := [B_{\mrv1}\bone,\,B_{\mrv2}\bone,\,...,\,B_{\mrv n_e}\bone]$, with 
\eqb{l}
B_{\mrv I} := \nabla N_I\cdot(\bv-\bv_\mrm)\,.
\eqe
The last term can also be used to rewrite the $\bL(\bv-\bv_\mrm)$ term as
\eqb{l}
\bL^h\,(\bv^h-\bv^h_\mrm) = \mB_\mrv\mv_e\,.
\eqe

\textit{4.1.1.3 Weak form contribution of a fluid element}

Given the above expressions, the contributions from element $\Omega^e$ to the fluid weak form \eqref{e:wfF} can be written as
\eqb{l}
G_\sF^e+G^e_\sG = \mw_e^\mrT\,\mf^e_\sF+\mq_e^\mrT\mg^e\,,
\label{e:GF}\eqe 
with the ($3n_e\times1$) FE force vector
\eqb{l}
\mf^e_\sF :=  \left\{
\begin{array}{ll}
\mf^e_{\sF\mathrm{in}}+\mf^e_{\sF\mathrm{int}}+\mf^e_\mathrm{supg}-\mf^e_{\sF\mathrm{ext}\bar f} ~& $for $\Omega^e\subset\sF^h\,, \\[2mm]
-\mf^e_{\sF\mathrm{ext}\bar t} ~& $for $\Omega^e\subset\partial_t\sF^h\,, \\[2mm]
-\mf^e_{\sF\mrs} ~& $for $\Omega^e\subset\sS^h\,,
\end{array}\right.
\label{e:f_eFd}
\eqe
and the ($n_e\times1$) FE pseudo force vector
\eqb{l}
\mg^e := \mg^e_\mrg + \mg^e_\mathrm{pspg}\,.
\eqe
They are composed of the FE forces
\eqb{lll}
\mf^e_{\sF\mathrm{in}} \dis \mm_e\,\mv'_e + \mf^e_\mathrm{con}\,, \\[3mm]
\mf^e_\mathrm{con} \dis \ds\int_{\Omega^e}\rho\,\mN^\mrT\mB_\mrv\mv_e\,\dif v\,, \\[4mm]
\mf^e_{\sF\mathrm{int}} \dis \mcc_e\,\mv_e - \md_e\,\mpp_e\,, \\[3mm]
\mf^e_\mathrm{supg} \dis \ds\int_{\Omega^e} \tau_\mrv\,\mB^\mrT_\mrf(\bv-\bv_\mrm)\,\dif v = \int_{\Omega^e} \tau_\mrv\,\mB^\mrT_\mrv\bff_{\!\mathrm{res}}\,\dif v\,, \\[4mm]
\mf^e_{\sF\mrs} \dis \ds\int_{\Omega^e}\mN^\mrT\,\bt\,\dif a \,, \\[4mm]
\mf^e_{\sF\mathrm{ext}\bar f} \dis \ds\int_{\Omega^e}\mN^\mrT\,\bar\bff\,\dif v\, \\[4mm] 
\mf^e_{\sF\mathrm{ext}\bar t} \dis \ds\int_{\Omega^e}\mN^\mrT\,\bar\bt\,\dif a\,,
\label{e:f_eF}\eqe
the FE pseudo forces
\eqb{lll}
\mg^e_\mrg \dis \md_e^\mrT\,\mv_e\,, \\[3mm]
\mg^e_\mathrm{pspg} \dis \ds\int_{\Omega^e}\tau_\mrp\,\mG^\mrT\bff_{\!\mathrm{res}}\,\dif v\,,
\label{e:f_eG}\eqe
and the elemental mass, damping and pressure-force matrices
\eqb{lll}
\mm_e \dis \ds\int_{\Omega^e}\rho\,\mN^\mrT\mN\,\dif v\,, \\[4mm]
\mcc_e \dis \ds\int_{\Omega^e}\mB^\mrT\,\bbC\,\mB\,\dif v\,, \\[4mm]
\md_e \dis \ds\int_{\Omega^e}\mD^\mrT\tilde\mN\,\dif v\,.
\label{e:mcd_e}\eqe
The tangent matrices of $\mf^e_\sF$ and $\mg^e$, needed for linearization, can be found in Appendix~\ref{s:FE_kF}.

\textbf{Remark 4.1}: One may simply change the sign of both $\mg^e_\mrg$ and $\mg^e_\mathrm{pspg}$ in order to highlight the symmetry between the second part of $\mf^e_{\sF\mathrm{int}}$ and $\mg^e_\mrg$. 

\textit{4.1.1.4 Stabilization terms}

In order to evaluate the residual $\bff_{\!\mathrm{res}}$ that appears in the stabilization terms $\mf^e_\mathrm{supg}$ and $\mg^e_\mathrm{pspg}$, we note that 
\eqb{l}
2\,\divz\bD^h = (v^h_{j,ij} + v^h_{i,jj})\,\be_i = (\mG^2+\mH)\,\mv_e\,,
\eqe
where $\mG^2 := [\mG^2_1,\,\mG^2_2,\,...,\,\mG^2_{n_e}]$, with
\eqb{l}
\mG^2_I := \nabla(\nabla N_I) = \left[\begin{matrix}
  N_{I,11} & N_{I,12} & N_{I,13} \\
  N_{I,21} & N_{I,22} & N_{I,23} \\
  N_{I,31} & N_{I,32} & N_{I,33}
\end{matrix}\right]
\eqe
and $\mH := [H_1\bone,\,H_2\bone,\,...,\,H_{n_e}\bone]$, with
\eqb{l}
H_I := \tr\mG^2_I = N_{I,11} + N_{I,22} + N_{I,33}\,.
\eqe
With this we can write
\eqb{l}
\divz\bsig^h = \eta\,\mF\,\mv_e - \mG\,\mpp_e\,, 
\eqe
where $\mF=\mG^2+\mH$.
Thus we obtain
\eqb{l}
\bff_\mathrm{\!res} \approx \bff^h_\mathrm{\!res} = \rho\,\mN\,\mv'_e + \rho\,\mB_\mrv\mv_e - \eta\,\mF\,\mv_e + \mG\,\mpp_e - \bar\bff\,.
\eqe
The stabilization parameters $\tau_\mrv$ and $\tau_\mrp$ appearing inside $\mf^e_\mathrm{supg}$ and $\mg^e_\mathrm{pspg}$ are computed from
\eqb{l}
\tau_\mrv = \tau_\mrp = \ds\Bigg[
\bigg(\frac{2}{\Delta t}\bigg)^2 + \bigg(\frac{2\norm{\bv}}{m_e\,h_e}\bigg)^2 +\bigg(\frac{4\nu}{m_e\,h_e^2}\bigg)^2\Bigg]^{-\frac{1}{2}}
\label{e:tau_vp}\eqe
\citep{shakib,tezduyar92,ESEflow}, where $\Delta t$ is the time step size, $h_e$ is the ``element length" in the local flow direction taken from
\eqb{l}
\ds\frac{1}{h_e} = \frac{1}{2}\sum_{I=1}^{n_e}\bigg|\nabla N_I\cdot\frac{\bv}{\norm{\bv}}\bigg|
\label{e:h_e}
\eqe
\citep{tezduyar92} and $m_e$ depends on the polynomical order of the shape functions. I.e.~for L1 (linear Lagrange) and L2 (quadratic Lagrange) elements we have $m_e=1/3$ and $m_e=1/12$, respectively.\footnote{In Eqs.~\eqref{e:tau_vp} and \eqref{e:h_e}, $\bv$ is taken from the previous time step in order to avoid the linearization of $\tau_\mrv$ and $\tau_\mrp$.} 
According to this, parameters $\tau_\mrv$ and $\tau_\mrp$ are local parameters that change from quadrature point to quadrature point. 

\textit{4.1.1.5 Transformation of derivatives}

In the above expressions $\nabla N_I$ denotes the gradient w.r.t.~the current configuration $\bx$, which is discretized by $\bx^h=\sum_IN_I\,\mx_{\mrm I}$, where $\mx_{\mrm I}$ are the nodal positions of the FE mesh.
Since it is convenient to define the shape functions on a master element in $\bxi=[\xi,\,\eta,\,\zeta]^\mrT$ space, where $\partial N_I/\partial\bxi$ is easily obtained, $\nabla N_I$ needs to be determined from
\eqb{l}
\nabla N_I = \ds\pa{N_I}{\bx} = \bj^{-\mrT}\,\pa{N_I}{\bxi}\,,
\label{e:N,x}\eqe
where
\eqb{l}
\bj = \ds\pa{\bx^h}{\bxi} = \ds\sum_{I=1}^{n_e}\bx_{\mrm I}\otimes\pa{N_I}{\bxi}
\eqe
denotes the Jacobian of the mapping $\bxi\rightarrow\bx$. 
Likewise, the second derivative $\mG^2_I=\nabla(\nabla N_I)$ is obtained from the formula
\eqb{l}
\mG^2_I = \ds\paqq{N_I}{\bx}{\bx} = \bj^{-\mrT}\Bigg[\sum_{J=1}^{n_e}\Big(\delta_{IJ} - \nabla N_I\cdot\bx_{\mrm J}\Big)\,\paqq{N_J}{\bxi}{\bxi}
\Bigg]\,\bj^{-1}
\label{e:N,xx}\eqe
that follows from differentiating \eqref{e:N,x}. 
Eq.~\eqref{e:N,xx} is equivalent to the expression given in \citet{dhatt}.

\subsubsection{Membrane deformation}

Following the notation of Eq.~\eqref{e:bvh}, the reference position and the current position within a membrane element are approximated by the interpolations
\eqb{lllll}
\bX \ais \bX^h \is \mN\,\mX_e\,, \\[1mm]
\bx \ais \bx^h \is \mN\,\mx_e\,,
\label{e:bxh}\eqe
where $\mX_e$ and $\mx_e$ are arranged just like $\mv_e$.
From this follows
\eqb{lllll}
\bA_\alpha \ais \bA^h_\alpha \is \mN_{,\alpha}\,\mX_e\,, \\[1mm] 
\ba_\alpha \ais \ba^h_\alpha \is \mN_{,\alpha}\,\mx_e\,,
\eqe 
where $\mN_{,\alpha}:=[N_{1,\alpha}\bone,\,N_{2,\alpha}\bone,\,...,\,N_{n_e,\alpha}\bone]$.
Likewise, 
\eqb{l}
\bw_{,\alpha} \approx \bw^h_{,\alpha} = \mN_{,\alpha}\,\mw_e
\eqe
follows from Eq.~\eqref{e:bwh}. 
Given $\bA_\alpha$ and $\ba_\alpha$, the metric tensor components $A^{\alpha\beta}$ and $a^{\alpha\beta}$ can be determined and the stress can be evaluated as discussed in Sec.~\ref{s:theo_s}.

Inserting the discretized expressions for $\dot\bv$, $\ba_\alpha$, $\bw$ and $\bw_\alpha$ into the membrane weak form \eqref{e:wfS}
yields the elemental weak form contribution
\eqb{l}
G_\sS^e = \mw_e^\mrT\,\mf^e_\sS\,,
\label{e:GS}\eqe 
with the ($3n_e\times1$) FE force vector
\eqb{l}
\mf^e_\sS :=  \left\{
\begin{array}{ll}
\mf^e_{\sS\mathrm{in}}+\mf^e_{\sS\mathrm{int}}+\mf^e_\mrc-\mf^e_{\sS\mrf}-\mf^e_{\sS\mathrm{ext}\bar f} ~& $for $\Omega^e\subset\sS^h\,, \\[2mm]
-\mf^e_{\sS\mathrm{ext}\bar t} & $for $\Omega^e\subset\partial_t\sS^h\,,
\end{array}\right.
\eqe
that is composed of
\eqb{lll}
\mf_{\sS\mathrm{in}}^e \dis \ds\int_{\Omega^e} \rho_\mrs\,\mN^\mrT\mN\,\dif v~\dot\mv_e\,, \\[4mm]
\mf_{\sS\mathrm{int}}^e \dis \ds\int_{\Omega^e}\sig^{\alpha\beta}\,\mN^\mrT_{,\alpha}\,\mN_{,\beta}\,\dif a~\mx_e\,,\\[4mm]
\mf^e_{\mrc} \dis - \ds\int_{\Omega^e} \mN^\mrT\,\bff_{\!\mrc}\,\dif a\,, \\[4mm]
\mf_{\sS\mrf}^e \dis \ds\int_{\Omega^e} \mN^\mrT\,\bff_{\!\mrf}\,\dif a\,, \\[4mm]
\mf_{\sS\mathrm{ext}\bar f}^e \dis \ds\int_{\Omega^e} \mN^\mrT\,\bar\bff_{\!\mrs}\,\dif a\,, \\[4mm]
\mf_{\sS\mathrm{ext}\bar t}^e \dis \ds\int_{\Omega^e}\mN_\mrt^\mrT\,\bar\bt_\mrs\,\dif s\,.
\label{e:f_icfe}\eqe
Using a quadrature-point-based contact formulation, the discretization of the contact traction $\bff_{\!\mrc}$ is straight forward (expression \eqref{e:fc} is simply evaluated at each quadrature point), but an active set strategy needs to be implemented in order to handle the state changes between contact and no contact \citep{wriggers-contact}.\\
The tangent matrix of $\mf^e_\sS$, needed for the linearization, can be found in Appendix~\ref{s:FE_kS}.

\subsubsection{Coupled system}

Combining contributions \eqref{e:GF} and \eqref{e:GS} yields the coupled weak form
\eqb{l}
G^e = \mw_e^\mrT\,\mf^e + \mq_e^\mrT\,\mg^e \,,
\label{e:G}\eqe 
with the ($3n_e\times1$) FE force vector
\eqb{l}
\mf^e :=  \mf^e_\sF + \mf^e_\sS = \left\{
\begin{array}{ll}
\mf^e_{\sF\mathrm{in}}+\mf^e_{\sF\mathrm{int}}+\mf^e_\mathrm{supg}-\mf^e_{\sF\mathrm{ext}\bar f} ~& $for $\Omega^e\subset\sF^h\,, \\[2mm]
-\mf^e_{\sF\mathrm{ext}\bar t} & $for $\Omega^e\subset\partial_t\sF^h\,, \\[2mm]
\mf^e_{\sS\mathrm{in}}+\mf^e_{\sS\mathrm{int}}+\mf^e_\mrc-\mf^e_{\sS\mathrm{ext}\bar f} ~& $for $\Omega^e\subset\sS^h\,, \\[2mm]
-\mf^e_{\sS\mathrm{ext}\bar t} & $for $\Omega^e\subset\partial_t\sS^h\,.
\end{array}\right.
\eqe
It can be seen that for a conforming FE discretization of surface $\sS$, such as is considered here, coupling condition \eqref{e:coupt} implies that the force vector $\mf^e_{\sS\mrf}$ of a membrane element cancels exactly with $\mf^e_{\sF\mrs}$ of the corresponding fluid boundary element.
In the coupled system, both $\mf^e_{\sS\mrf}$ and $\mf^e_{\sF\mrs}$ therefore do not appear anymore.

\subsubsection{Double pressure nodes}\label{s:2xp}

Since the membrane is described here as a 2D surface that is discretized by 2D surface finite elements, the membrane nodes carry a special role.
Unless the membrane is located at the boundary of the fluid, it is surround by fluid on both sides and generally supports pressure jumps.
A finite element node on $\sS^h$ therefore must carry two pressure dofs. 
One for each side of the membrane.
Otherwise, the formulation does not properly account for pressure jumps. 
This is especially important for flexible membranes, where pressure jumps tend to become large.
In practice, each FE node on $\sS^h$ that is not located at  boundary $\partial\sS^h$ (where both fluid sides connect), is assigned two pressure dofs.\footnote{\citet{Tezduyar:2007eb} propose to also use double pressure dofs at the boundary of $\partial\sS^h$ in order to provide additional numerical stability.} 
When the elemental connectivity is then set up, care has to be taken in order to connect the element on each side of $\sS^h$ with the correct dofs.
\\
As long as a no-slip condition is considered on both sides of $\sS$, as is done here, the velocity field is continuous across $\sS$ and no extra velocity degrees of freedom are needed on $\sS^h$.

\subsection{Temporal discretization}\label{s:temp}

The elemental force vectors $\mf^e$ and $\mg^e$ are assembled into the global vectors
\eqb{l}
\mf = \mf_{\sF\mathrm{in}} + \mf_{\sS\mathrm{in}} + \mf_{\sF\mathrm{int}}+\mf_{\sS\mathrm{int}}+\mf_\mrc+\mf_\mathrm{supg}-\mf_\mathrm{ext}
\eqe
and
\eqb{l}
\mg = \mg_\mrg+\mg_\mathrm{pspg}\,,
\eqe
where $\mf_\mathrm{ext}:=\mf_{\sF\mathrm{ext}\bar f}+\mf_{\sF\mathrm{ext}\bar t}+\mf_{\sS\mathrm{ext}\bar f}+\mf_{\sS\mathrm{ext}\bar t}$.
The former can be written as $\mf=[\mf_\mathrm{br}^\mrT,\,\mf_\mrr^\mrT]^\mrT$, where $\mf_\mathrm{br}$ are the boundary reactions of the nodes on $\partial_{\hat x}\sF$ and $\partial_x\sS$, and $\mf_\mrr$ are the residual forces of all the remaining nodes.
Accordingly, the global residual vector
\eqb{l}
\mr := \left[\begin{matrix}
  \mf_\mrr \\[1mm]
  \mg
\end{matrix}\right],
\eqe
can be defined. 
The finite element forces are in equilibrium if $\mr=\mathbf{0}$. 
In general, $\mr=\mathbf{0}$ is a coupled system of ordinary differential equations for the unknown nodal positions 
$\mx := [\bx_I]$, velocities $\mv:=[\bv_I]$, accelerations $\ma:=[\bv'_I]$ (for fixed $\mx$) and pressures $\mpp:=[p_I]$, for $I=1,...,n_\mathrm{no}$, that are all functions of time. 
The generalized-$\alpha$ scheme \citep{chung93,jansen99,cottrell} is used to discretize $\mr=\mathbf{0}$ in time.
Instead of solving for the functions $\mx(t)$, $\mv(t)$, $\ma(t)$ and $\mpp(t)$, the approximations $\mx^n\approx\mx(t_n)$, $\mv^n\approx\mv(t_n)$, $\ma^n\approx\ma(t_n)$ and $\mpp^n\approx\mpp(t_n)$ are determined at discrete time steps $t_n$, $n=0,...,n_t$.
This is based on the Newmark update formulas for step $t_n\rightarrow t_{n+1}$
\eqb{lll}
\mx^{n+1} \is \mx^n + \Delta t\,\mv^n + \ds\frac{\Delta t^2}{2}\,\big((1-2\beta)\,\ma^n + 2\beta\,\ma^{n+1} \big)\,,\\[3mm]
\mv^{n+1} \is \mv^n + \Delta t\,\big((1-\gamma)\,\ma^n + \gamma\,\ma^{n+1} \big)\,, 
\label{e:Newmark}\eqe
where $\beta$ and $\gamma$ are non-dimensional parameters.\footnote{They should not be confused with the physical parameters $\beta$ and $\gamma$ used for the surface inclination and surface tension in other sections.} 
According to the generalized-$\alpha$ scheme, $\mr$ is then evaluated for $\mpp^{n+1}$ and
\eqb{lllll}
\mx^{n+\alpha_\mrf} \is \mx^n \plus \alpha_\mrf\,(\mx^{n+1}-\mx^n)\,, \\[2mm]
\mv^{n+\alpha_\mrf} \is \mv^n \plus \alpha_\mrf\,(\mv^{n+1}-\mv^n)\,, \\[2mm]
\ma^{n+\alpha_\mrm} \is \ma^n \plus \alpha_\mrm\,(\ma^{n+1}-\ma^n)\,,
\label{e:gen-a}\eqe
where $0<\alpha_\mrm\leq1$ and $0<\alpha_\mrf\leq1$ are chosen parameters.\footnote{Note that the $\alpha$ introduced by \citet{chung93} corresponds to $1-\alpha$ here.}
The global force vectors thus take the form
\eqb{lll}
\mf  
\is \mf_{\sF\mathrm{in}}\big(\ma^{n+\alpha_\mrm},\mv^{n+\alpha_\mrf}\big) + \mf_{\sS\mathrm{in}}\big(\ma^{n+\alpha_\mrm})
     + \mf_{\sF\mathrm{int}}\big(\mv^{n+\alpha_\mrf},\mpp^{n+1}\big) 
     + \mf_{\sS\mathrm{int}}\big(\mx^{n+\alpha_\mrf}\big) \\[2mm]
\plus \mf_\mrc\big(\mx^{n+\alpha_\mrf}\big)+ \mf_\mathrm{supg}\big(\ma^{n+\alpha_\mrm},\mv^{n+\alpha_\mrf},\mpp^{n+1}\big)
     - \mf_\mathrm{ext}\,, \\[3mm]
\mg
\is \mg_\mrg\big(\mv^{n+\alpha_\mrf}\big) + \mg_\mathrm{pspg}\big(\ma^{n+\alpha_\mrm},\mv^{n+\alpha_\mrf},\mpp^{n+1}\big)\,.     
\label{e:globsys}\eqe
The temporal inconsistency that is introduced if $\alpha_\mrm\neq\alpha_\mrf\neq1$ is a deliberate feature of the generalized-$\alpha$ method.
The system $\mr=\mathbf{0}$ thus reduces to a system of algebraic equations that can be solved for $\mx^{n+1}$, $\mv^{n+1}$, $\ma^{n+1}$ and $\mpp^{n+1}$ given the previous values $\mx^n$, $\mv^n$, $\ma^n$ and $\mpp^n$. 
One option is to pick $\muu:=[\mv,\,\mpp]$ as the primary unknowns, solve $\mr=\mathbf{0}$ for $\muu^{n+1}$, and then obtain $\ma^{n+1}$ and $\mx^{n+1}$ (which is really only needed for the membrane nodes) from \eqref{e:Newmark}.
Since the system $\mr=\mathbf{0}$ is non-linear, the Newton-Raphson method is used.\footnote{A direct sparse solver is used in all subsequent examples apart from the finest droplet discretization in Sec.~5.2, which uses the conjugate gradient method preconditioned by an incomplete LU factorization.}
This requires the tangent matrix $\mk$ that is assembled from the elemental entries
\eqb{l}
\mk^e := \ds\pa{\mr^e}{\muu^{n+1}_e}\,.
\label{e:kFE}\eqe
It is given in Appendix~\ref{s:FE_kt} for the considered fluid and membrane elements.
In the following computations, the Newmark parameters are taken as \citep{chung93}
\eqb{lll}
\gamma \is \ds\frac{1}{2} - \alpha_\mrf + \alpha_\mrm\,, \\[3mm]
\beta \is \ds\frac{1}{4}\big(1-\alpha_\mrf+\alpha_\mrm\big)^2
\eqe
using the generalized-$\alpha$ parameters\footnote{They are obtained taking a spectral radius of $\rho_\infty = \frac{1}{2}$ for the first order system, see \citet{jansen99}.}
\eqb{l}
\alpha_\mrf = \ds\frac{2}{3}\,,\quad  \alpha_\mrm = \ds\frac{5}{6}\,.
\eqe
This choice ensures second order accuracy in time and unconditional stability (for linear problems).

\subsection{Normalization}\label{s:norm} 

In order to implement the above expressions within a computer code\footnote{In this work a self-written parallel Matlab %(www.mathworks.com) 
code is used on a 12-core Apple workstation (2x 2.66 GHz 6-Core Intel Xeon, 64 GB DDR3 RAM).} they have to be normalized. 
The normalization can also help to improve the conditioning of the monolithic system of equations.
We therefore chose a length scale $L_0$, time scale $T_0$ and force $F_0$, and use those to normalize all lengths, times and forces in the system.
Velocities, masses, fluid densities, fluid viscosities, fluid pressures, membrane densities and membrane stresses are then normalized by the scales
\eqb{l}
v_0 := \ds\frac{L_0}{T_0}\,,\quad
m_0 := \ds\frac{F_0T_0^2}{L_0}\,,\quad
\rho_0 := \ds\frac{m_0}{L_0^3}\,,\quad
\eta_0 := \ds\frac{F_0T_0}{L_0^2}\,,\quad
p_0 := \ds\frac{F_0}{L_0^2}\,,\quad
\rho^\mrs_0 := \ds\frac{m_0}{L_0^2}\,,\quad
\gamma_0 := \ds\frac{F_0}{L_0}\,.
\eqe
System \eqref{e:globsys} can then be expressed in the normalized form
\eqb{lll}
\bar\mf(\bar\muu^{n+1}) \is \bar\mf_{\sF\mathrm{in}} + \bar\mf_{\sS\mathrm{in}} + \bar\mf_{\sF\mathrm{int}} + \bar\mf_{\sS\mathrm{int}} + \bar\mf_\mrc + \bar\mf_\mathrm{supg} - \bar\mf_\mathrm{ext}\,, \\[3mm]
\bar\mg(\bar\muu^{n+1}) \is \bar\mg_\mrg + \bar\mg_\mathrm{pspg}\,,
\label{e:globsysbar}\eqe
where a bar denotes normalization with the corresponding scale from above, e.g.
\eqb{l}
\bar\mf^e_{\sF\mathrm{in}} = \bar\mm_e\,\bar\ma_e + \bar\mf^e_\mathrm{con}\,,
\eqe
with
\eqb{lll}
\bar\mm_e \dis \ds\int_{\bar\Omega^e}\bar\rho\,\mN^T\mN\,\dif\bar v\,,\\[4mm]
\bar\mf^e_\mathrm{con} \dis \ds\int_{\bar\Omega^e}\bar\rho\,\mN^T\bar\mB_\mrv\,\bar\mv_e\,\dif\bar v\,,
\eqe
and $\bar\rho = \rho/\rho_0$, $\dif\bar v = \dif v/L_0^3$, $\bar\mB_\mrv=\mB_\mrv T_0$, $\bar\mv_e=\mv_e/v_0$ and $\bar\ma_e=\ma_e\,T_0/v_0$.
All the other quantities appearing in \eqref{e:globsysbar} are normalized in the same fashion.
Solving \eqref{e:globsysbar} then gives the normalized unknowns $\bar\mv=\mv/v_0$ and $\bar\mpp=\mpp/p_0$, while \eqref{e:Newmark} can be solved for $\bar\mx=\mx/L_0$ and $\bar\ma=\ma\,T_0/v_0$.

\subsection{Mesh motion}

Apart from the unknown material velocity $\mv$ and pressure $\mpp$, the discrete mesh velocity $\mv_\mrm$ can also be regarded as an unknown.
In that case suitable (differential) equations have to be formulated for $\mv_\mrm$.
A simpler approach is to determine the mesh velocity from the membrane velocity using linear interpolation:
On the membrane surface the mesh motion is considered Lagrangian, i.e.~$\mv_\mrm=\mv$, whereas it is treated Eulerian ($\mv_\mrm=\mathbf{0}$) beyond a certain distance from the membrane. 
In-between, simple linear interpolation is used.
Details of this are reported in the following examples.
Linear interpolation, and ALE in general, does not work for some FSI problems.
An example are solids revolving within the fluid. 
For such cases, other techniques need to be considered.

\section{Numerical examples}\label{s:ex}

This section presents three numerical examples that range from very low to quite large Reynolds numbers.
The first example considers a solid membrane (with no bending resistance), the second example considers a liquid membrane, and the third example considers a solid shell with low bending resistance. 
The examples exhibit large membrane deformations that lead to strong FSI coupling.

\subsection{Fluid-inflated cylinder}\label{s:ex1}

The first numerical example considers the radial inflation of a cylindrical membrane due to radial inflow.
The numerical solution will be compared to the analytical solution derived in Sec.~\ref{s:ana_infl}.
The initial inner radius of the cylinder $R_\mathrm{in}$, the maximum inflow velocity $v_0$ and the fluid density $\rho$ are used for normalization, such that $L_0=R_\mathrm{in}$, $T_0=R_\mathrm{in}/v_0$ and $\rho_0=\rho$.
The outer radius of the membrane at initialization time $t=0$ is taken as $R_\mrs=2L_0$.
Computationally, only a quarter of the cylindrical domain is modelled with a chosen height of $H=L_0$.
Sliding wall conditions\footnote{The normal velocity and the tangential traction are set to zero.} are applied to all fluid boundaries except the membrane surface, where coupling conditions apply, and the inflow boundary, where the radial inflow velocity
\begin{equation}
v_\mathrm{in}(t) = v_0\left\{\begin{array}{ll}
\big(1-\cos(\pi t/T_0)\big)/2~ & \textnormal{for}~t < T_0 \\[1mm]
1 & \textnormal{else}
\end{array}\right.
\end{equation}
is prescribed.
The Reynolds number, $Re=\rho\,v_\mathrm{in} \,L_0/\eta$, is chosen as $Re=100$ guaranteeing a purely laminar flow.
For water at room temperature ($\rho\approx 1000\,$kg/m$^3$, $\eta=1.00\,$mNs/m$^2$) this implies $v_0=10\,$m/s.
The membrane is modelled as a massless, incompressible Neo-Hookean, rubber-like material according to \eqref{e:sig_sol}.
The membrane's nondimensional shear stiffness is taken as $\bar\mu = 0.1$.
The fluid domain is discretized by $N_\mrf=n_r\times n_\theta\times 1$ quadratic volume elements in $\be_r$, $\be_\theta$ and $\be_3$ direction (see Fig.~\ref{f:infl_cyl}), while the membrane domain is discretized by $N_\mrs=n_\theta\times 1$ quadratic surface elements along $\be_\theta$ and $\be_3$.
Tab.~\ref{t:ec_g_conv} shows the considered meshes.
%-------------------------------------------------------------------------------------------------------------------------------
\begin{table}[h]
\centering
\begin{tabular}{|r|r|r|r|r|}
  \hline
   total elements & fluid elements & membrane elements & nodes & dofs \\[0.5mm] \hline 
   & & & & \\[-3.5mm]   
   7   &  $6\times1\times1$ & $1\times1$ &   117 &   495 \\[1mm] 
   42  & $13\times3\times1$ & $3\times1$ &   567 & 2,331 \\[1mm] 
   100 & $24\times4\times1$ & $4\times1$ & 1,323 & 5,373 \\[1mm] 
   \hline
\end{tabular}
\caption{Fluid-inflated cylinder: Considered FE meshes based on quadratic Lagrange elements.}
\label{t:ec_g_conv}
\end{table}
%-------------------------------------------------------------------------------------------------------------------------------
The time step is chosen as $\Delta\bar t =0.0025$ for all cases.
The radial mesh velocity at time step $t_{n+1}$ is defined by the linear interpolation
\eqb{l}
v_\mrm\big(R,t_{n+1}\big) = \ds\frac{R-R_\mathrm{in}}{R_\mrs-R_\mathrm{in}}\,v_\mrs(t_n)\,, 
\eqe
where $v_\mrs(t_n)$ is the cylinder's radial velocity at the previous time step.
\par
\begin{figure}[!ht]
 \begin{center}
  \subfigure[$\bar t=0$]{
  \includegraphics[scale=0.115,angle=0]{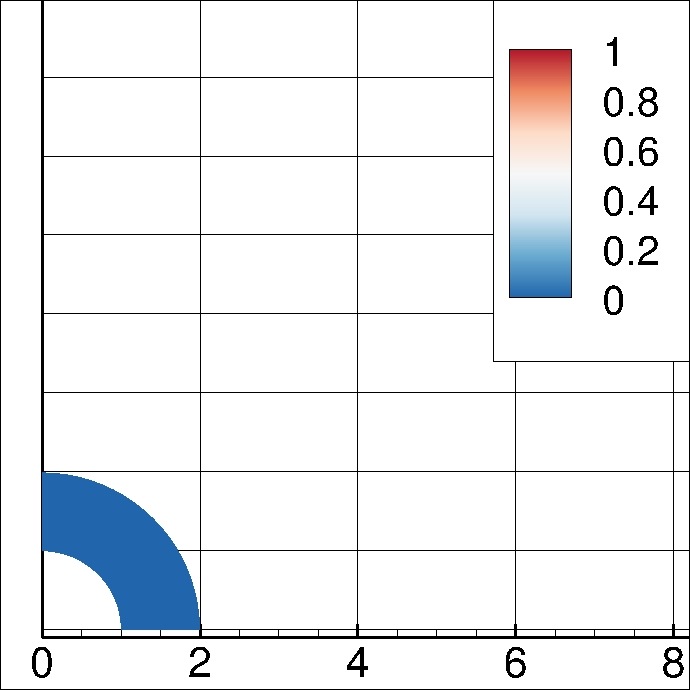}}
  \subfigure[$\bar t=1$]{
  \includegraphics[scale=0.115,angle=0]{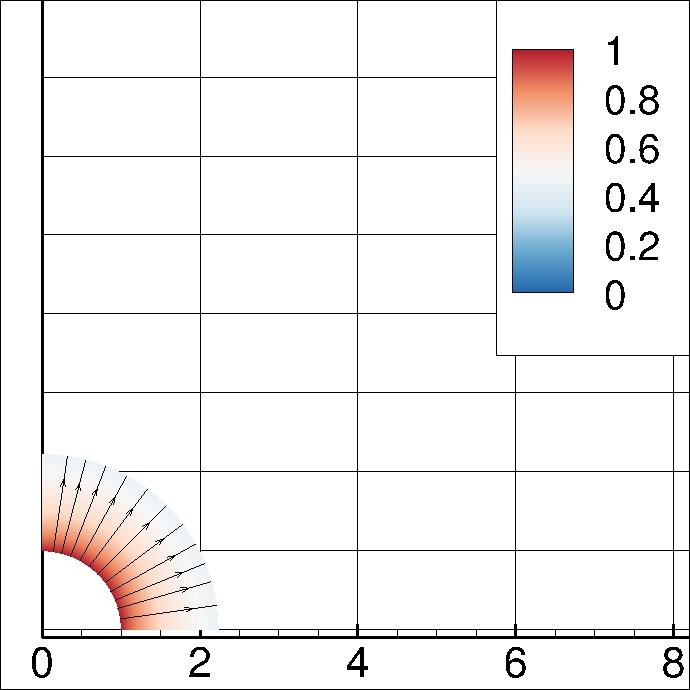}}
  \subfigure[$\bar t=6$]{
  \includegraphics[scale=0.115,angle=0]{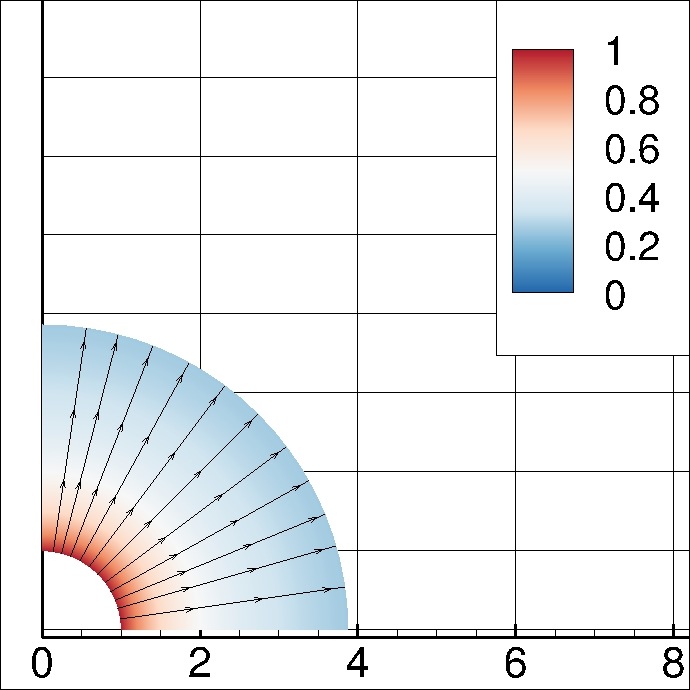}}
  \subfigure[$\bar t=11$]{
  \includegraphics[scale=0.115,angle=0]{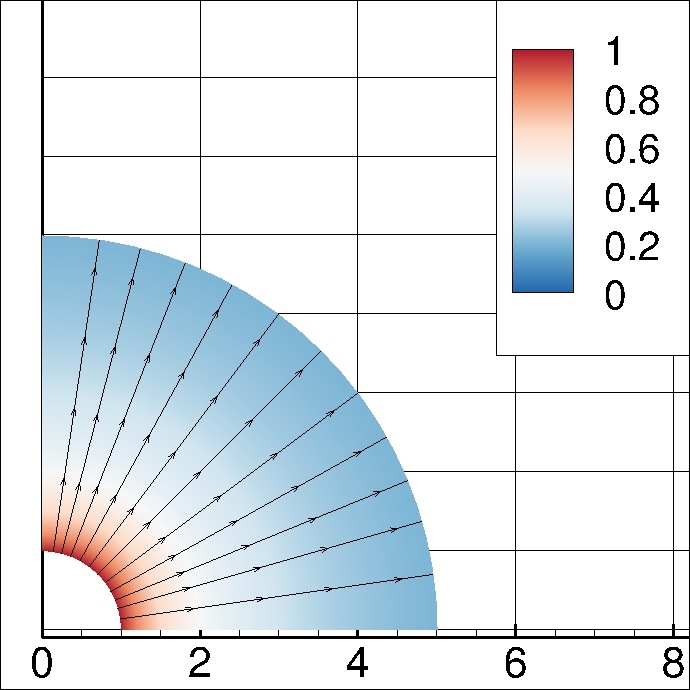}}
  \subfigure[$\bar t=21$]{
  \includegraphics[scale=0.115,angle=0]{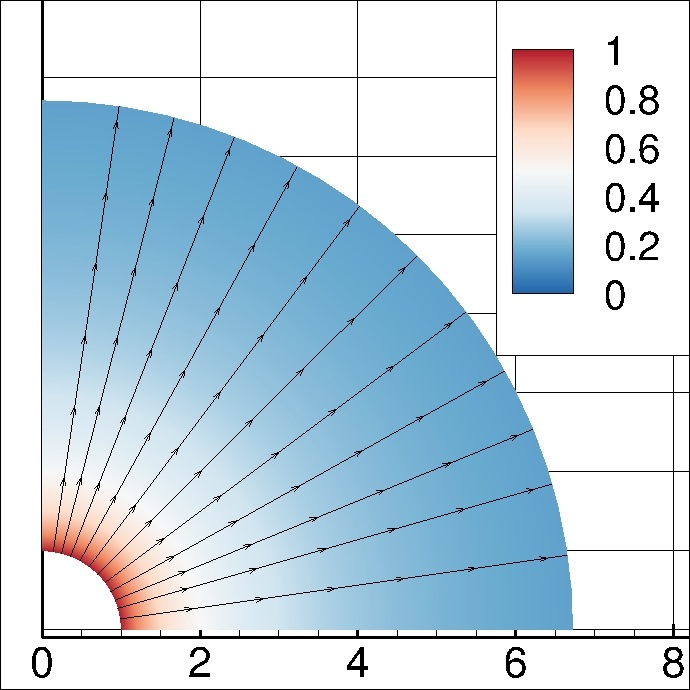}}
 \end{center}
 \vspace{-6mm}
 \caption{Fluid-inflated cylinder: Radial flow field $\bar v=v/v_0$ and cylinder expansion at various time steps. 
 Computationally, only a quarter of the system is modelled.}
 \label{f:ec_res1}
\end{figure}
\begin{figure}[!ht]
 \begin{center}
  \subfigure[$r(t)$ at $R=R_\mrs$]{ 
  \includegraphics[scale=1,angle=0]{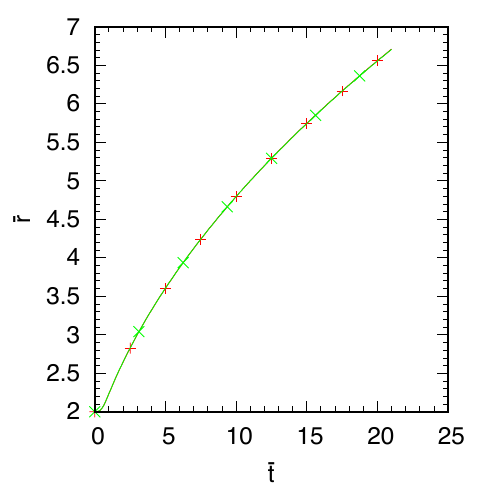}}
  \subfigure[Convergence]{
  \includegraphics[scale=1,angle=0]{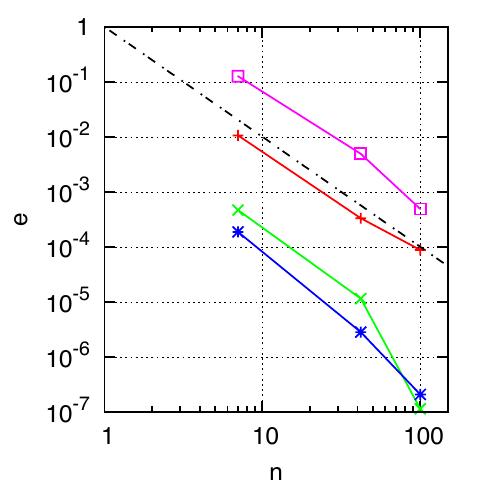}}
 \end{center}
 \vspace{-6mm}
 \caption{Fluid-inflated cylinder: (a) Membrane position $\bar r=r/L_0$ vs.~time $\bar t=t/T_0$. (Analytical result: green~$\times$, FE solution: red~$+$).
(b) Numerical error (L$^2$-norm) vs.~total number of L2 elements (radius $r$:~red~$+$, velocity $v$: green~$\times$, acceleration $a$: blue~$\star$, pressure $p$: magenta~$\square$) at $R=R_\mrs$ and $\bar{t}=21$.
The dash-dotted line marks quadratic convergence behavior.}
 \label{f:ec_res_xr}
\end{figure}
\begin{figure}[!ht]
 \begin{center}
  \subfigure[$v(t)$ at $r=r_\mrs$]{
  \includegraphics[scale=1,angle=0]{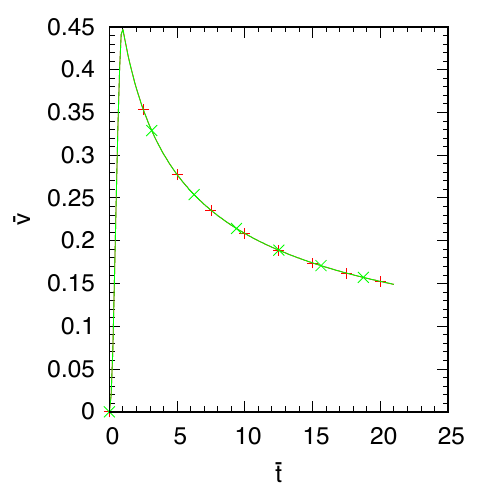}}
  \subfigure[$v(t)$ at $\bar t=21$]{
  \includegraphics[scale=1,angle=0]{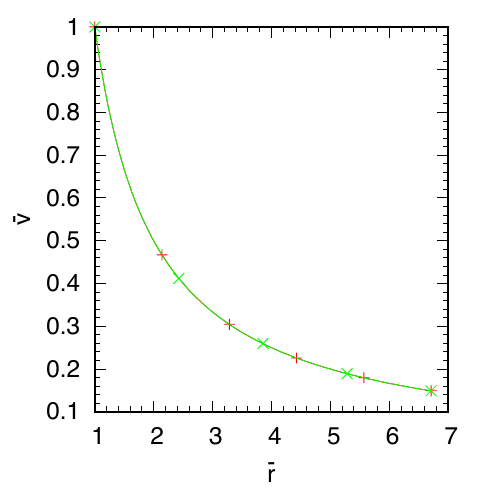}}
 \end{center}
 \vspace{-6mm}
 \caption{Fluid-inflated cylinder: (a) Normalized membrane velocity vs.~time; (b) Normalized fluid velocity vs.~radial position at $t=21\,T_0$.
(Analytical result: green~$\times$, FE solution: red~$+$)}
 \label{f:ec_res_vr}
\end{figure}
\begin{figure}[!ht]
 \begin{center}
  \subfigure[$p(t)$ at $r=r_\mrs$]{
  \includegraphics[scale=1,angle=0]{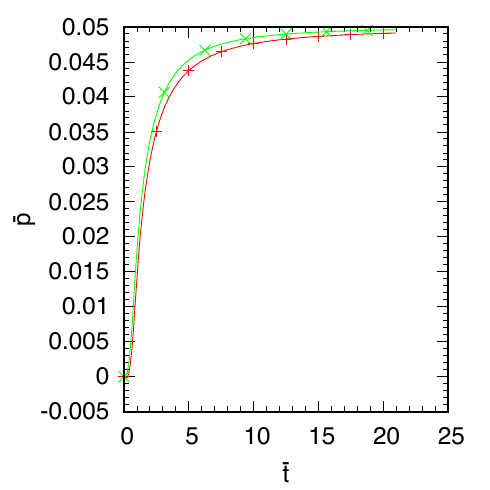}}
  \subfigure[$p(t)$ at $\bar t=21$]{
  \includegraphics[scale=1,angle=0]{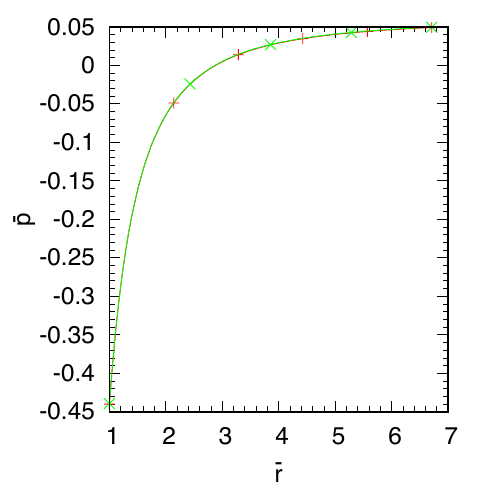}}
 \end{center}
  \vspace{-6mm}
 \caption{Fluid-inflated cylinder: (a) Normalized membrane pressure vs.~time; (b) Normalized fluid pressure vs.~radial position at $t=21\,T_0$.
(Analytical result: green~$\times$, FE solution: red~$+$)}
 \label{f:ec_res_p}
\end{figure}

Fig.~\ref{f:ec_res1} shows the radial flow field and the membrane displacement due to the cylinder inflation at different time steps.
%-------------------------------------------------------------------------------------------------------------------------------
%-------------------------------------------------------------------------------------------------------------------------------
The solid membrane is stretched by more than a factor of 3.
For the membrane displacement (Fig.~\ref{f:ec_res_xr}) and velocity (Fig.~\ref{f:ec_res_vr}) the numerical result is in perfect agreement with the analytical solution derived in Sec.~\ref{s:ana_infl}; see Eqs.~(\ref{e:ana1_v}) \& (\ref{e:ana1_r}).
%-------------------------------------------------------------------------------------------------------------------------------
%-------------------------------------------------------------------------------------------------------------------------------
%-------------------------------------------------------------------------------------------------------------------------------
For the pressure shown in Fig.~\ref{f:ec_res_p} we observe deviations from the analytical result \eqref{e:ana1_p} during the transient part and again nearly perfect agreement at the final simulation time.
%-------------------------------------------------------------------------------------------------------------------------------
%-------------------------------------------------------------------------------------------------------------------------------
The numerical results improve for a higher mesh resolution.
The finite element discretization and its implementation shows quadratic convergence behavior as expected, see Fig.~\ref{f:ec_res_xr}b.

\subsection{Rolling droplet}\label{s:ex2}

The second example simulates rolling contact of a liquid droplet on an inclined substrate considering a low Reynolds number and a contact angle of 180$^\circ$.
As we expect the motion to come close to the spinning solution of Sec.~\ref{s:ana_spin}, a purely Lagrangian FE description is chosen ($\bv_\mrm=\bv$). 
This also allows to use a classical contact description between droplet and substrate.
\\
There is earlier computational work on rolling droplets \citep{rasool12, Li13, Thampi13, Wind14}.
But it is either 2D, or non-FE. 
So the present study seems to be the first 3D FE simulation of rolling droplets.
Novel is also the way contact is treated here -- by using a computational contact algorithm with an active-set strategy. 
Within that, a no-slip (sticking) condition is assumed on the contact surface, i.e.~\eqref{e:bgc}.
If slip occurs, a stick-slip algorithm is needed for the droplet \citep{dropslide}.

\begin{figure}[h]
\begin{center} \unitlength1cm
\begin{picture}(0,5.7)
\put(-8.05,-.1){\includegraphics[height=58mm]{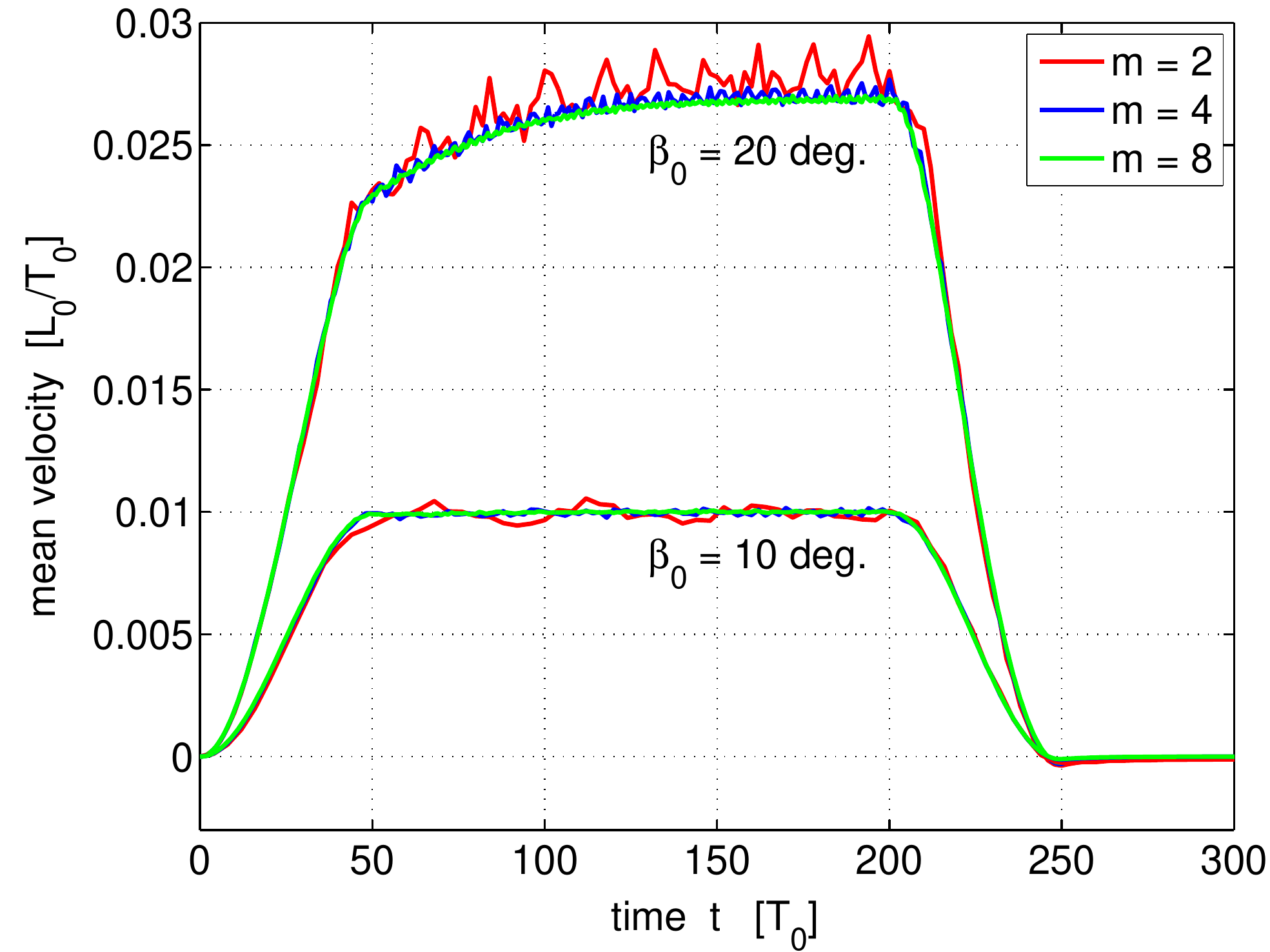}}
\put(0.15,-.1){\includegraphics[height=58mm]{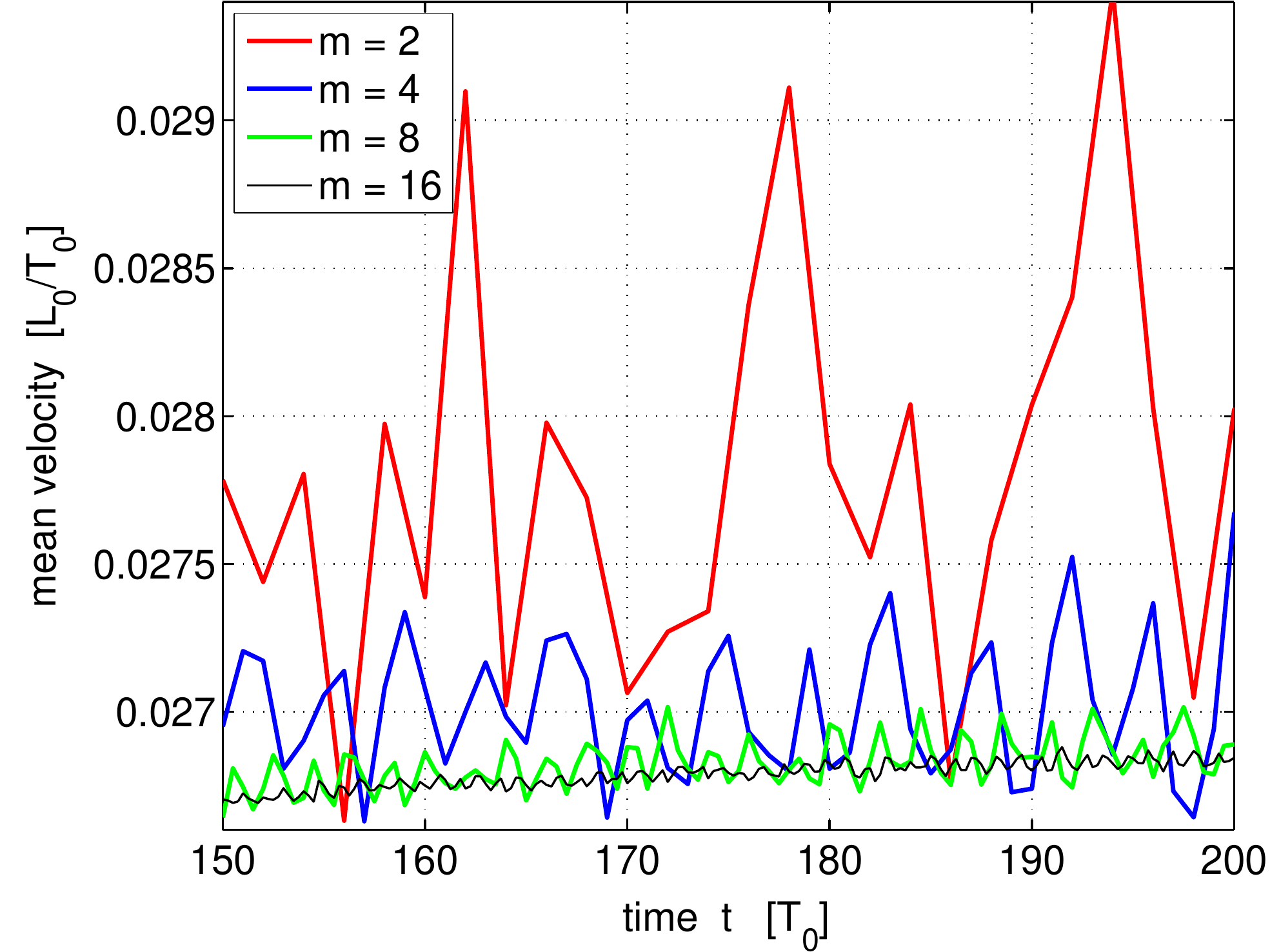}}
\end{picture}
\caption{Rolling droplet: Mean droplet velocity vs.~time for $\beta_0=10^\circ$ and $\beta_0=20^\circ$ using the meshes from Tab.~\ref{t:rdrop}. The right hand side shows an enlargment for $\beta_0=20^\circ$. As seen, the FE results converge upon mesh refinement.} 
\label{f:rdrop_vt}
\end{center}
\end{figure}

\begin{figure}[h]
\begin{center} \unitlength1cm
\begin{picture}(0,5.7)
\put(-8.05,-.1){\includegraphics[height=58mm]{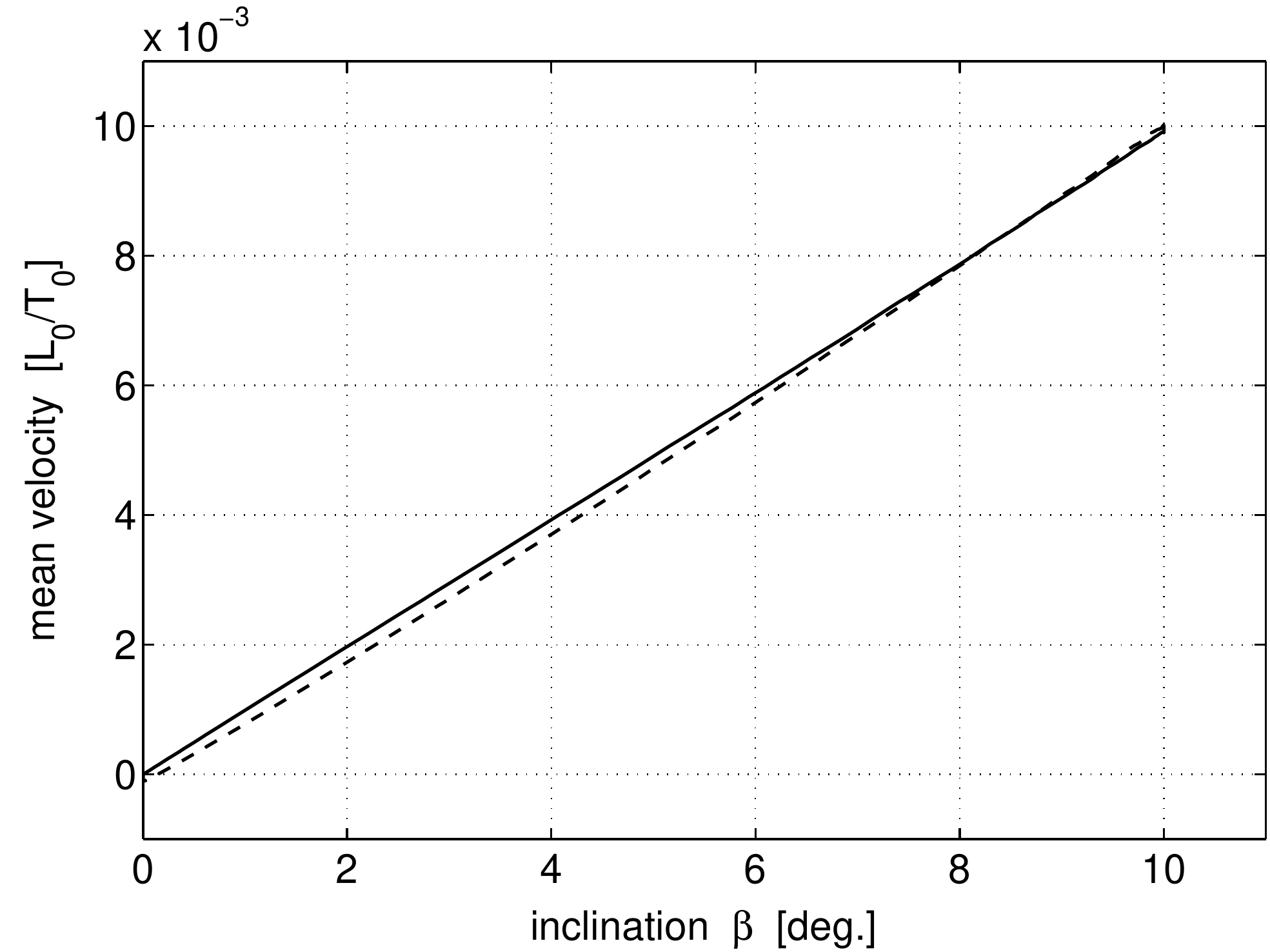}}
\put(0.15,-.1){\includegraphics[height=58mm]{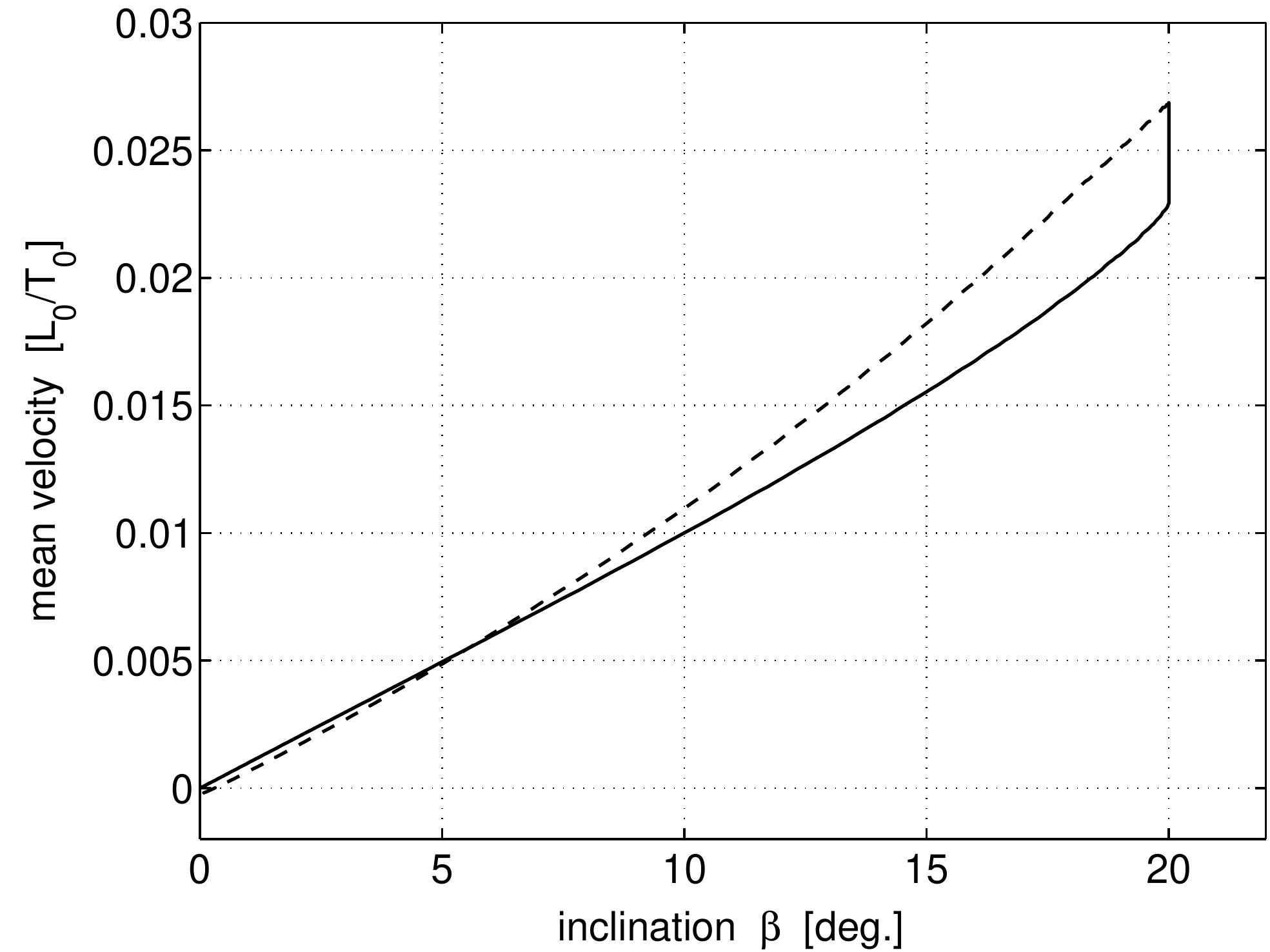}}
\put(-7.85,-.1){a.}
\put(.35,-.1){b.}
\end{picture}
\caption{Rolling droplet: Mean droplet velocity vs.~$\beta$ for $\beta_0=10^\circ$ (a) and $\beta_0=20^\circ$ (b) using $m=16$. The return branch (for decreasing $\beta$) is marked by a dashed line.}  
\label{f:rdrop_vb}
\end{center}
\end{figure}

\begin{figure}[h]
\begin{center} \unitlength1cm
\begin{picture}(0,4)
%\put(-8.1,1.5){\includegraphics[height=25mm]{../../codes/cFEAR/Input/iFSI/Droplet/oDropRoll/roll07/roll20_m8_te200_dt0p5_000v.jpg}}
%\put(-4.95,1.5){\includegraphics[height=25mm]{../../codes/cFEAR/Input/iFSI/Droplet/oDropRoll/roll07/roll20_m8_te200_dt0p5_100v.jpg}}
%\put(-1.8,1.5){\includegraphics[height=25mm]{../../codes/cFEAR/Input/iFSI/Droplet/oDropRoll/roll07/roll20_m8_te200_dt0p5_200v.jpg}}
%\put(1.35,1.5){\includegraphics[height=25mm]{../../codes/cFEAR/Input/iFSI/Droplet/oDropRoll/roll15/roll20_m8_te350_dtp5_400v.jpg}}
%\put(4.65,1.5){\includegraphics[height=25mm]{../../codes/cFEAR/Input/iFSI/Droplet/oDropRoll/roll15/roll20_m8_te350_dtp5_700v.jpg}}
\put(-8.1,1.5){\includegraphics[height=25mm]{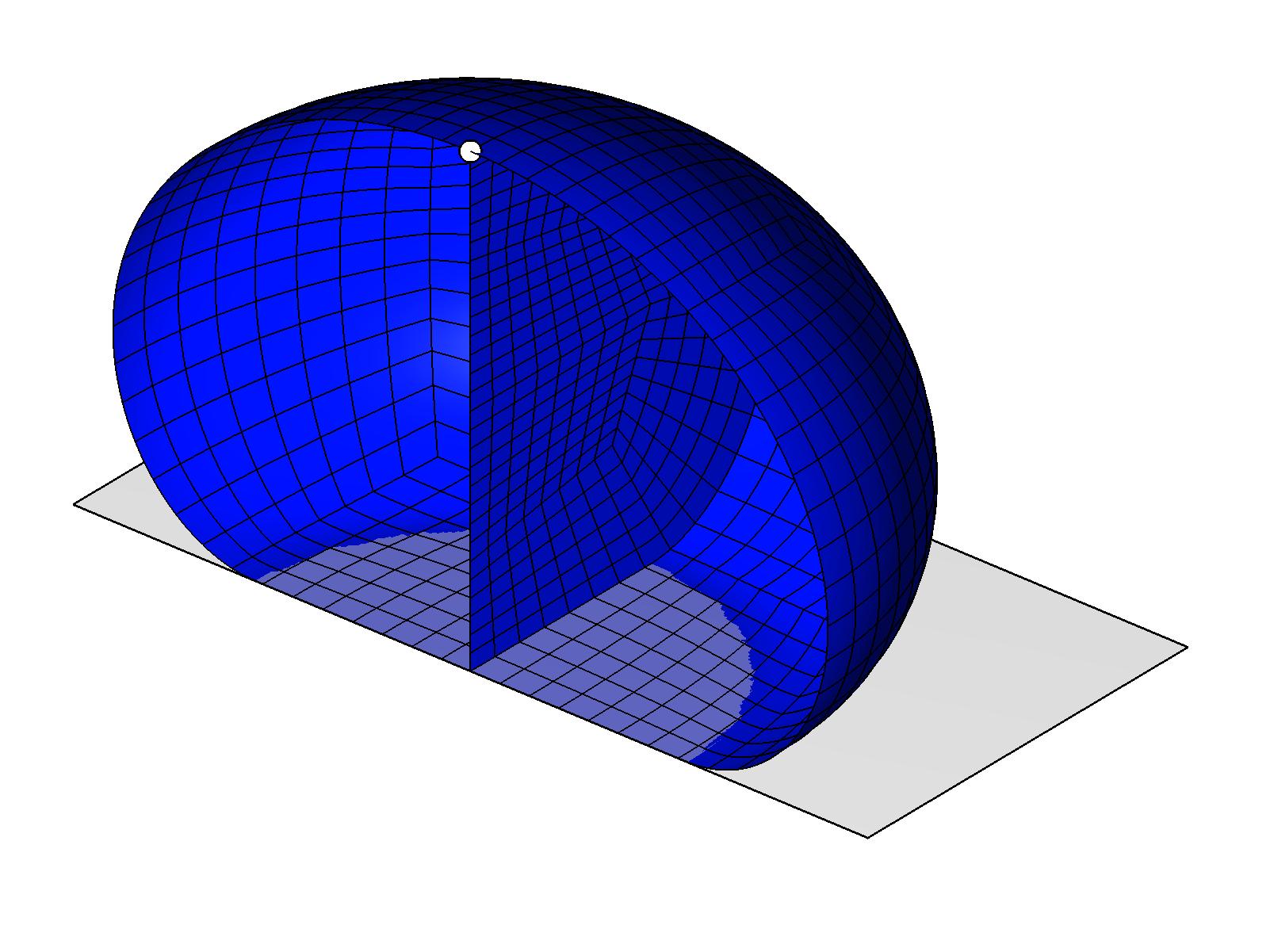}}
\put(-4.95,1.5){\includegraphics[height=25mm]{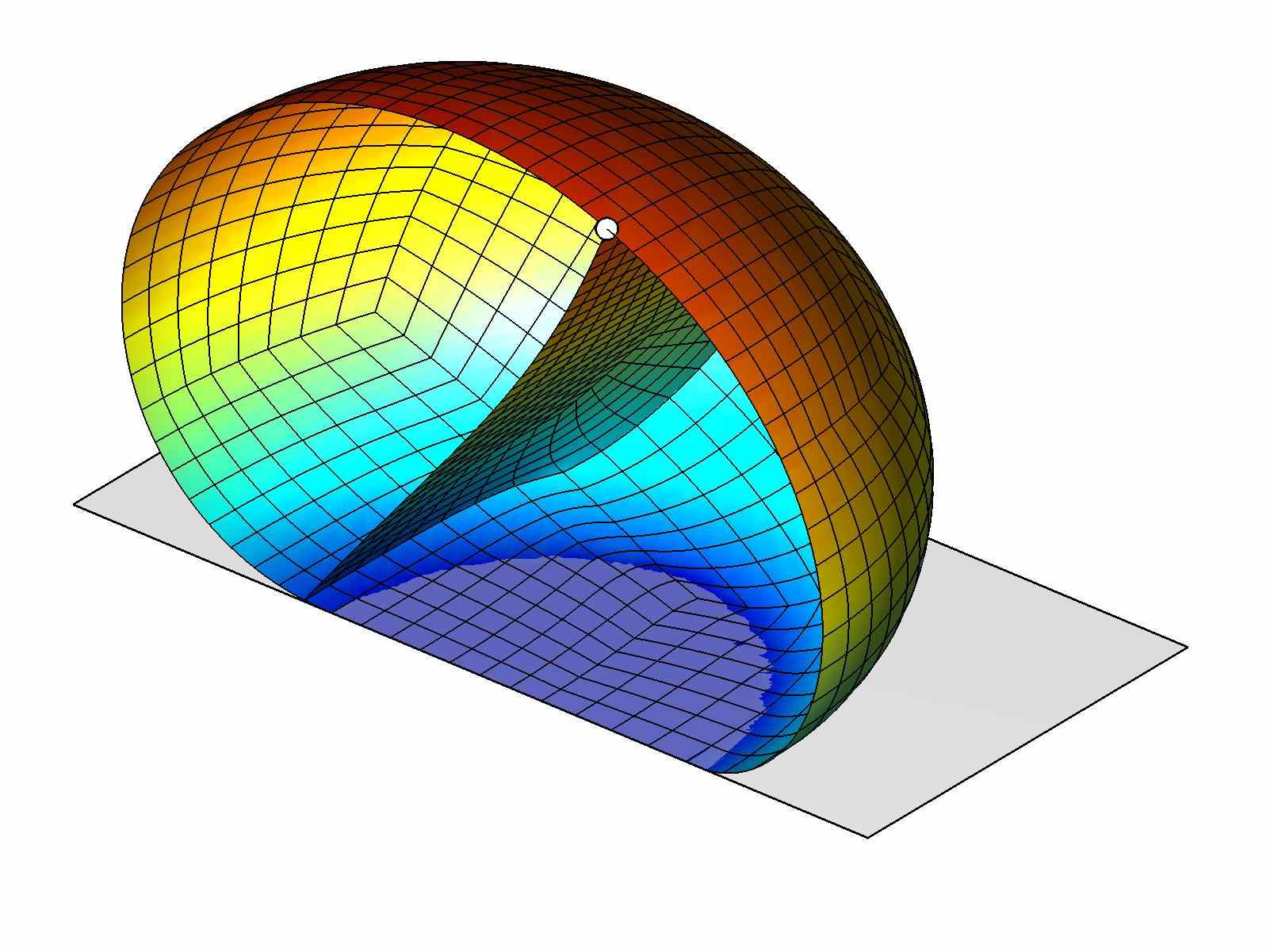}}
\put(-1.8,1.5){\includegraphics[height=25mm]{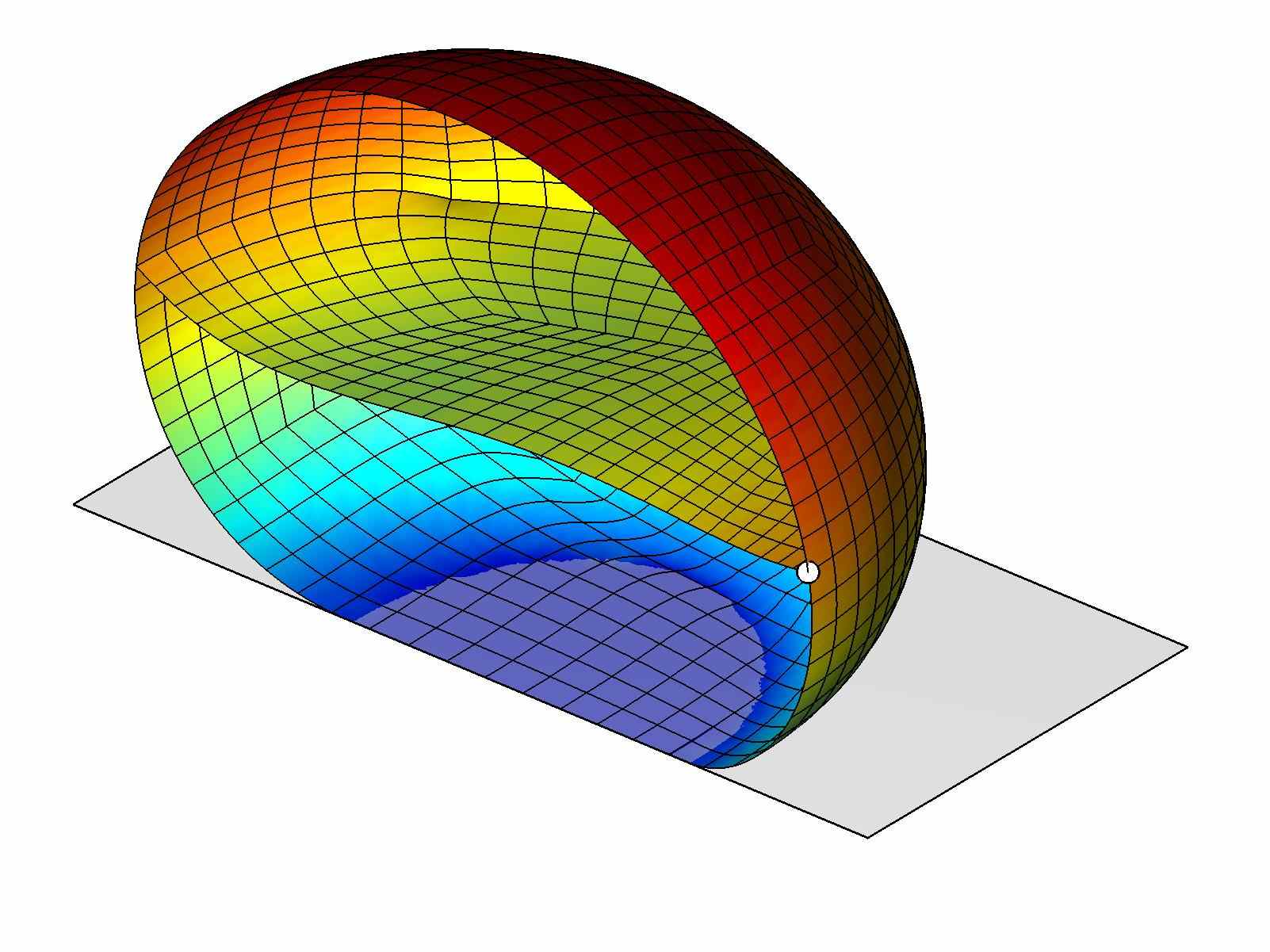}}
\put(1.35,1.5){\includegraphics[height=25mm]{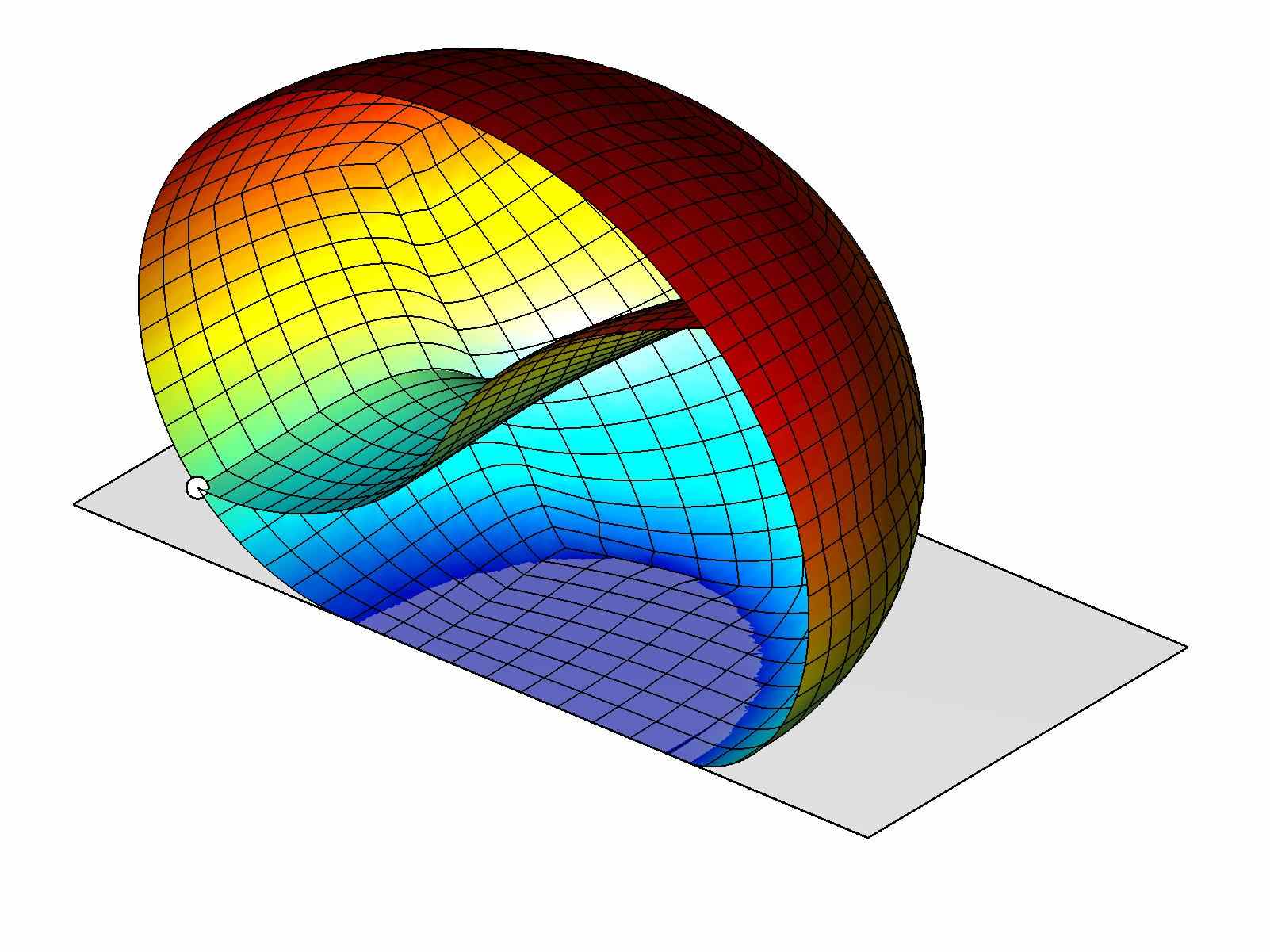}}
\put(4.65,1.5){\includegraphics[height=25mm]{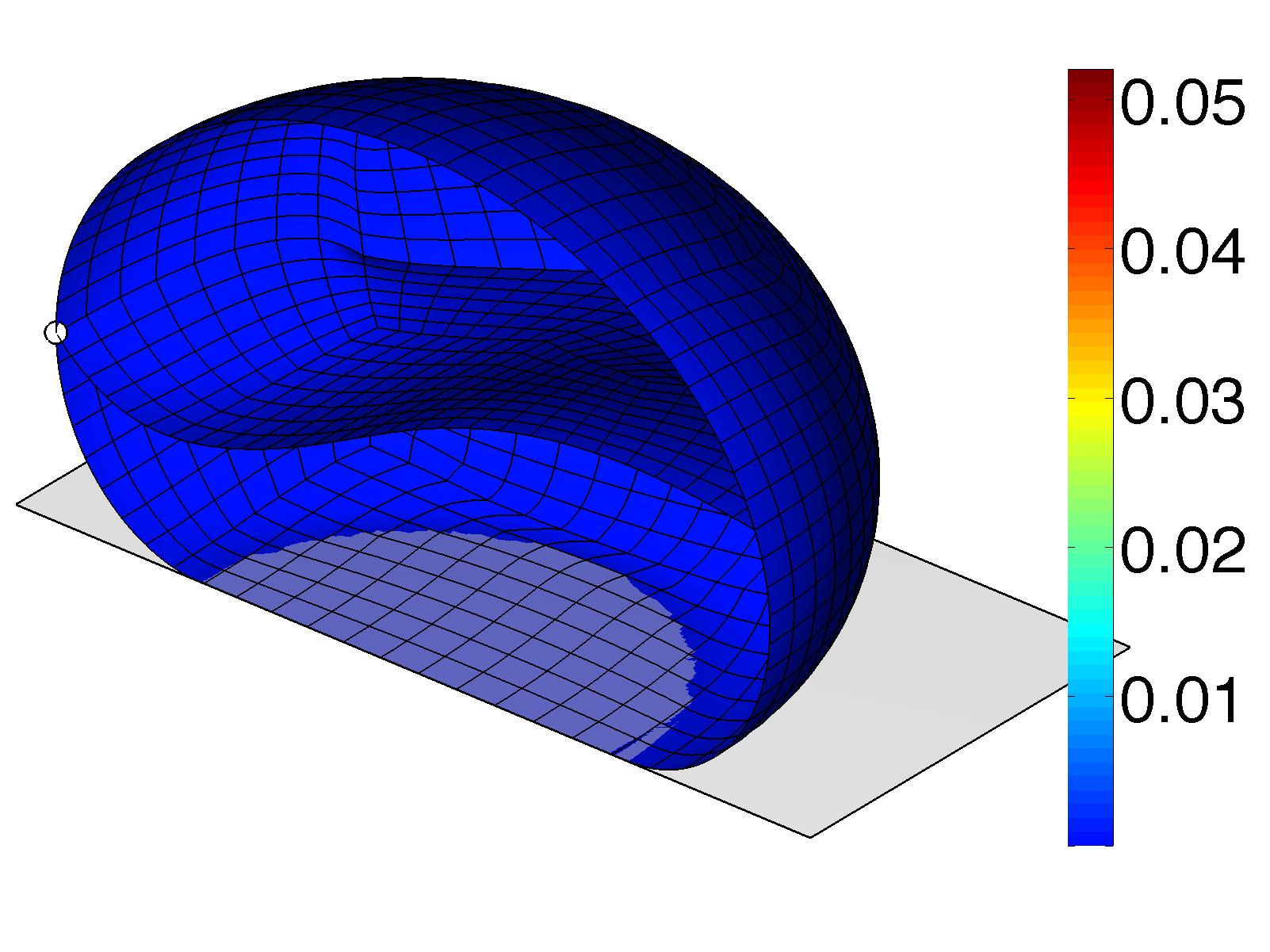}}
%\put(-7.98,-.2){\includegraphics[height=18.6mm]{../../codes/cFEAR/Input/iFSI/Droplet/oDropRoll/roll07/roll20_m8_te200_dt0p5_ba000v.jpg}}
%\put(-4.83,-.2){\includegraphics[height=18.6mm]{../../codes/cFEAR/Input/iFSI/Droplet/oDropRoll/roll07/roll20_m8_te200_dt0p5_ba100v.jpg}}
%\put(-1.68,-.2){\includegraphics[height=18.6mm]{../../codes/cFEAR/Input/iFSI/Droplet/oDropRoll/roll07/roll20_m8_te200_dt0p5_ba200v.jpg}}
%\put(1.47,-.2){\includegraphics[height=18.6mm]{../../codes/cFEAR/Input/iFSI/Droplet/oDropRoll/roll15/roll20_m8_te350_dtp5_ba400v.jpg}}
%\put(4.62,-.2){\includegraphics[height=18.6mm]{../../codes/cFEAR/Input/iFSI/Droplet/oDropRoll/roll15/roll20_m8_te350_dtp5_ba700v.jpg}}
\put(-7.98,-.2){\includegraphics[height=18.6mm]{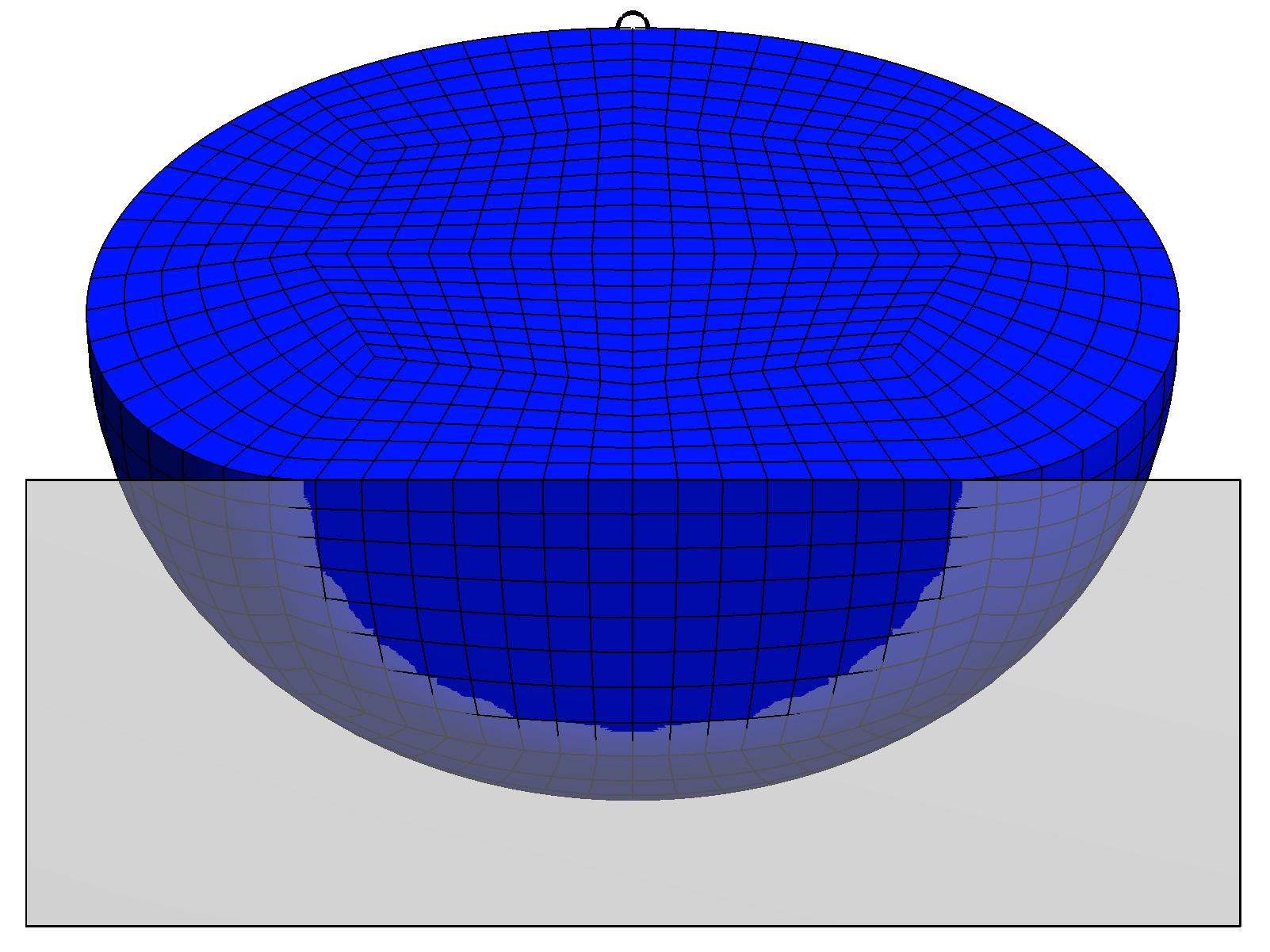}}
\put(-4.83,-.2){\includegraphics[height=18.6mm]{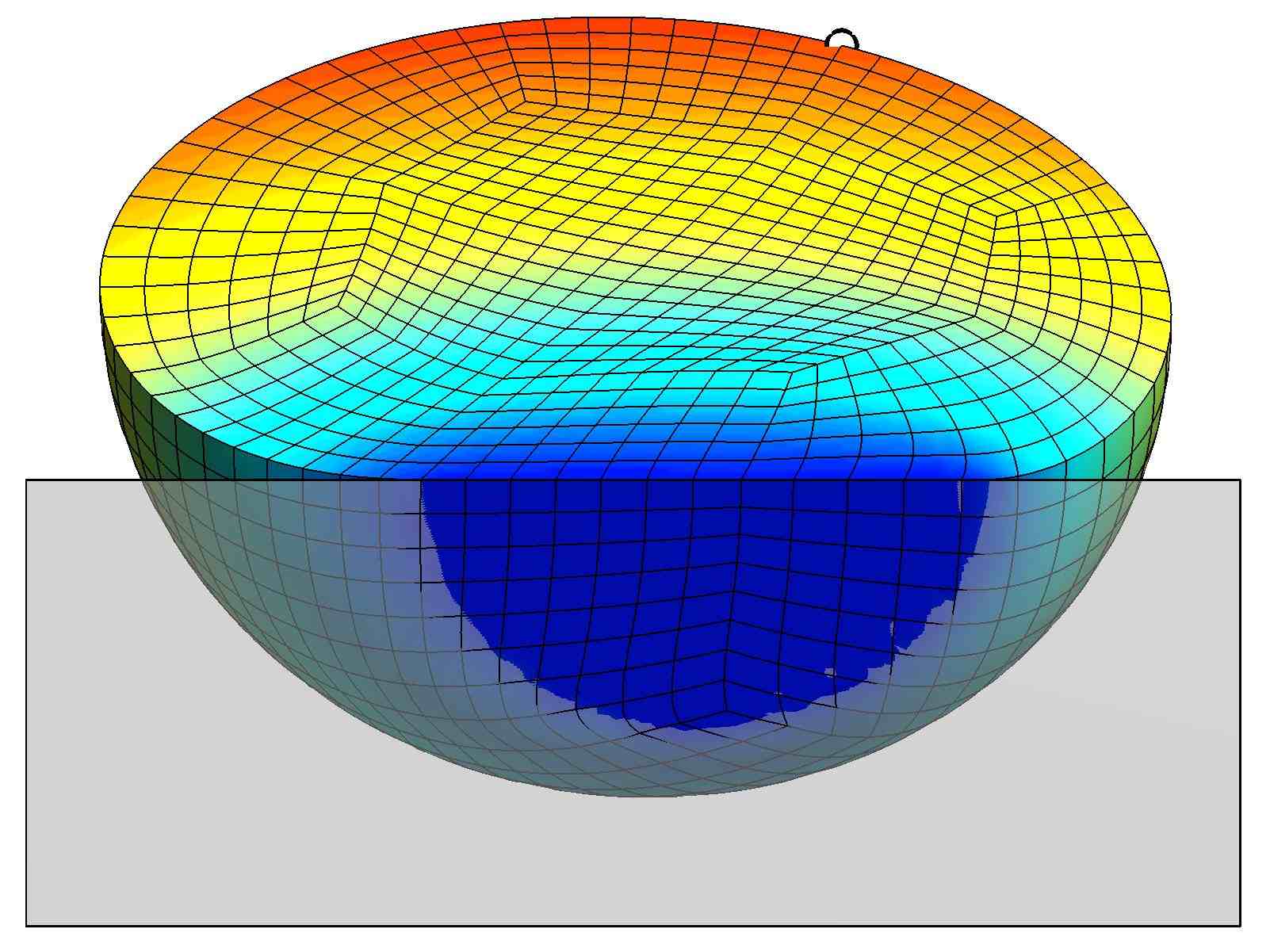}}
\put(-1.68,-.2){\includegraphics[height=18.6mm]{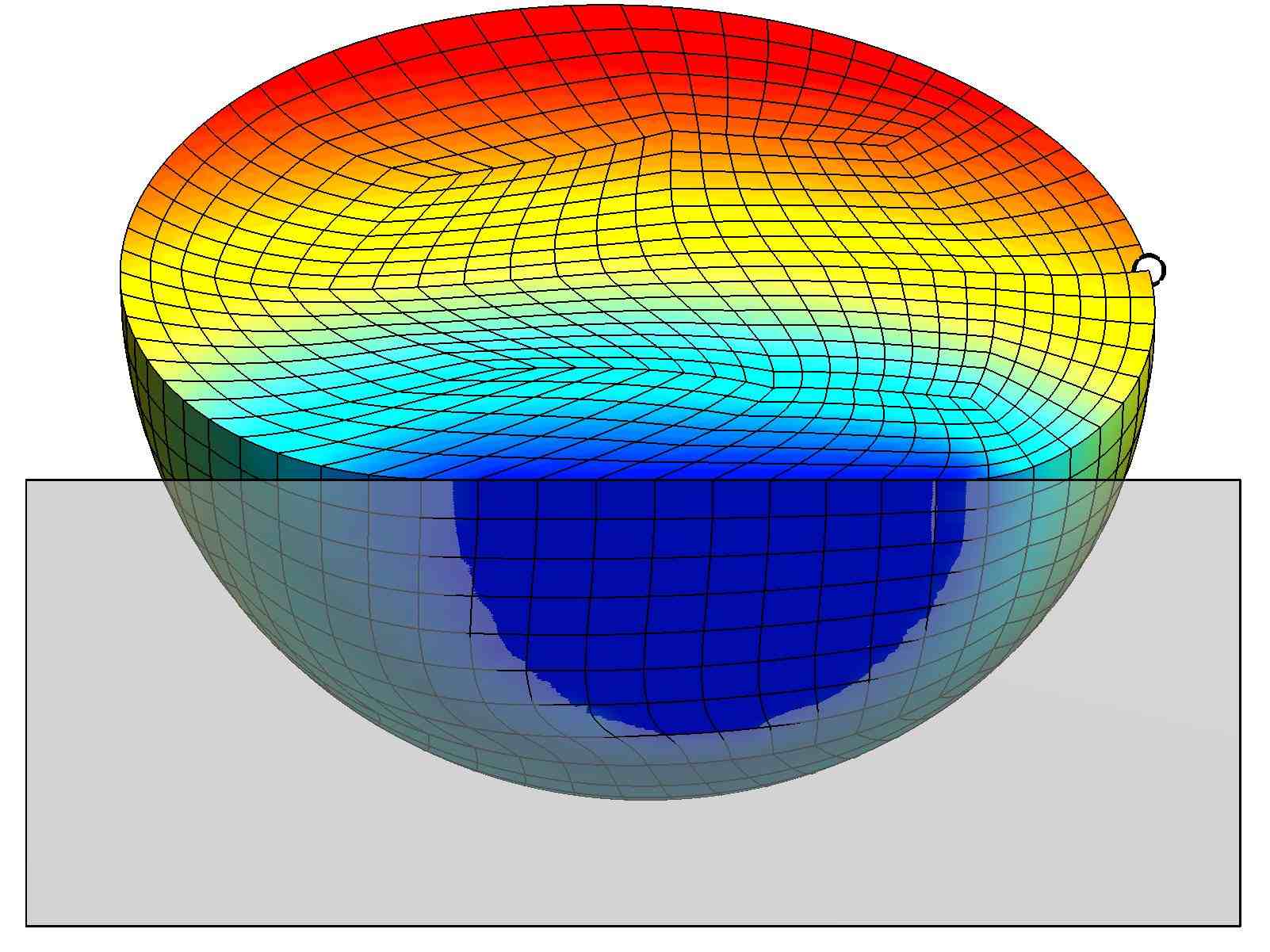}}
\put(1.47,-.2){\includegraphics[height=18.6mm]{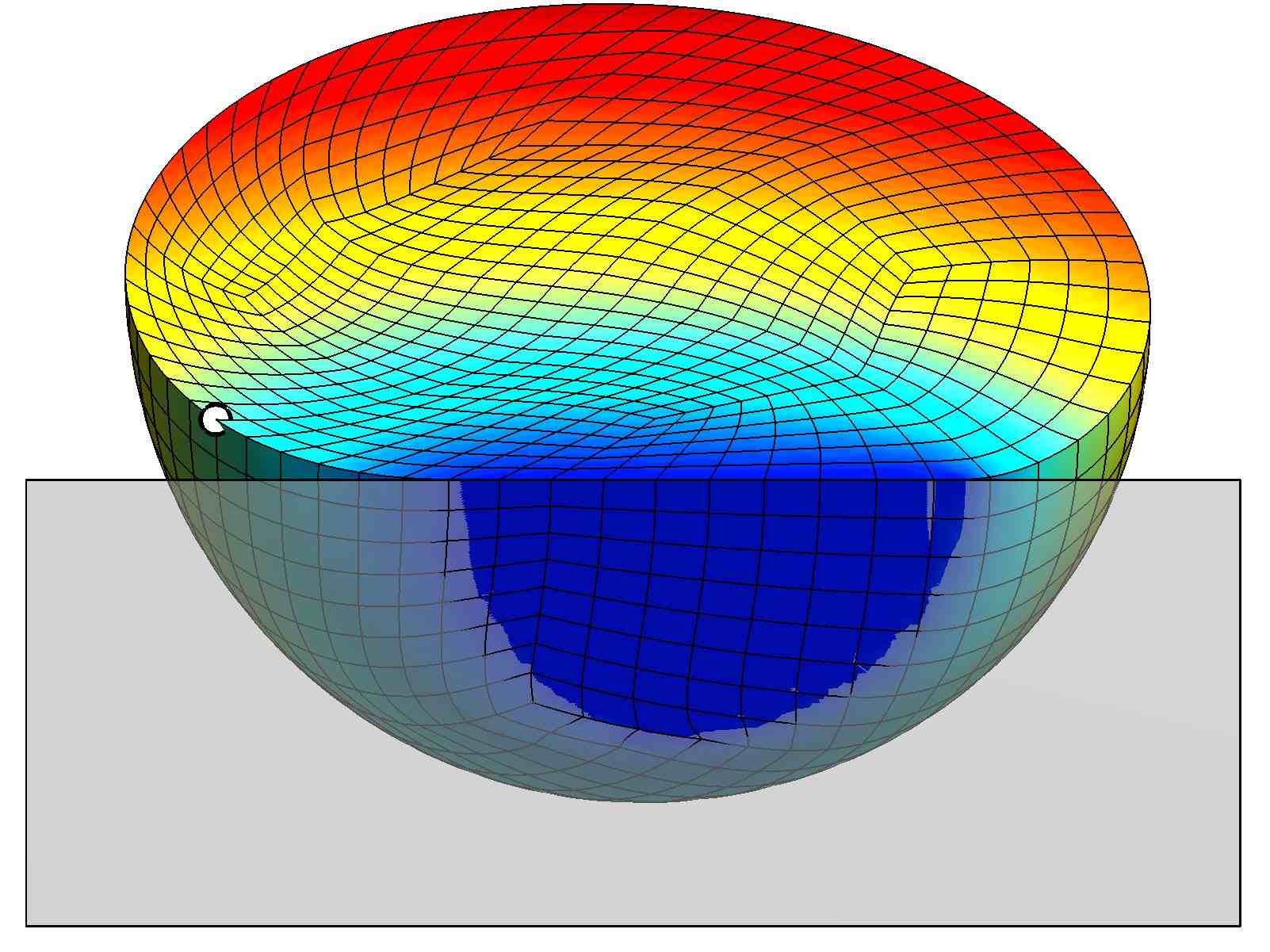}}
\put(4.62,-.2){\includegraphics[height=18.6mm]{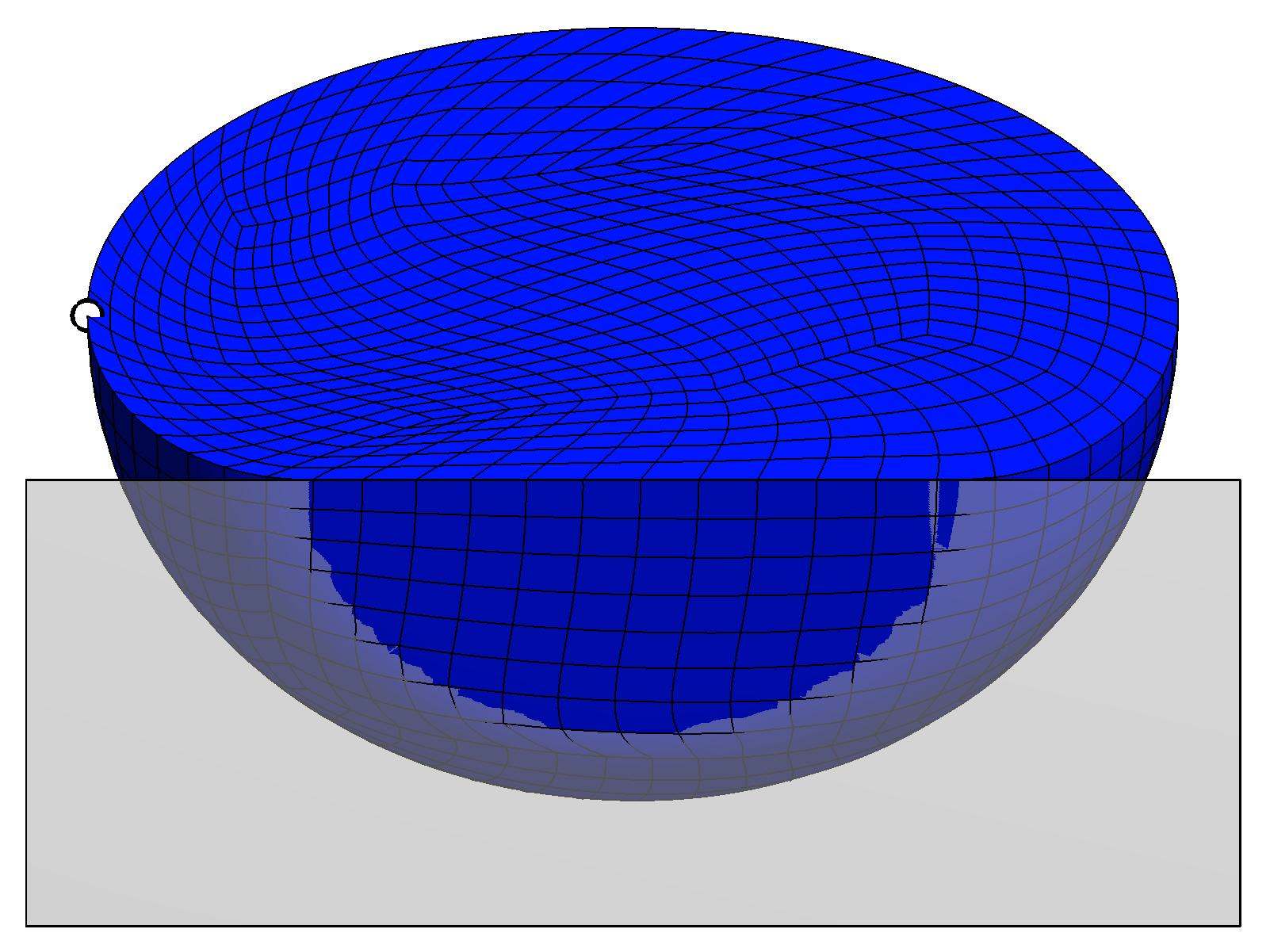}}
\end{picture}
\caption{Rolling droplet: Velocity magnitude $\norm{\bv}/v_0$ at $t=0$, $t=50\,T_0$, $t=100\,T_0$, $t=200\,T_0$ and $t=350\,T_0$ (left to right) for $\beta_0=20^\circ$ and $m=8$. 
Only half of the symmetric droplet is shown.
In the top panel the symmetry surface is removed and instead a selected material plane is tracked during deformation.
A single fluid particle is marked by `$\circ$'.}  
\label{f:rdrop_v}
\end{center}
\end{figure}

\begin{figure}[h]
\begin{center} \unitlength1cm
\begin{picture}(0,4)
%\put(-8.1,1.5){\includegraphics[height=25mm]{../../codes/cFEAR/Input/iFSI/Droplet/oDropRoll/roll07/roll20_m8_te200_dt0p5_000p.jpg}}
%\put(-4.95,1.5){\includegraphics[height=25mm]{../../codes/cFEAR/Input/iFSI/Droplet/oDropRoll/roll07/roll20_m8_te200_dt0p5_100p.jpg}}
%\put(-1.8,1.5){\includegraphics[height=25mm]{../../codes/cFEAR/Input/iFSI/Droplet/oDropRoll/roll07/roll20_m8_te200_dt0p5_200p.jpg}}
%\put(1.35,1.5){\includegraphics[height=25mm]{../../codes/cFEAR/Input/iFSI/Droplet/oDropRoll/roll15/roll20_m8_te350_dtp5_400p.jpg}}
%\put(4.65,1.5){\includegraphics[height=25mm]{../../codes/cFEAR/Input/iFSI/Droplet/oDropRoll/roll15/roll20_m8_te350_dtp5_700p.jpg}}
\put(-8.1,1.5){\includegraphics[height=25mm]{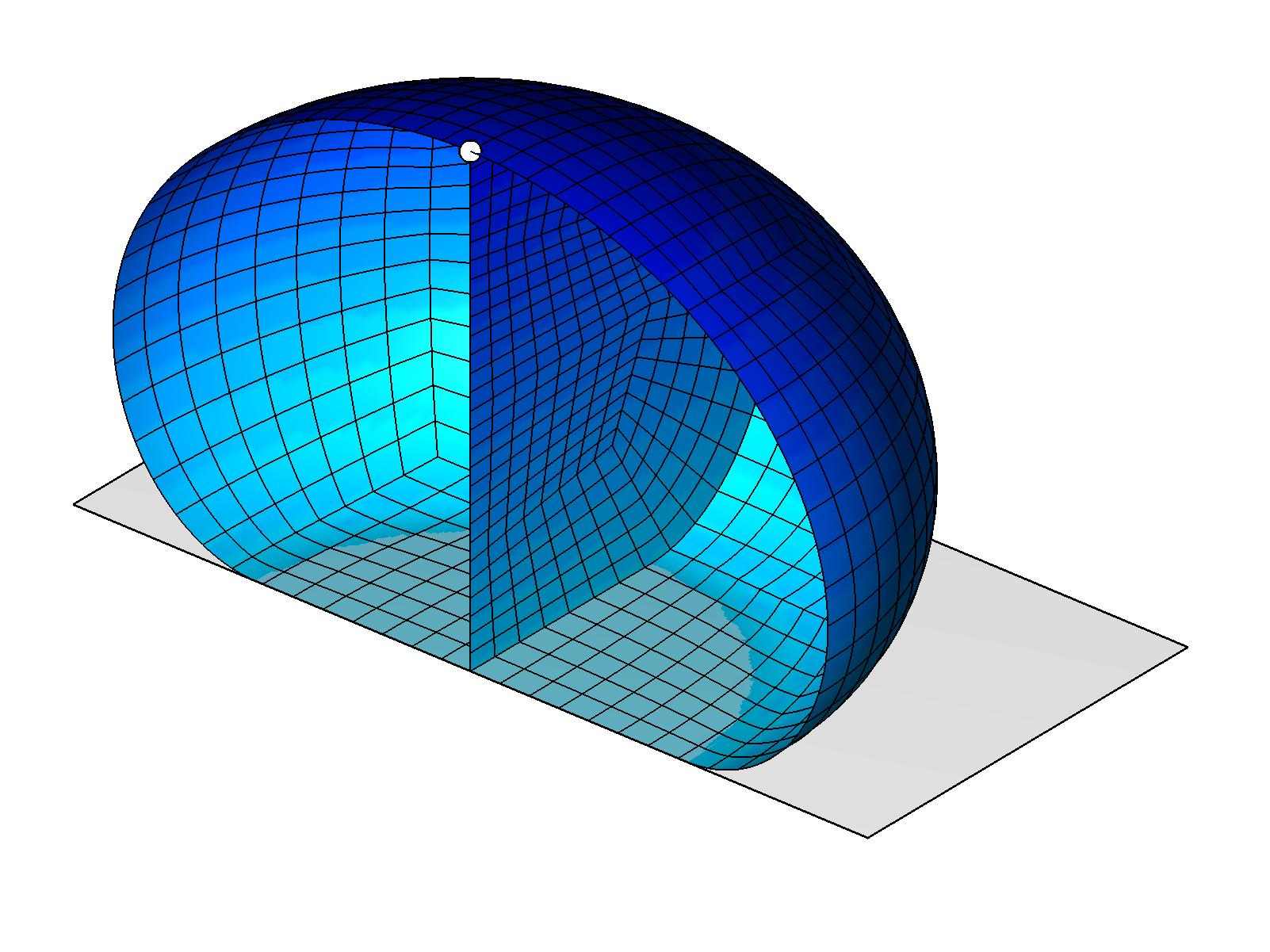}}
\put(-4.95,1.5){\includegraphics[height=25mm]{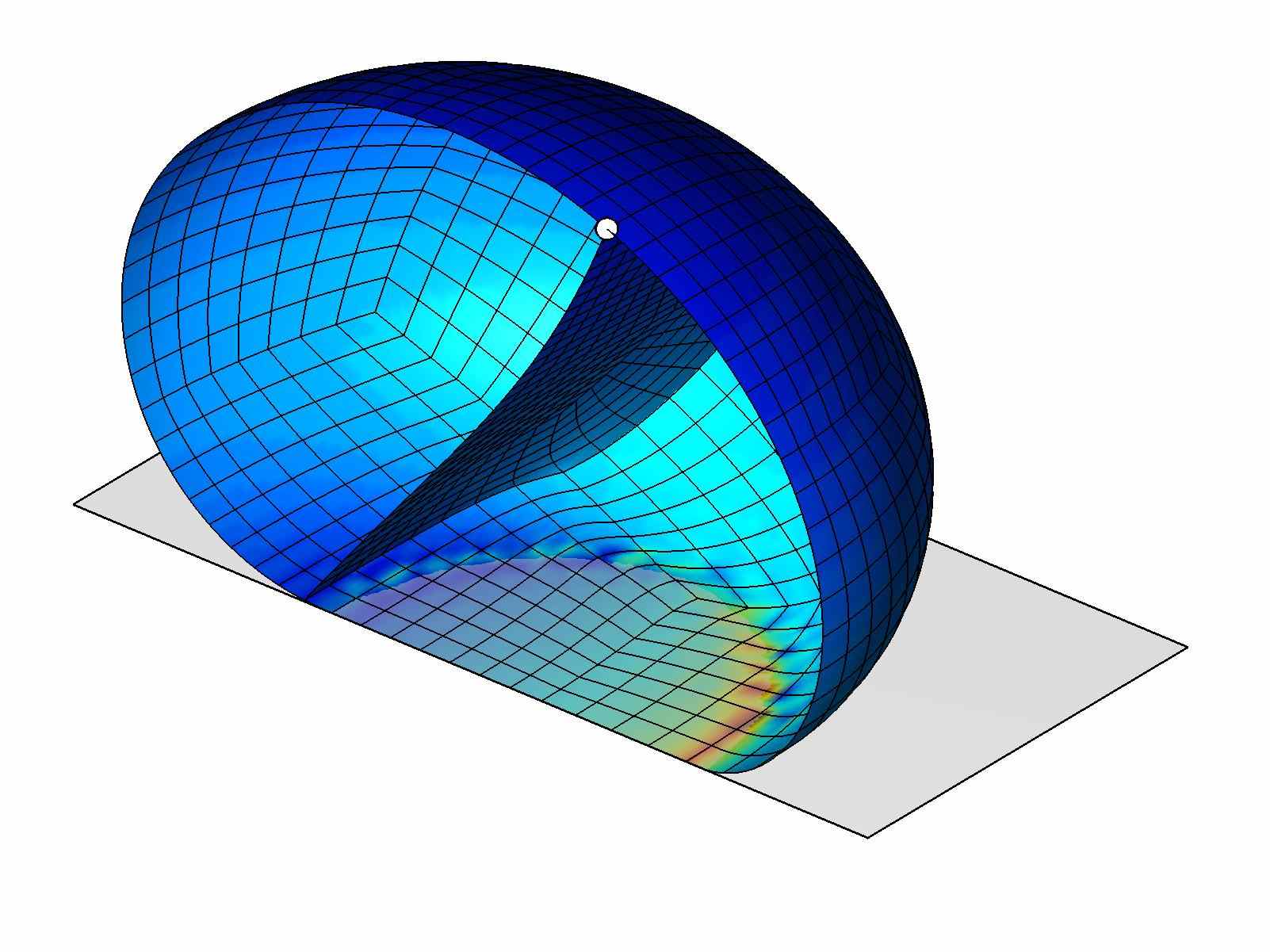}}
\put(-1.8,1.5){\includegraphics[height=25mm]{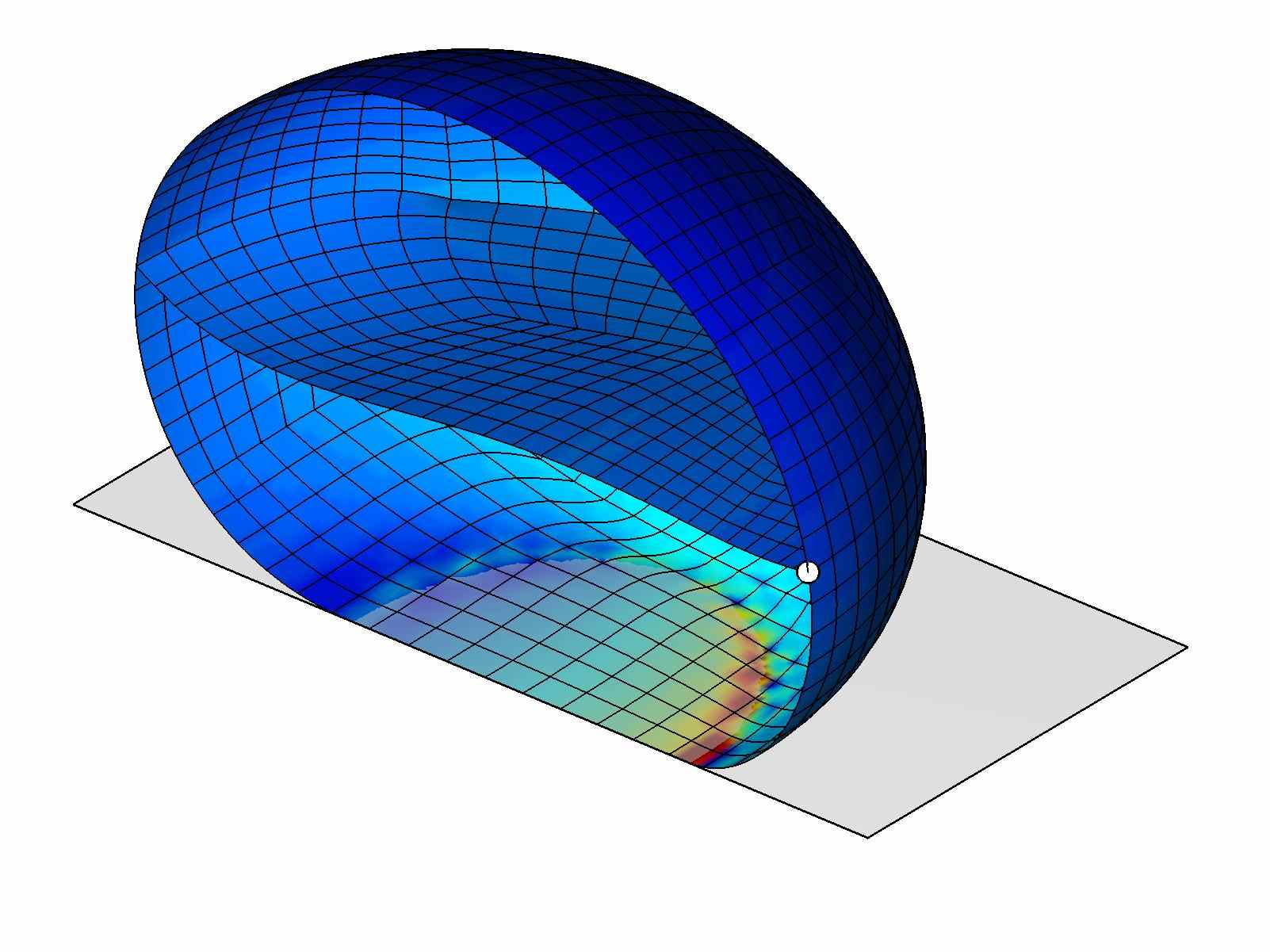}}
\put(1.35,1.5){\includegraphics[height=25mm]{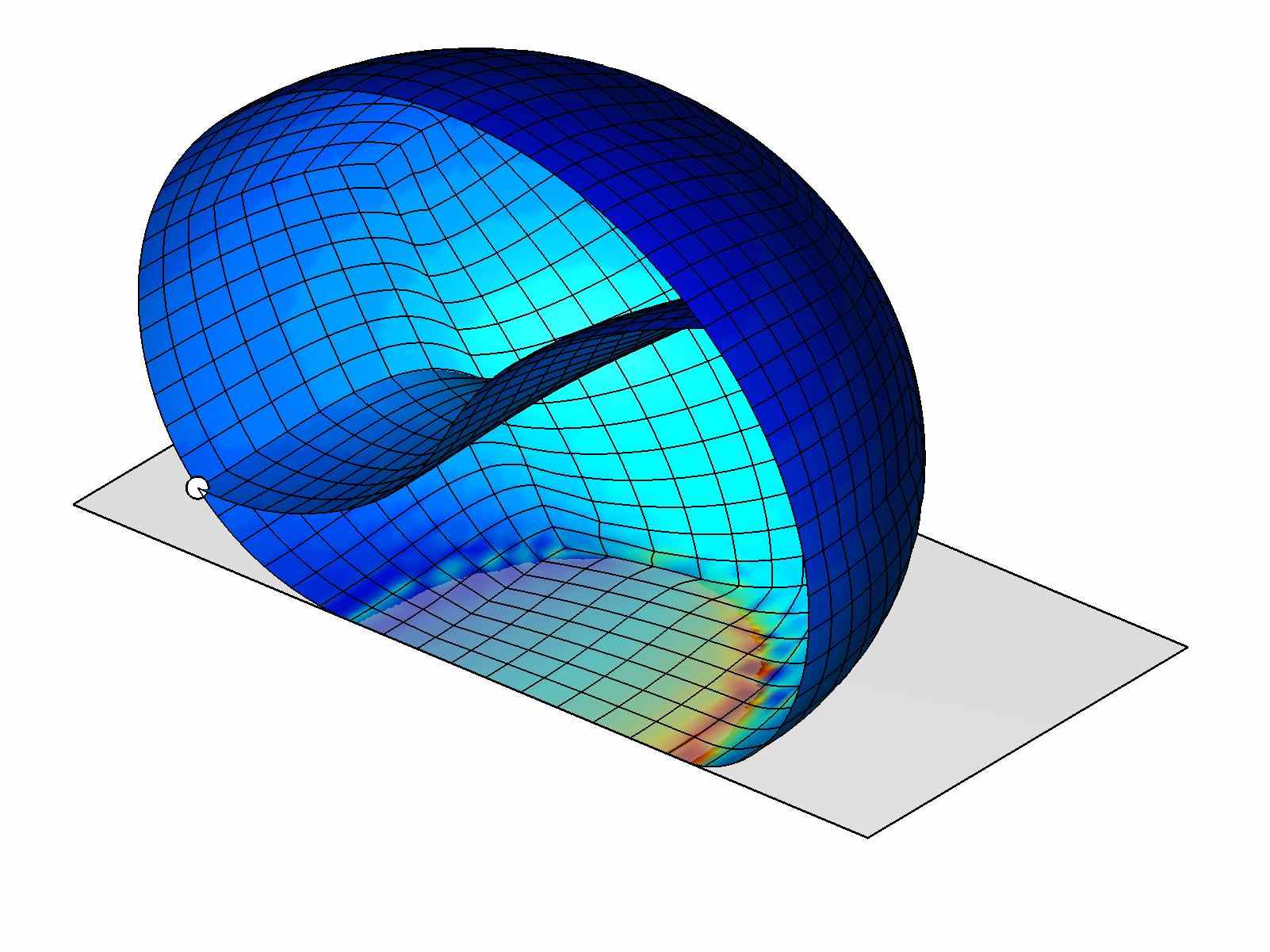}}
\put(4.65,1.5){\includegraphics[height=25mm]{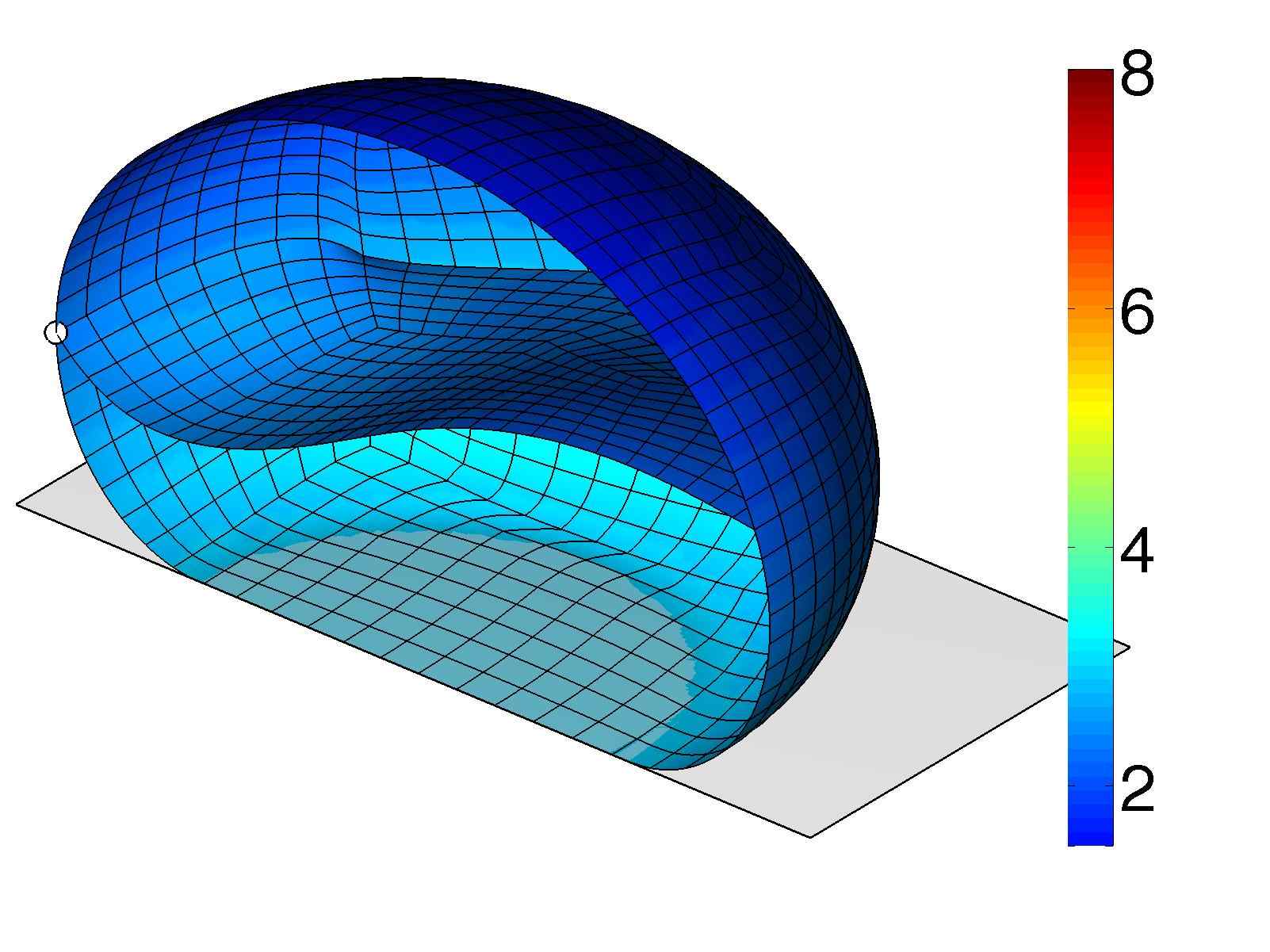}}
%\put(-7.98,-.2){\includegraphics[height=18.6mm]{../../codes/cFEAR/Input/iFSI/Droplet/oDropRoll/roll07/roll20_m8_te200_dt0p5_ba000p.jpg}}
%\put(-4.83,-.2){\includegraphics[height=18.6mm]{../../codes/cFEAR/Input/iFSI/Droplet/oDropRoll/roll07/roll20_m8_te200_dt0p5_ba100p.jpg}}
%\put(-1.68,-.2){\includegraphics[height=18.6mm]{../../codes/cFEAR/Input/iFSI/Droplet/oDropRoll/roll07/roll20_m8_te200_dt0p5_ba200p.jpg}}
%\put(1.47,-.2){\includegraphics[height=18.6mm]{../../codes/cFEAR/Input/iFSI/Droplet/oDropRoll/roll15/roll20_m8_te350_dtp5_ba400p.jpg}}
%\put(4.62,-.2){\includegraphics[height=18.6mm]{../../codes/cFEAR/Input/iFSI/Droplet/oDropRoll/roll15/roll20_m8_te350_dtp5_ba700p.jpg}}
\put(-7.98,-.2){\includegraphics[height=18.6mm]{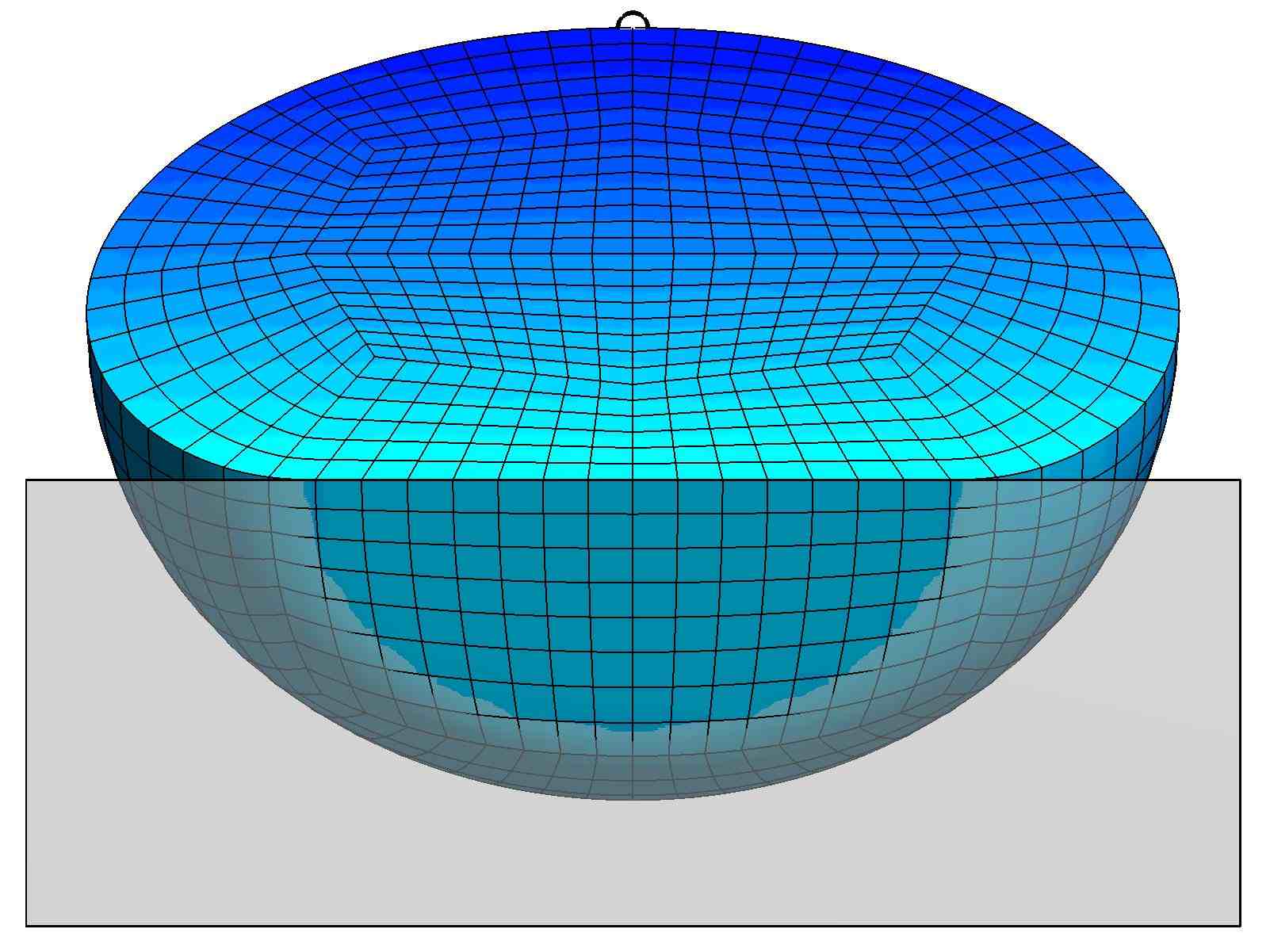}}
\put(-4.83,-.2){\includegraphics[height=18.6mm]{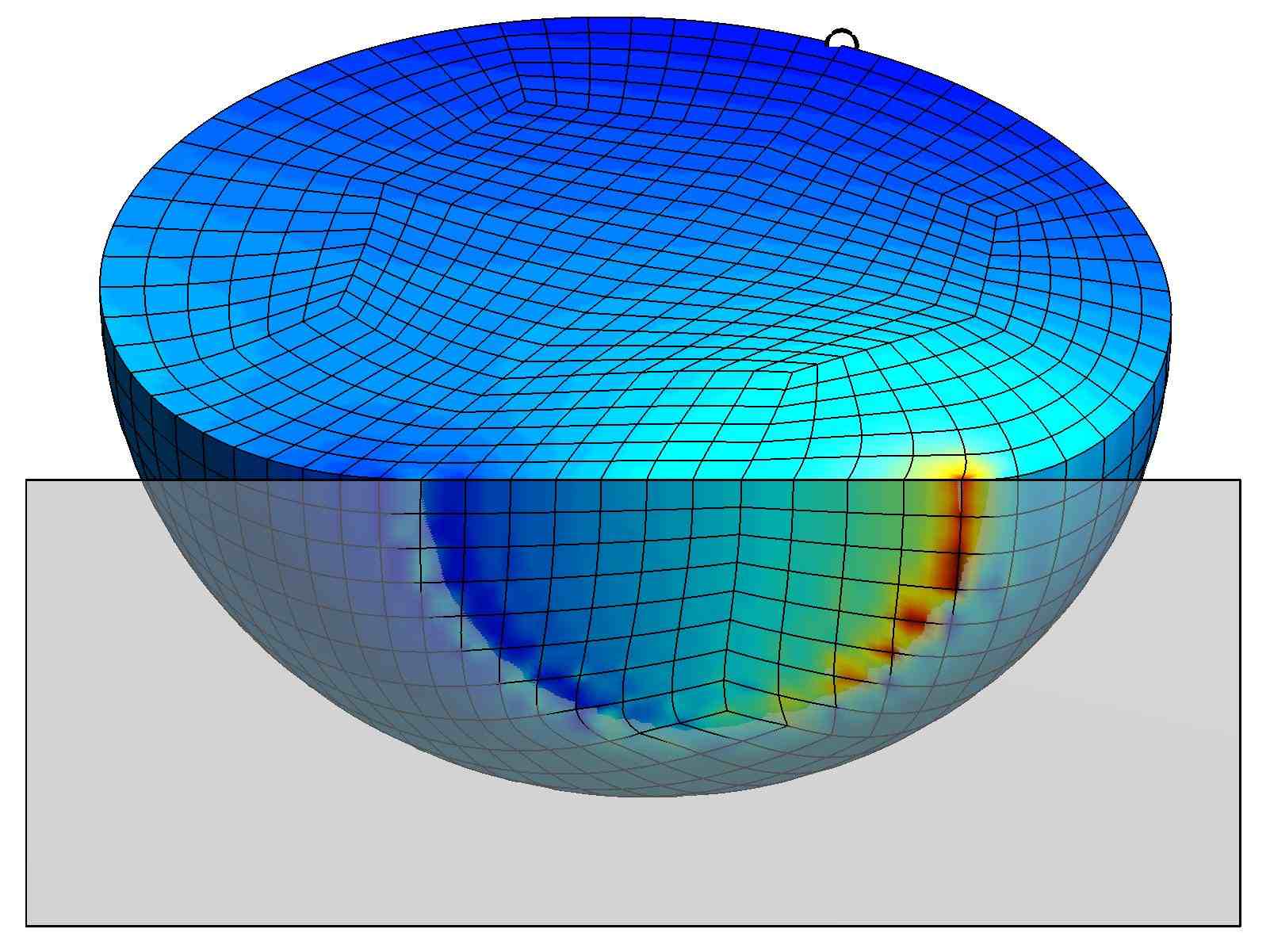}}
\put(-1.68,-.2){\includegraphics[height=18.6mm]{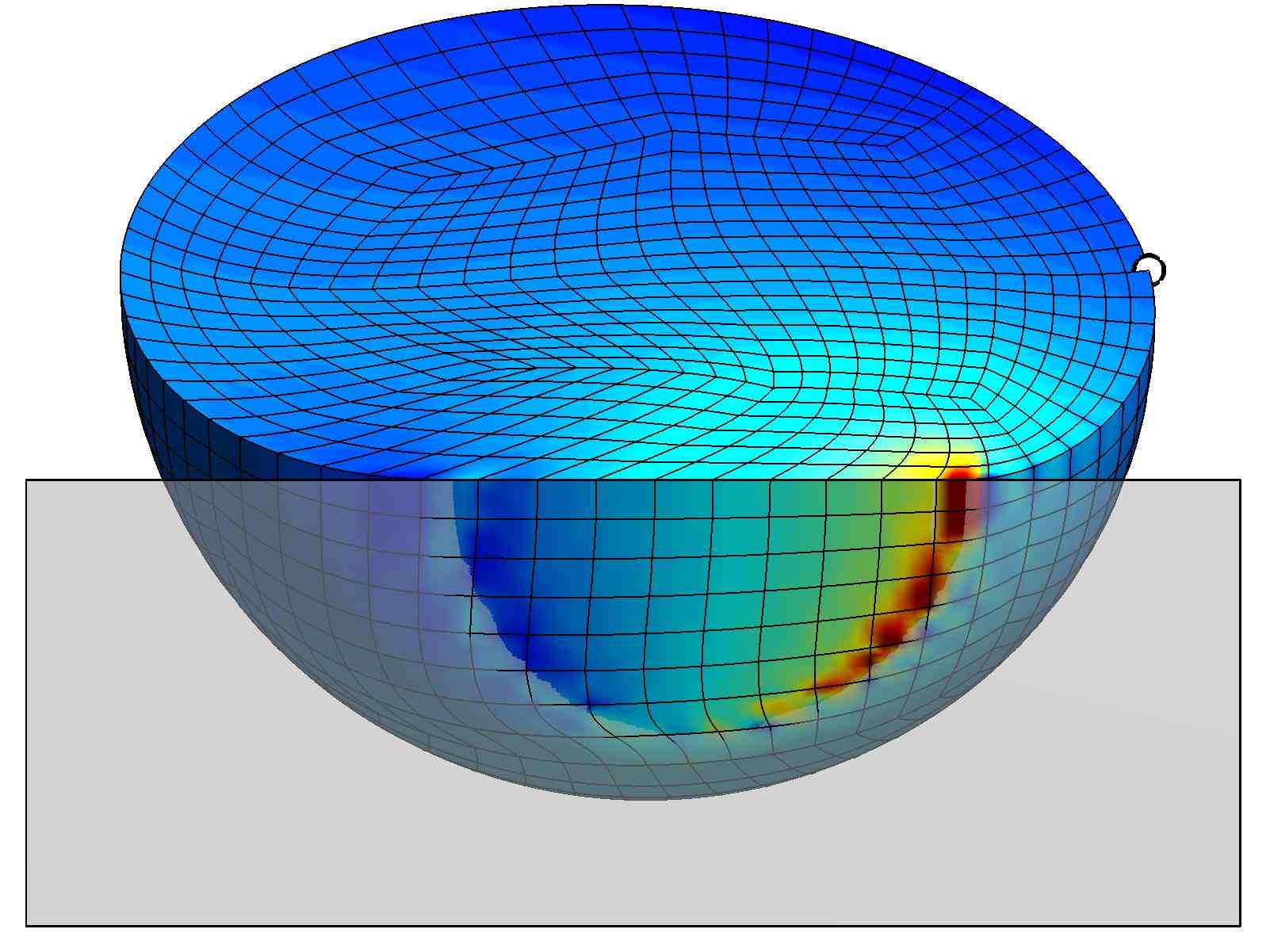}}
\put(1.47,-.2){\includegraphics[height=18.6mm]{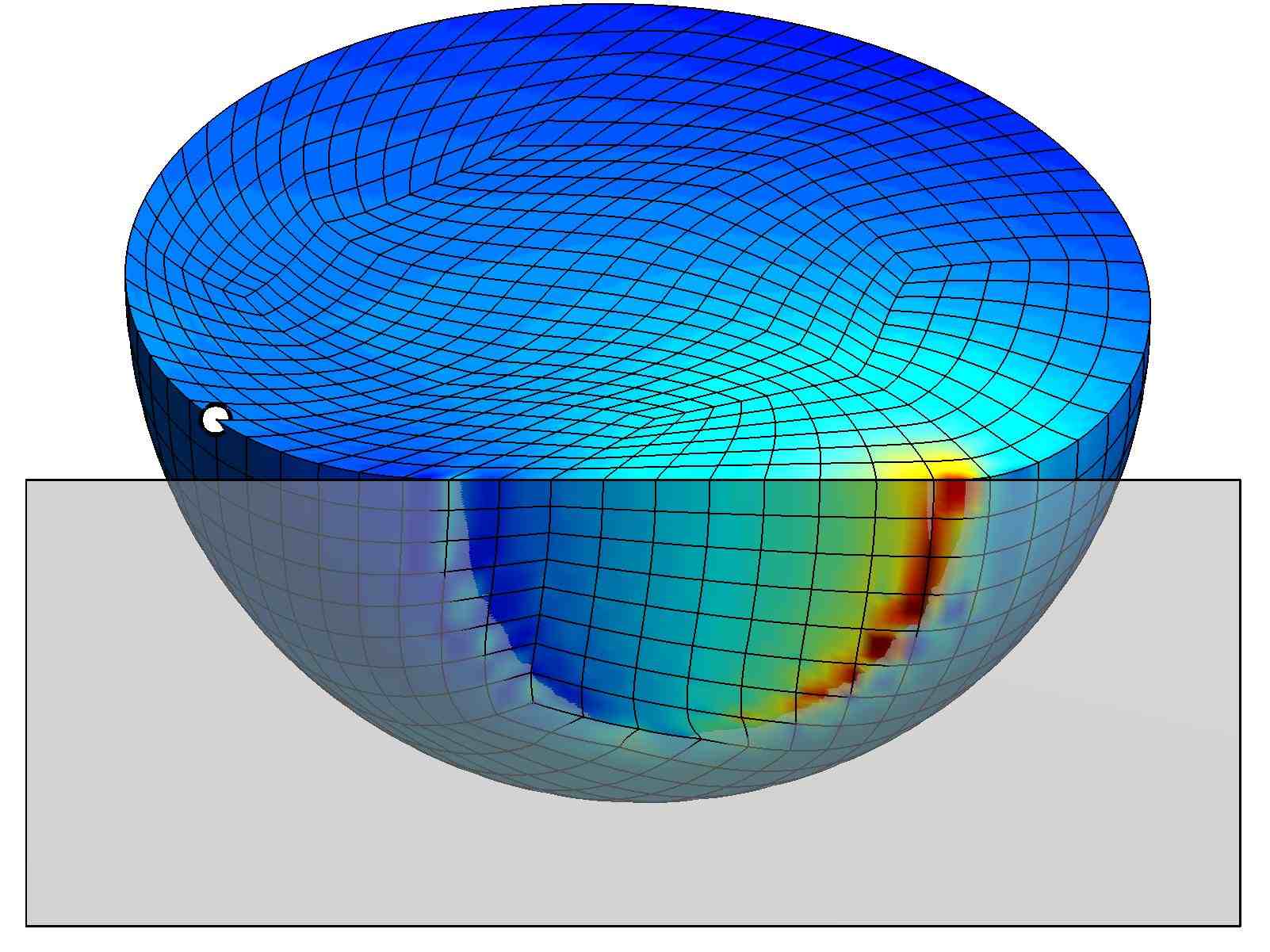}}
\put(4.62,-.2){\includegraphics[height=18.6mm]{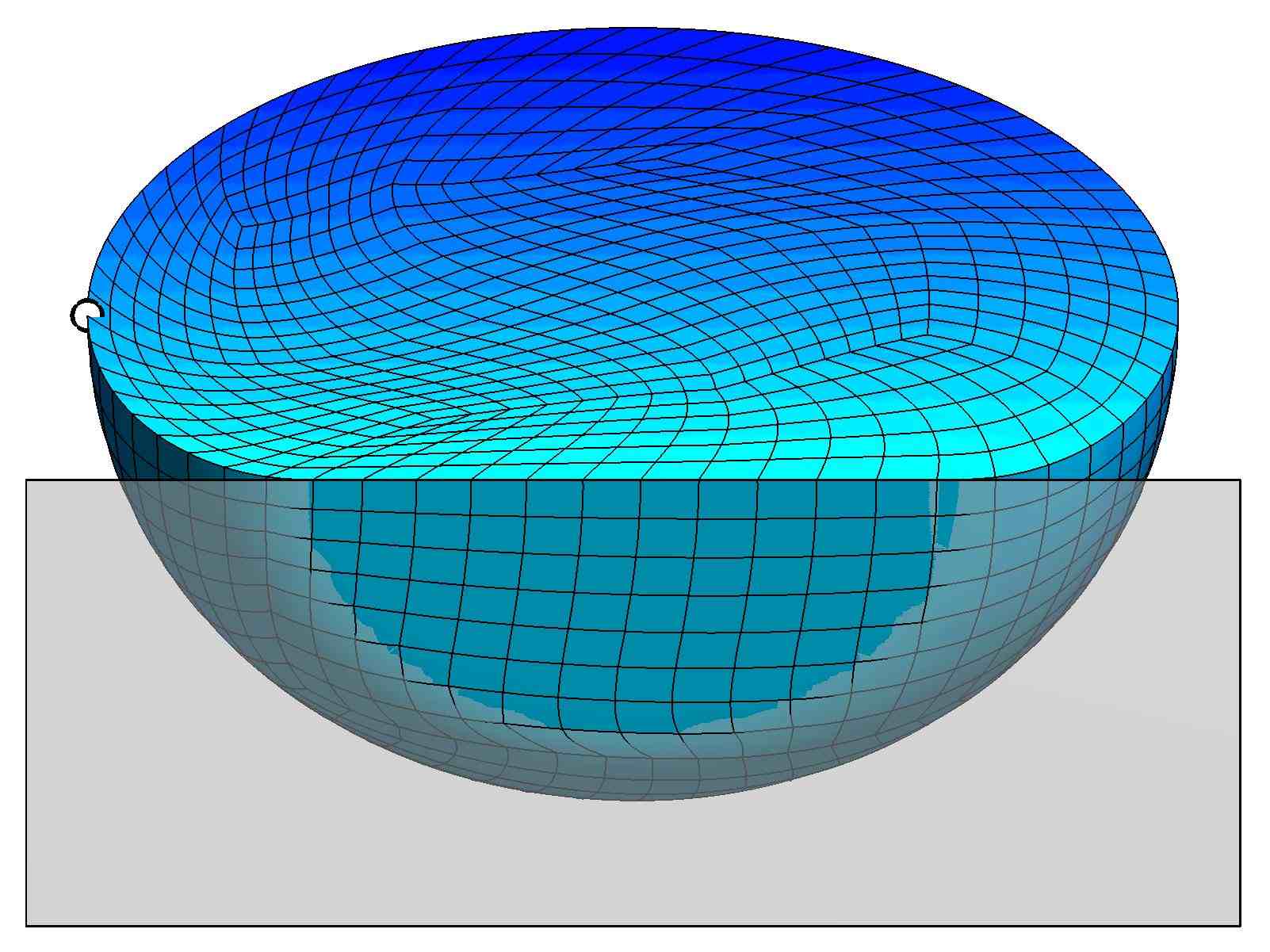}}
\end{picture}
\caption{Rolling droplet: pressure field $p/p_0$ at $t=0$, $t=50\,T_0$, $t=100\,T_0$, $t=200\,T_0$ and $t=350\,T_0$ (left to right) for $\beta_0=20^\circ$ and $m=8$}  
\label{f:rdrop_p}
\end{center}
\end{figure}

\begin{figure}[h]
\begin{center} \unitlength1cm
\begin{picture}(0,4)
%\put(-8.1,1.5){\includegraphics[height=25mm]{../../codes/cFEAR/Input/iFSI/Droplet/oDropRoll/roll07/roll20_m8_te200_dt0p5_000ps.jpg}}
%\put(-4.95,1.5){\includegraphics[height=25mm]{../../codes/cFEAR/Input/iFSI/Droplet/oDropRoll/roll07/roll20_m8_te200_dt0p5_100ps.jpg}}
%\put(-1.8,1.5){\includegraphics[height=25mm]{../../codes/cFEAR/Input/iFSI/Droplet/oDropRoll/roll07/roll20_m8_te200_dt0p5_200ps.jpg}}
%\put(1.35,1.5){\includegraphics[height=25mm]{../../codes/cFEAR/Input/iFSI/Droplet/oDropRoll/roll15/roll20_m8_te350_dtp5_400ps.jpg}}
%\put(4.65,1.5){\includegraphics[height=25mm]{../../codes/cFEAR/Input/iFSI/Droplet/oDropRoll/roll15/roll20_m8_te350_dtp5_700ps.jpg}}
\put(-8.1,1.5){\includegraphics[height=25mm]{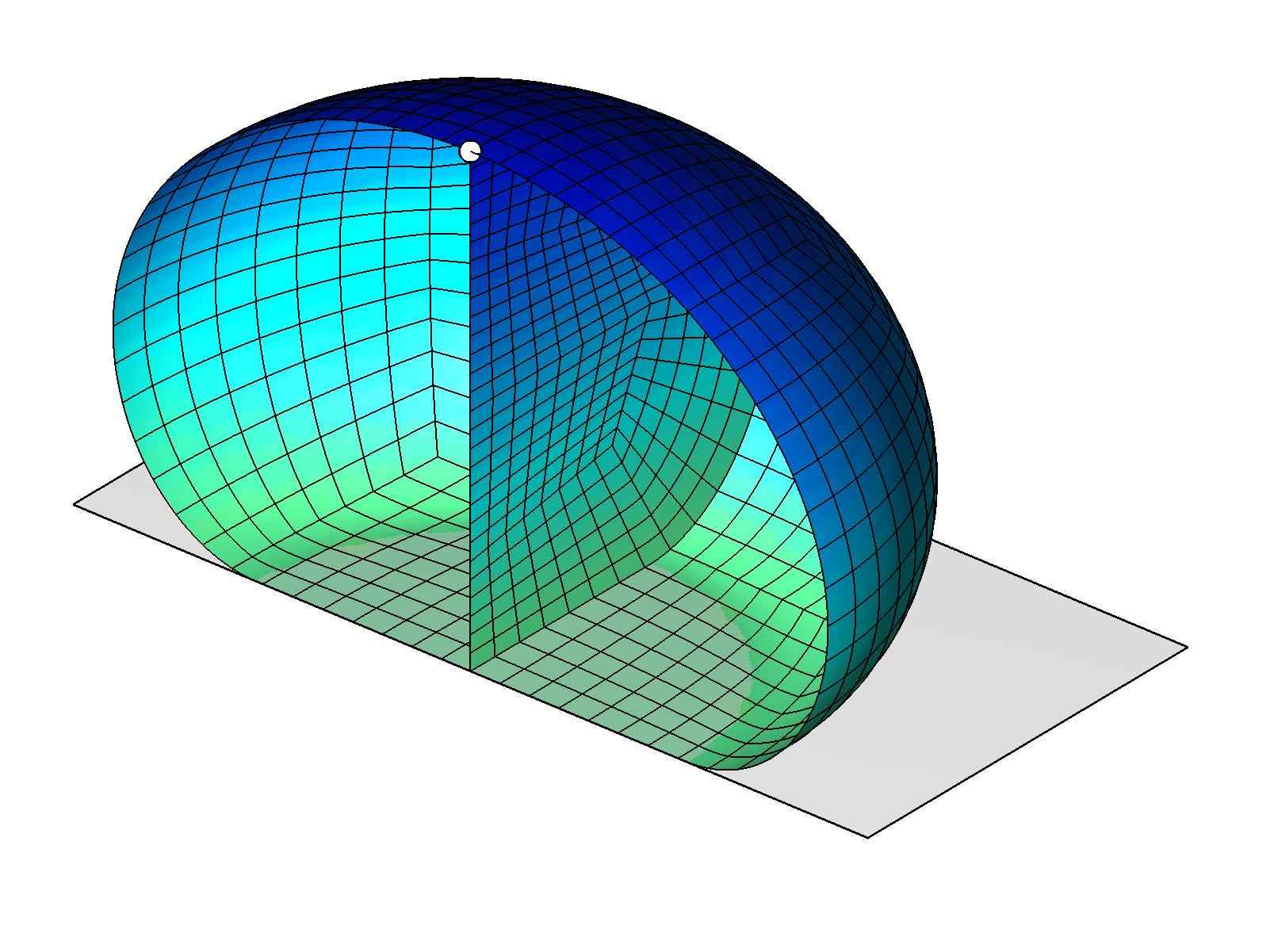}}
\put(-4.95,1.5){\includegraphics[height=25mm]{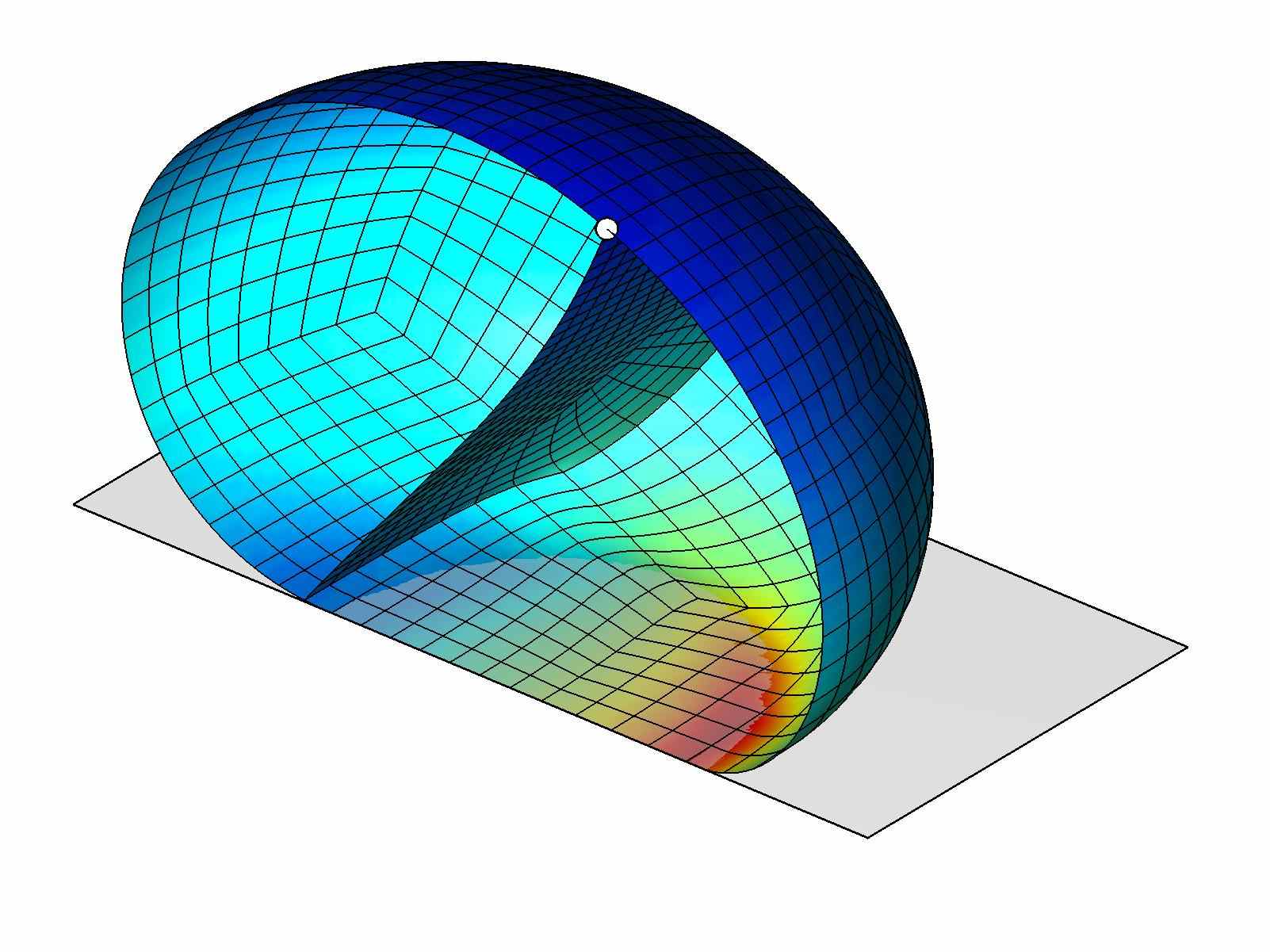}}
\put(-1.8,1.5){\includegraphics[height=25mm]{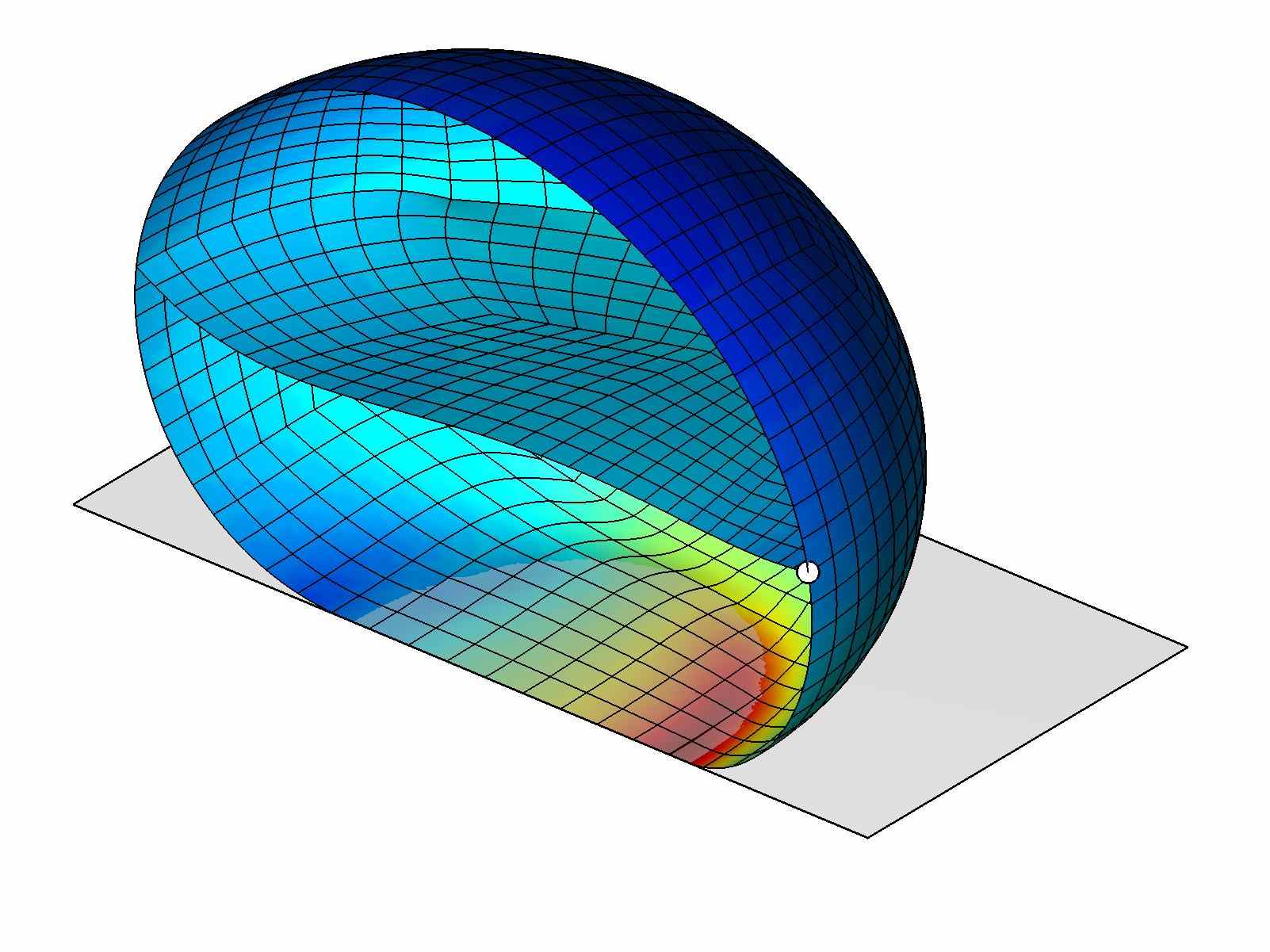}}
\put(1.35,1.5){\includegraphics[height=25mm]{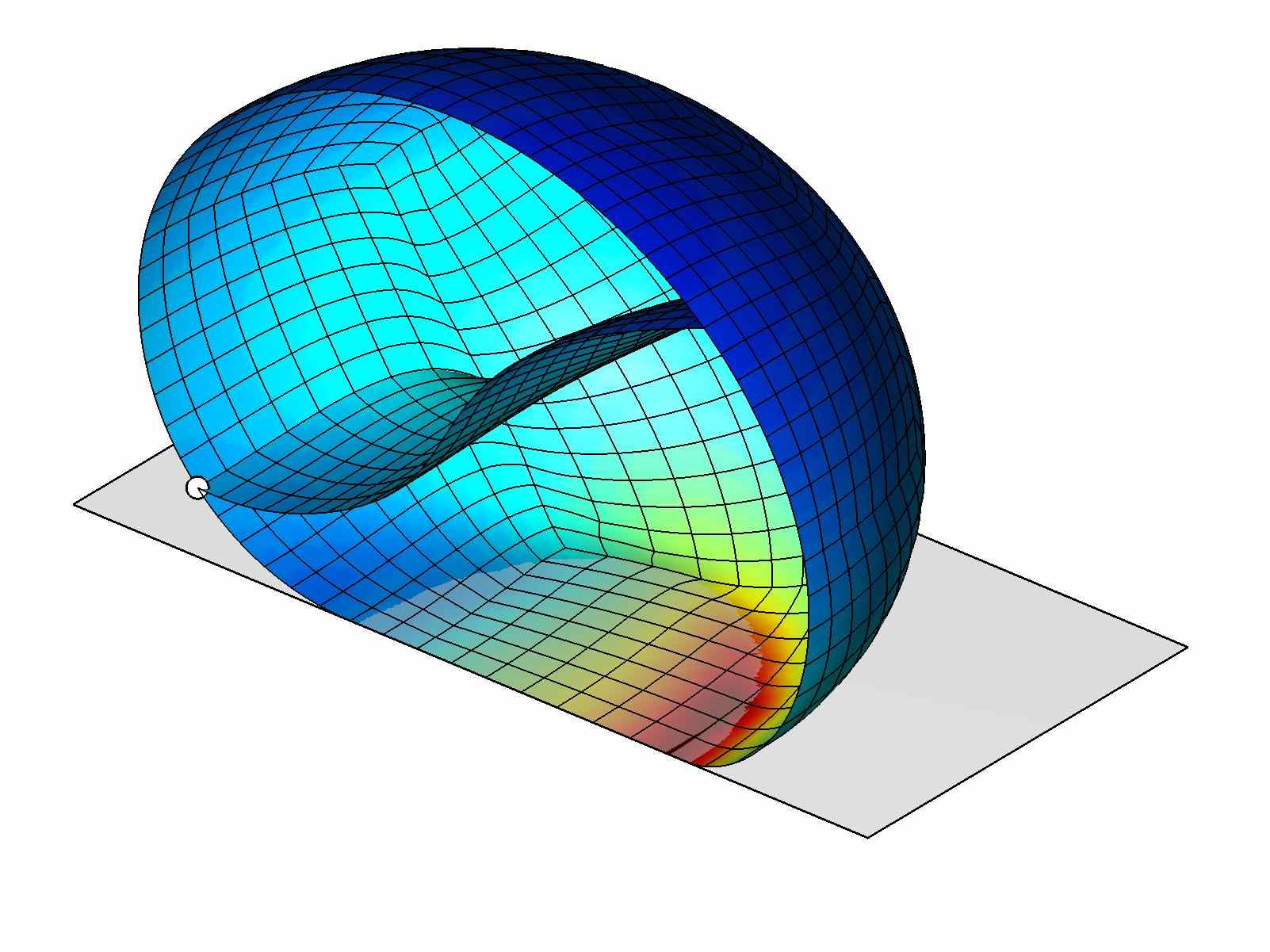}}
\put(4.65,1.5){\includegraphics[height=25mm]{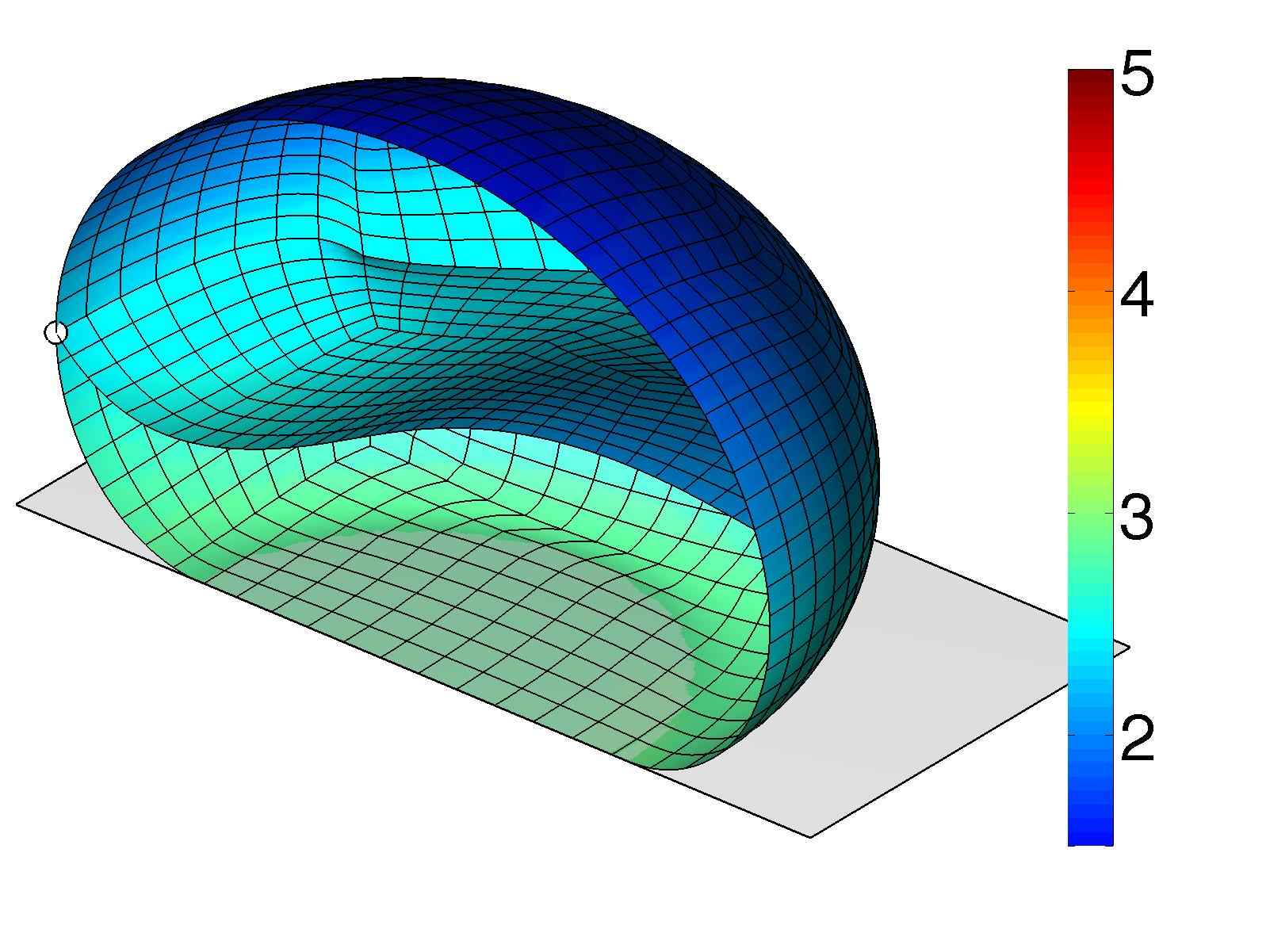}}
%\put(-7.98,-.2){\includegraphics[height=18.6mm]{../../codes/cFEAR/Input/iFSI/Droplet/oDropRoll/roll07/roll20_m8_te200_dt0p5_ba000ps.jpg}}
%\put(-4.83,-.2){\includegraphics[height=18.6mm]{../../codes/cFEAR/Input/iFSI/Droplet/oDropRoll/roll07/roll20_m8_te200_dt0p5_ba100ps.jpg}}
%\put(-1.68,-.2){\includegraphics[height=18.6mm]{../../codes/cFEAR/Input/iFSI/Droplet/oDropRoll/roll07/roll20_m8_te200_dt0p5_ba200ps.jpg}}
%\put(1.47,-.2){\includegraphics[height=18.6mm]{../../codes/cFEAR/Input/iFSI/Droplet/oDropRoll/roll15/roll20_m8_te350_dtp5_ba400ps.jpg}}
%\put(4.62,-.2){\includegraphics[height=18.6mm]{../../codes/cFEAR/Input/iFSI/Droplet/oDropRoll/roll15/roll20_m8_te350_dtp5_ba700ps.jpg}}
\put(-7.98,-.2){\includegraphics[height=18.6mm]{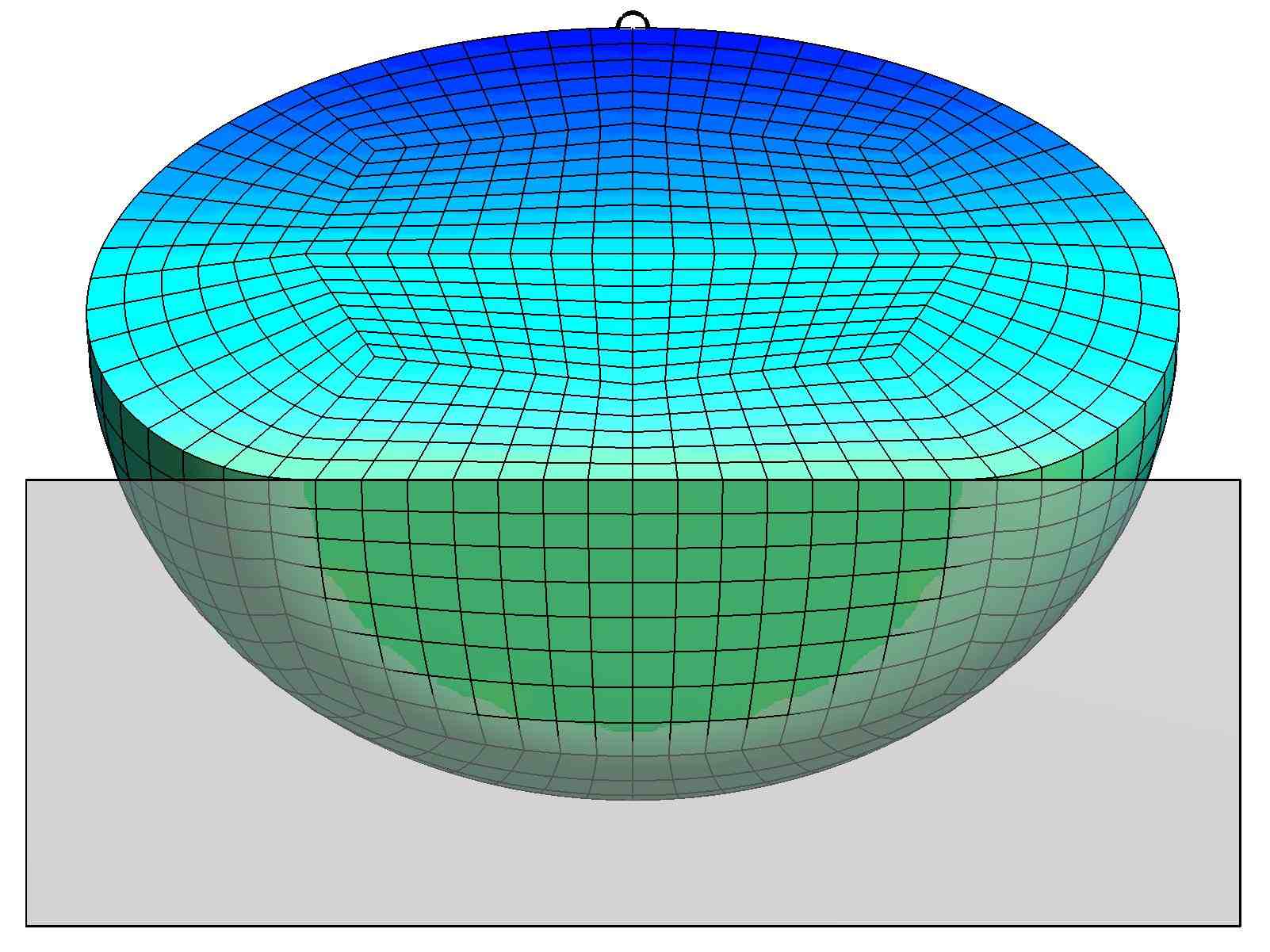}}
\put(-4.83,-.2){\includegraphics[height=18.6mm]{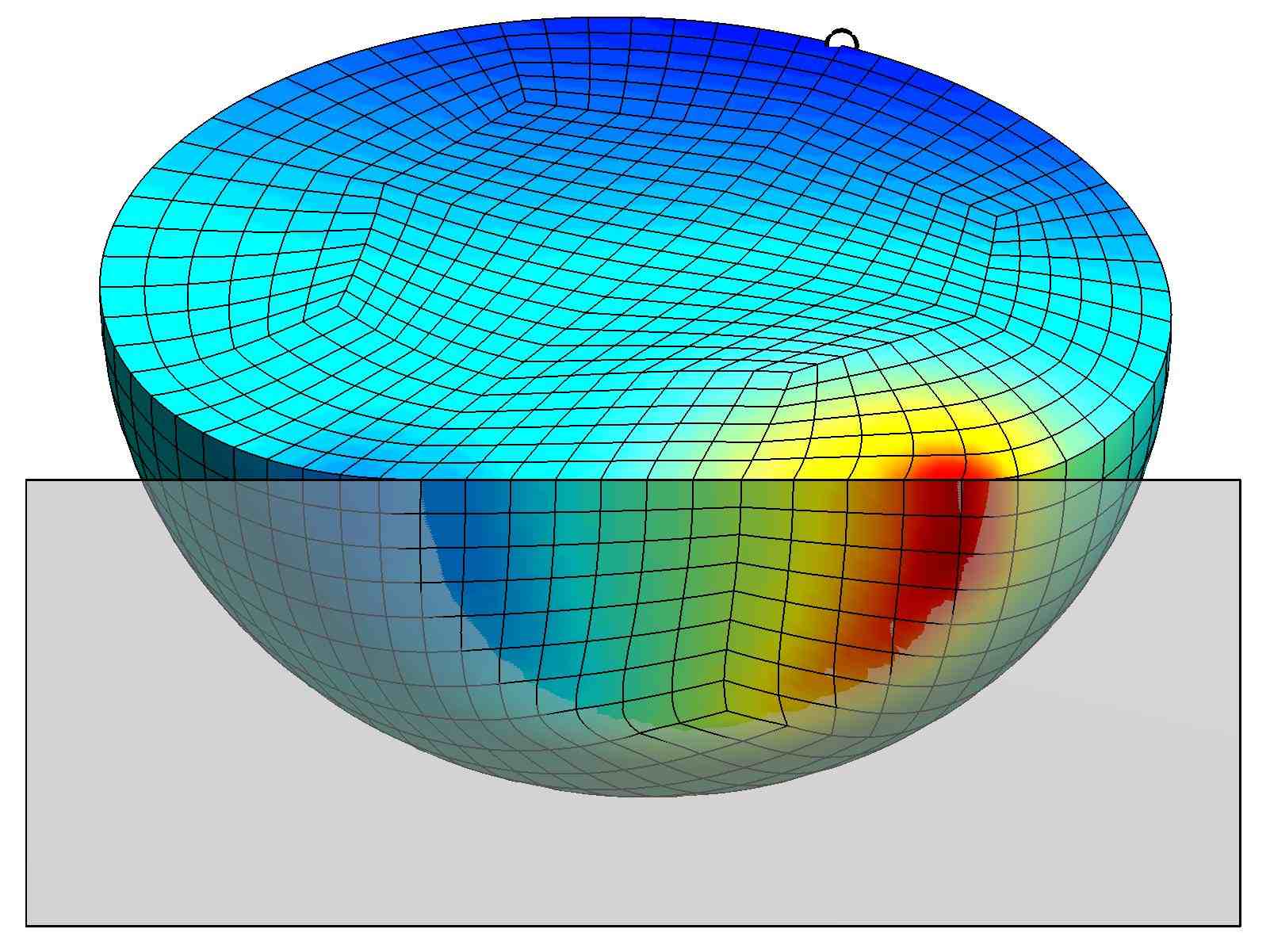}}
\put(-1.68,-.2){\includegraphics[height=18.6mm]{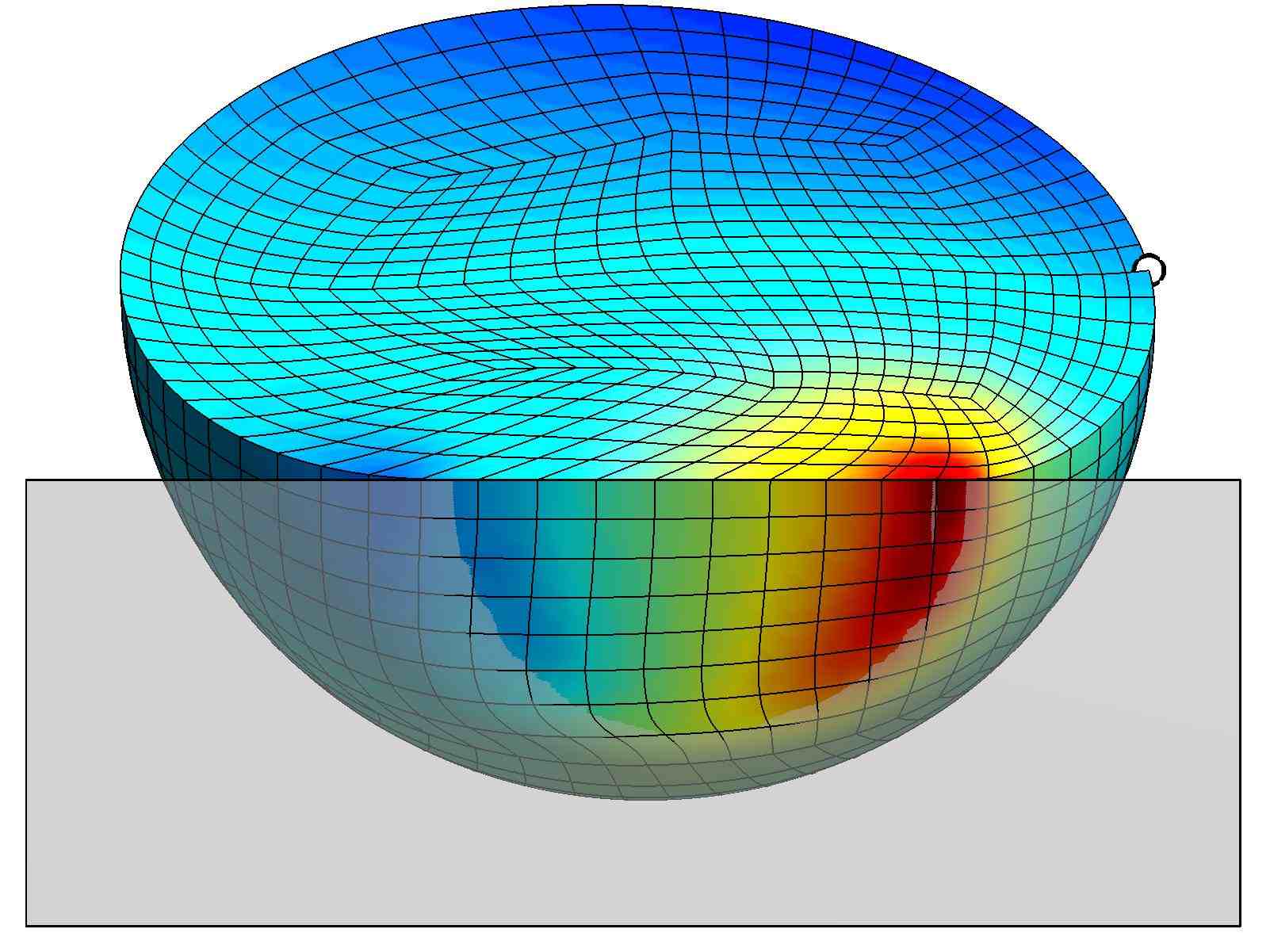}}
\put(1.47,-.2){\includegraphics[height=18.6mm]{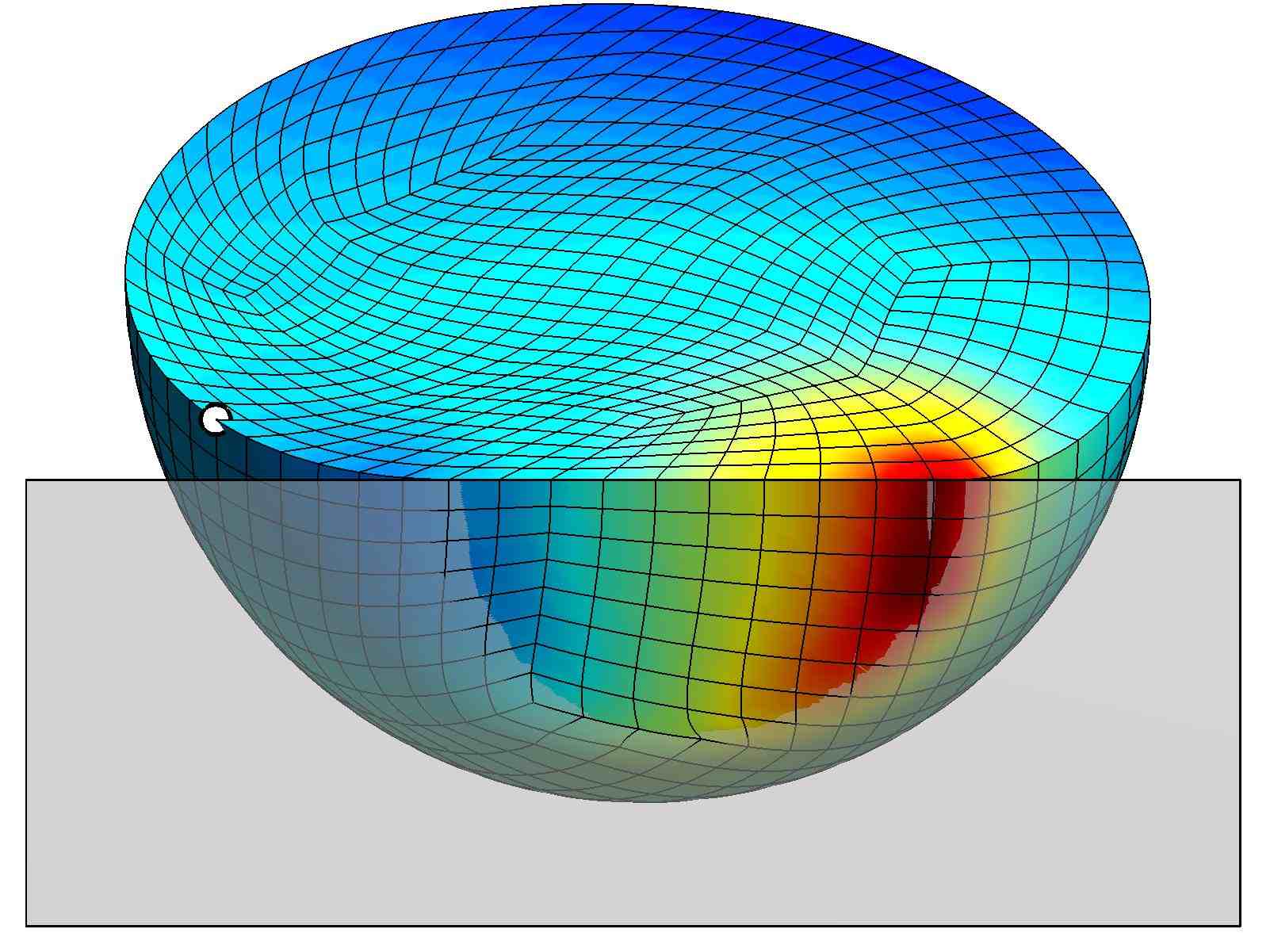}}
\put(4.62,-.2){\includegraphics[height=18.6mm]{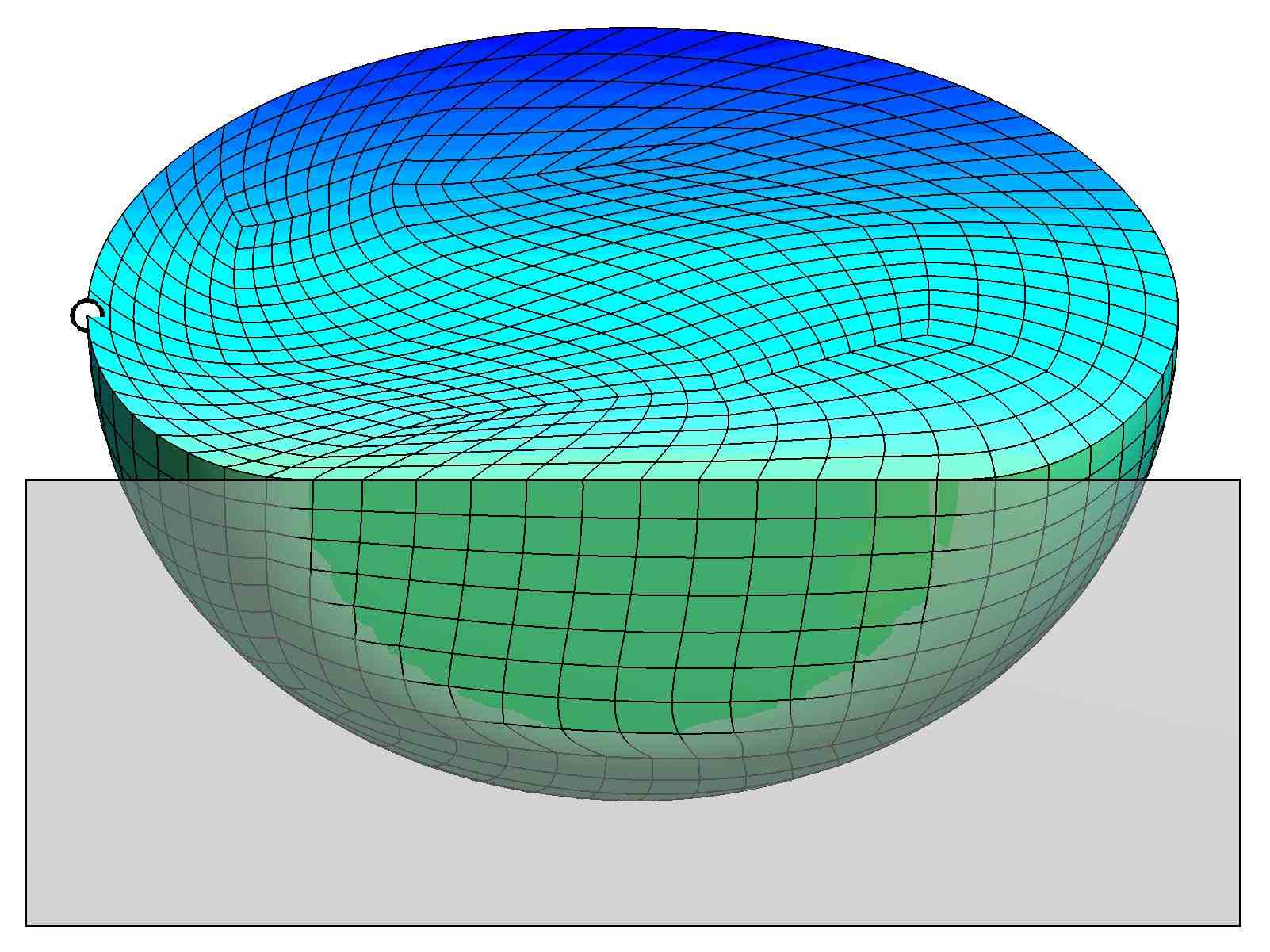}}
\end{picture}
\caption{Rolling droplet: smoothed pressure field at $t=0$, $t=50\,T_0$, $t=100\,T_0$, $t=200\,T_0$ and $t=350\,T_0$ (left to right) for $\beta_0=20^\circ$ and $m=8$. 
See also supplementary movie file \texttt{drop\underline{ }roll\underline{ }p.mpg}.} 
\label{f:rdrop_ps}
\end{center}
\end{figure}

\begin{figure}[!ht]
\begin{center} \unitlength1cm
\begin{picture}(0,4)
%\put(-7.2,1.5){\includegraphics[height=25mm]{../../codes/cFEAR/Input/iFSI/Droplet/oDropRoll/roll07/roll20_m8_te200_dt0p5_100Ws.jpg}}
%\put(-3.85,1.5){\includegraphics[height=25mm]{../../codes/cFEAR/Input/iFSI/Droplet/oDropRoll/roll15/roll20_m8_te350_dtp5_400Ws_c.jpg}}
%\put(-7.08,-.2){\includegraphics[height=18.6mm]{../../codes/cFEAR/Input/iFSI/Droplet/oDropRoll/roll07/roll20_m8_te200_dt0p5_ba100Ws.jpg}}
%\put(-3.93,-.2){\includegraphics[height=18.6mm]{../../codes/cFEAR/Input/iFSI/Droplet/oDropRoll/roll15/roll20_m8_te350_dtp5_ba400Ws.jpg}}
\put(-7.2,1.5){\includegraphics[height=25mm]{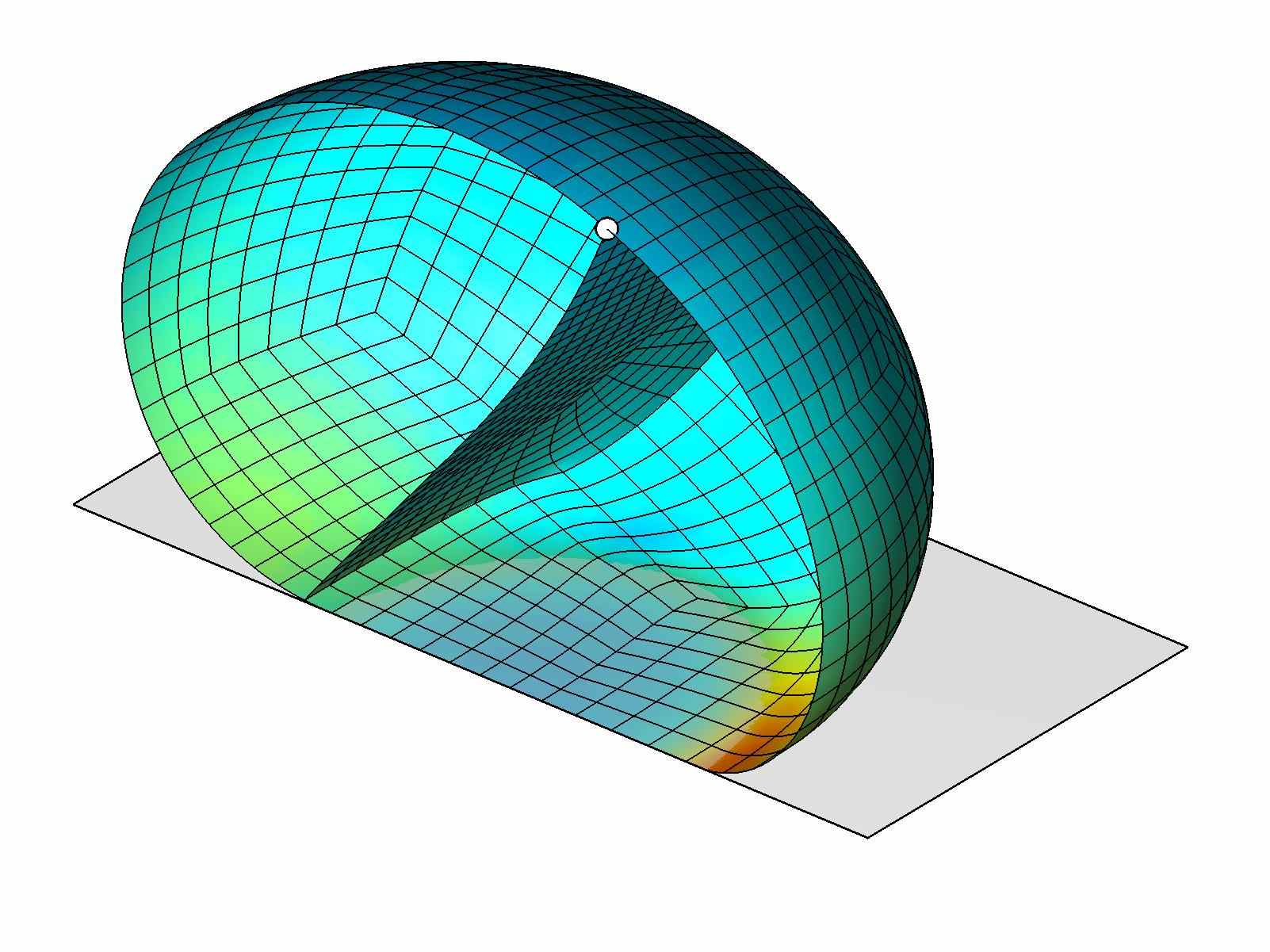}}
\put(-3.85,1.5){\includegraphics[height=25mm]{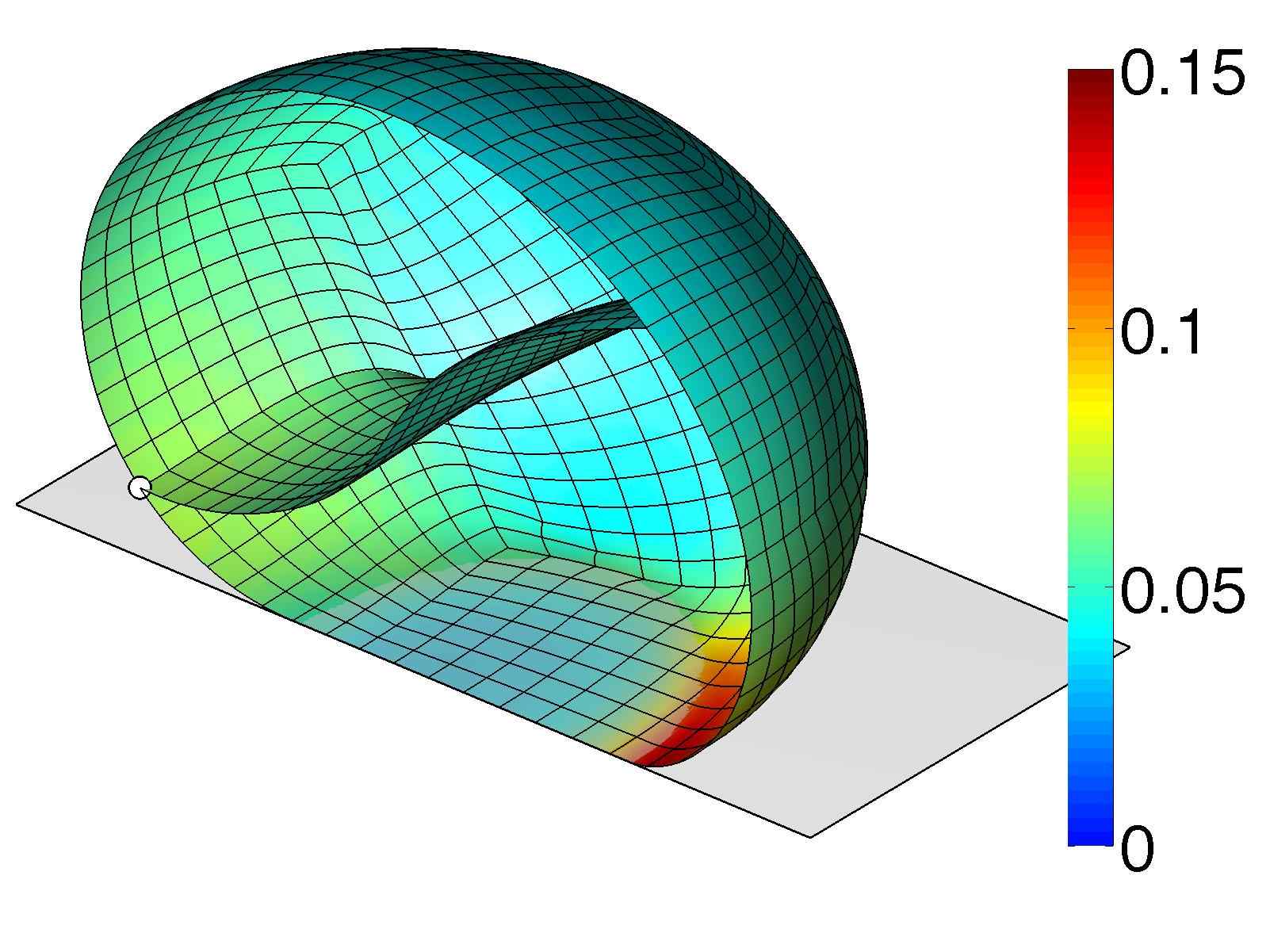}}
\put(-7.08,-.2){\includegraphics[height=18.6mm]{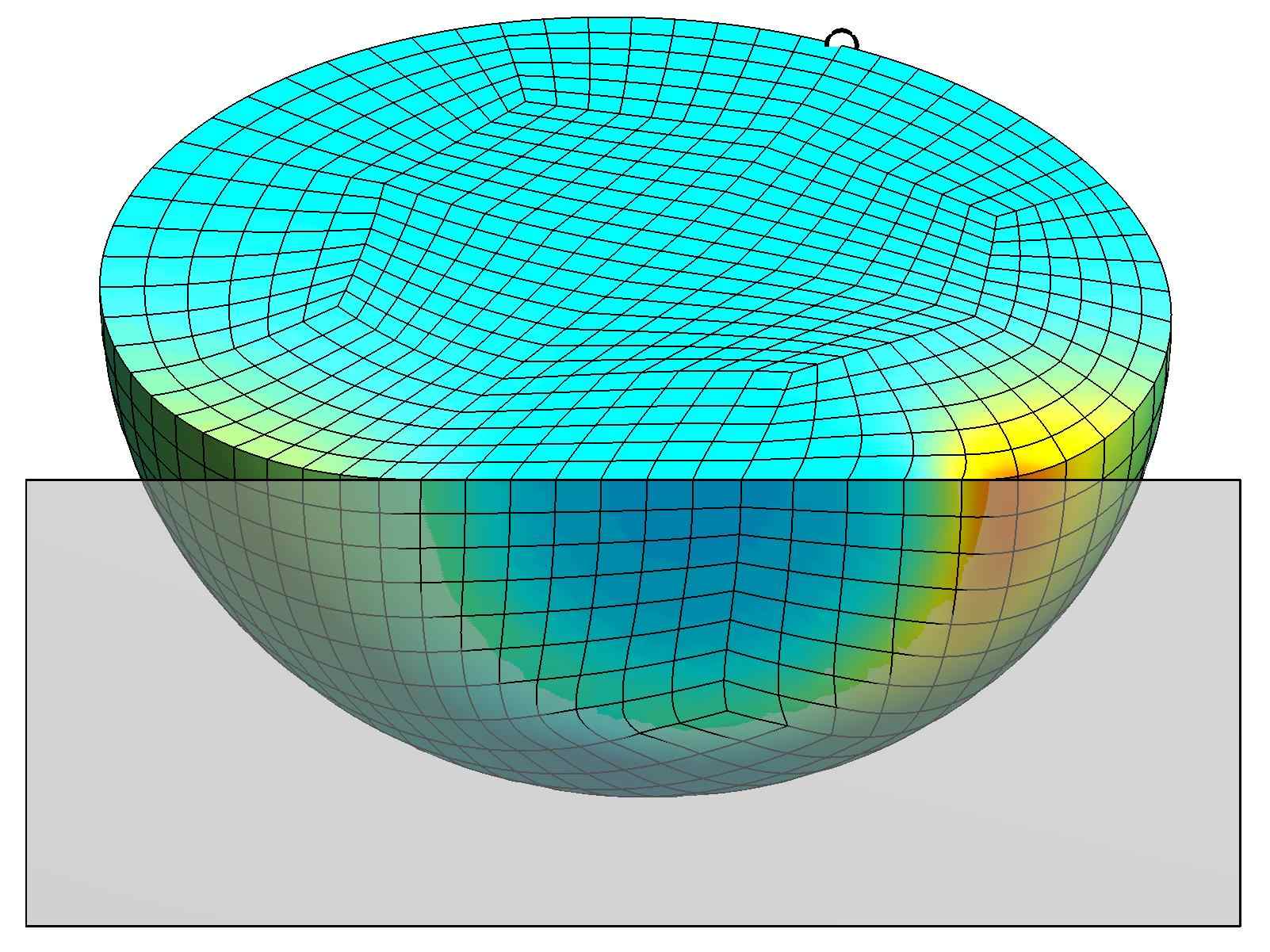}}
\put(-3.93,-.2){\includegraphics[height=18.6mm]{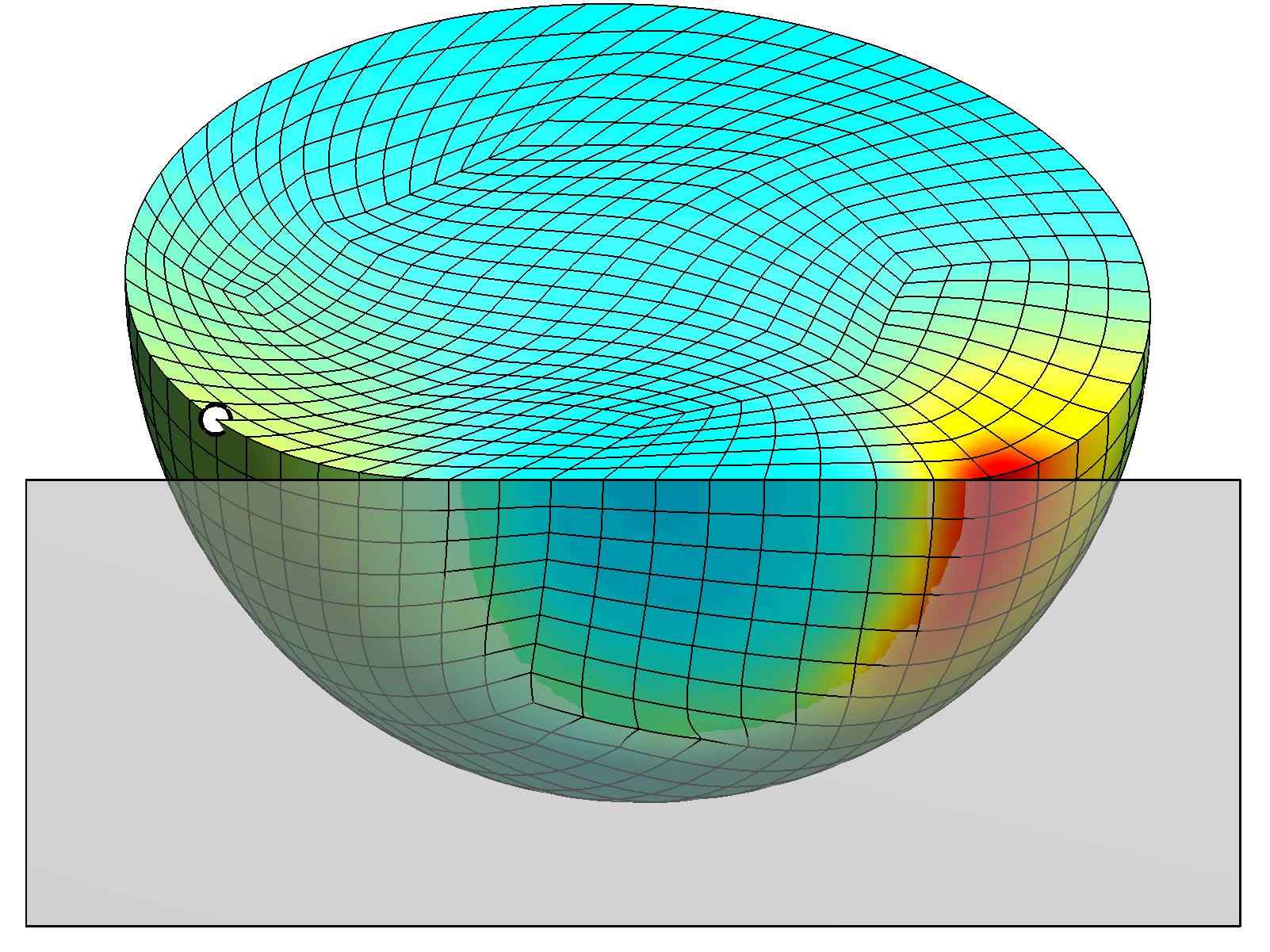}}
%
%\put(0.95,1.5){\includegraphics[height=25mm]{../../codes/cFEAR/Input/iFSI/Droplet/oDropRoll/roll07/roll20_m8_te200_dt0p5_100Disss.jpg}}
%\put(4.25,1.5){\includegraphics[height=25mm]{../../codes/cFEAR/Input/iFSI/Droplet/oDropRoll/roll15/roll20_m8_te350_dtp5_400Disss_c.jpg}}
%\put(1.07,-.2){\includegraphics[height=18.6mm]{../../codes/cFEAR/Input/iFSI/Droplet/oDropRoll/roll07/roll20_m8_te200_dt0p5_ba100Disss.jpg}}
%\put(4.22,-.2){\includegraphics[height=18.6mm]{../../codes/cFEAR/Input/iFSI/Droplet/oDropRoll/roll15/roll20_m8_te350_dtp5_ba400Disss.jpg}}
\put(0.95,1.5){\includegraphics[height=25mm]{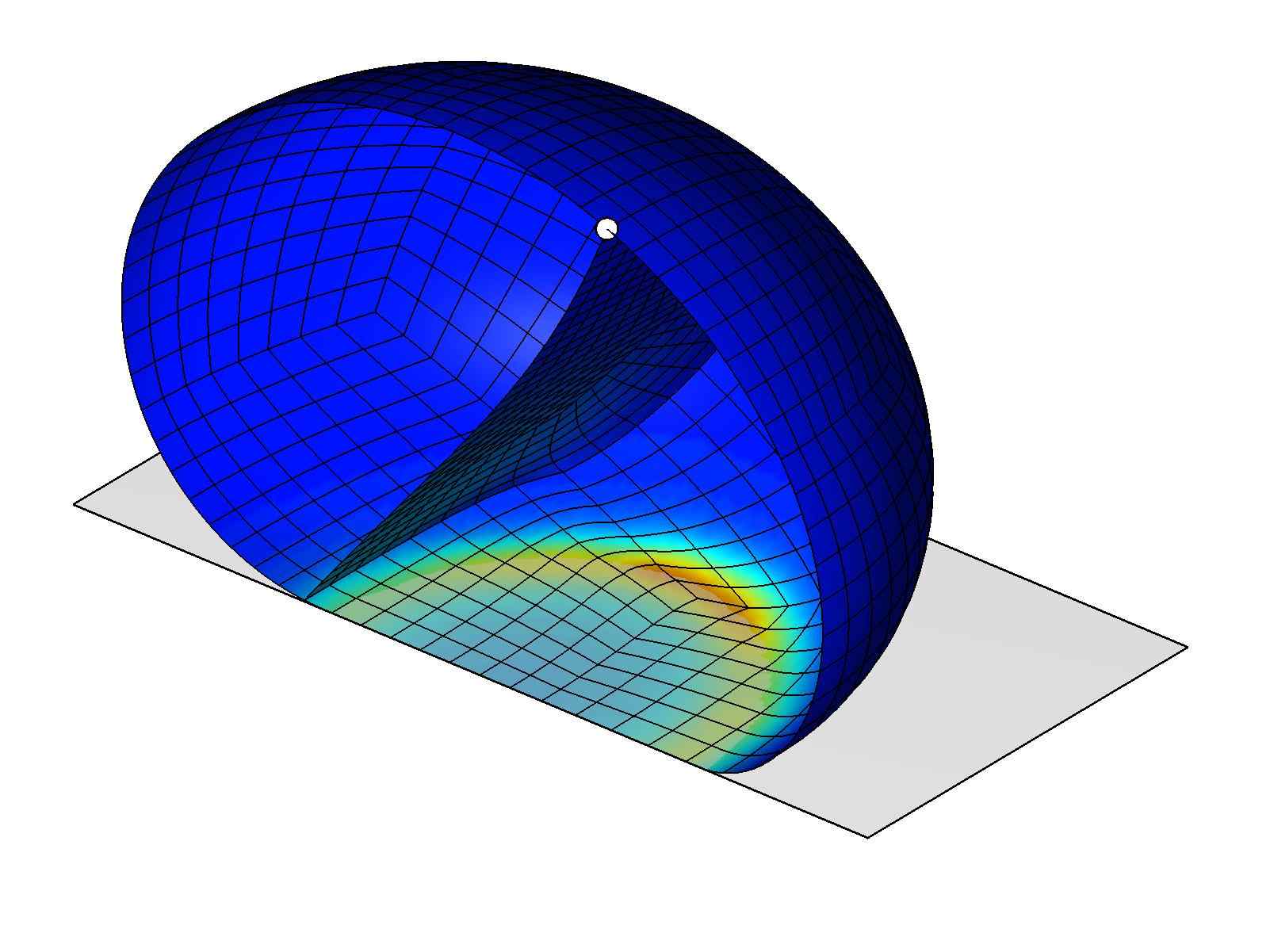}}
\put(4.25,1.5){\includegraphics[height=25mm]{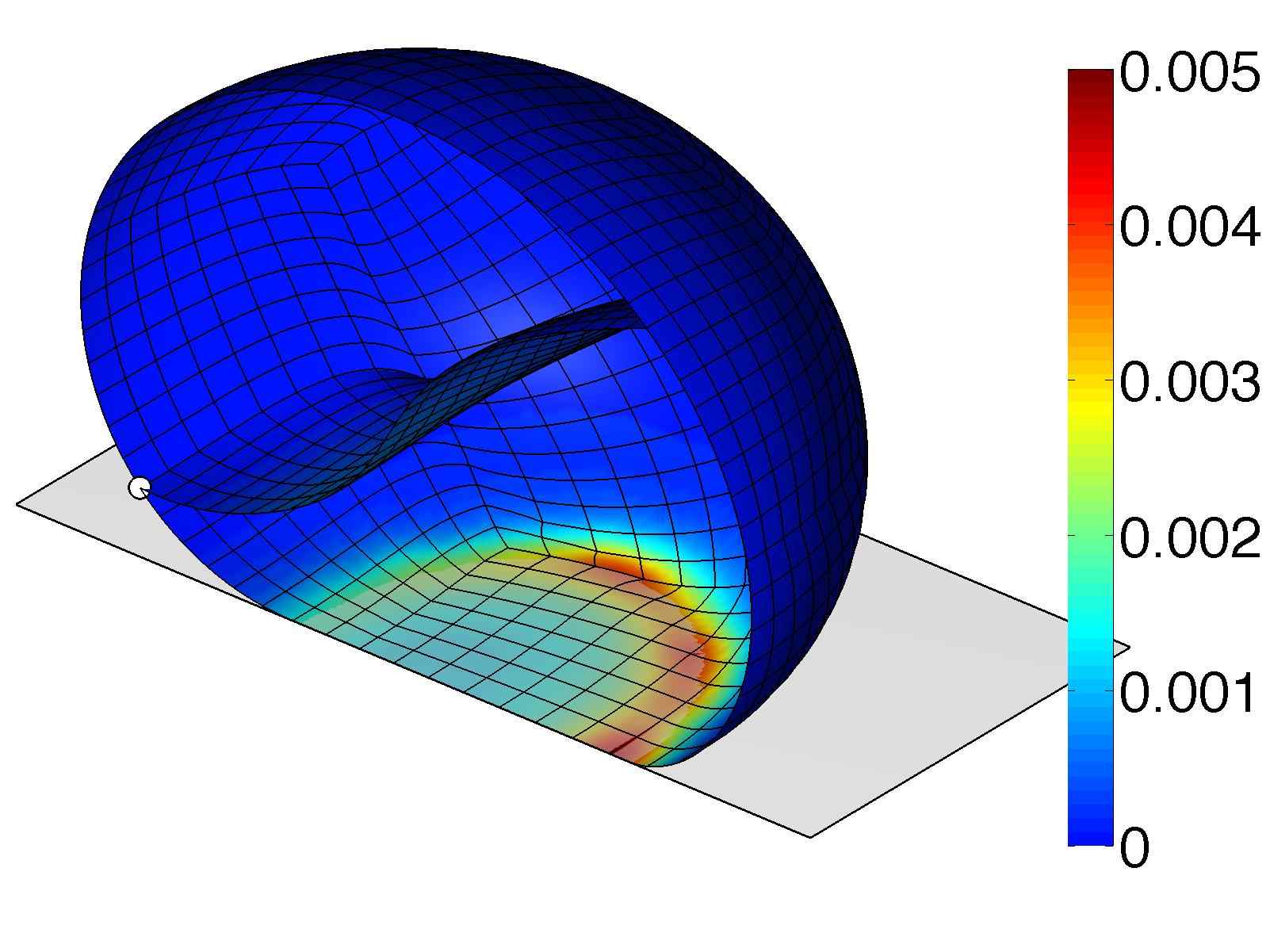}}
\put(1.07,-.2){\includegraphics[height=18.6mm]{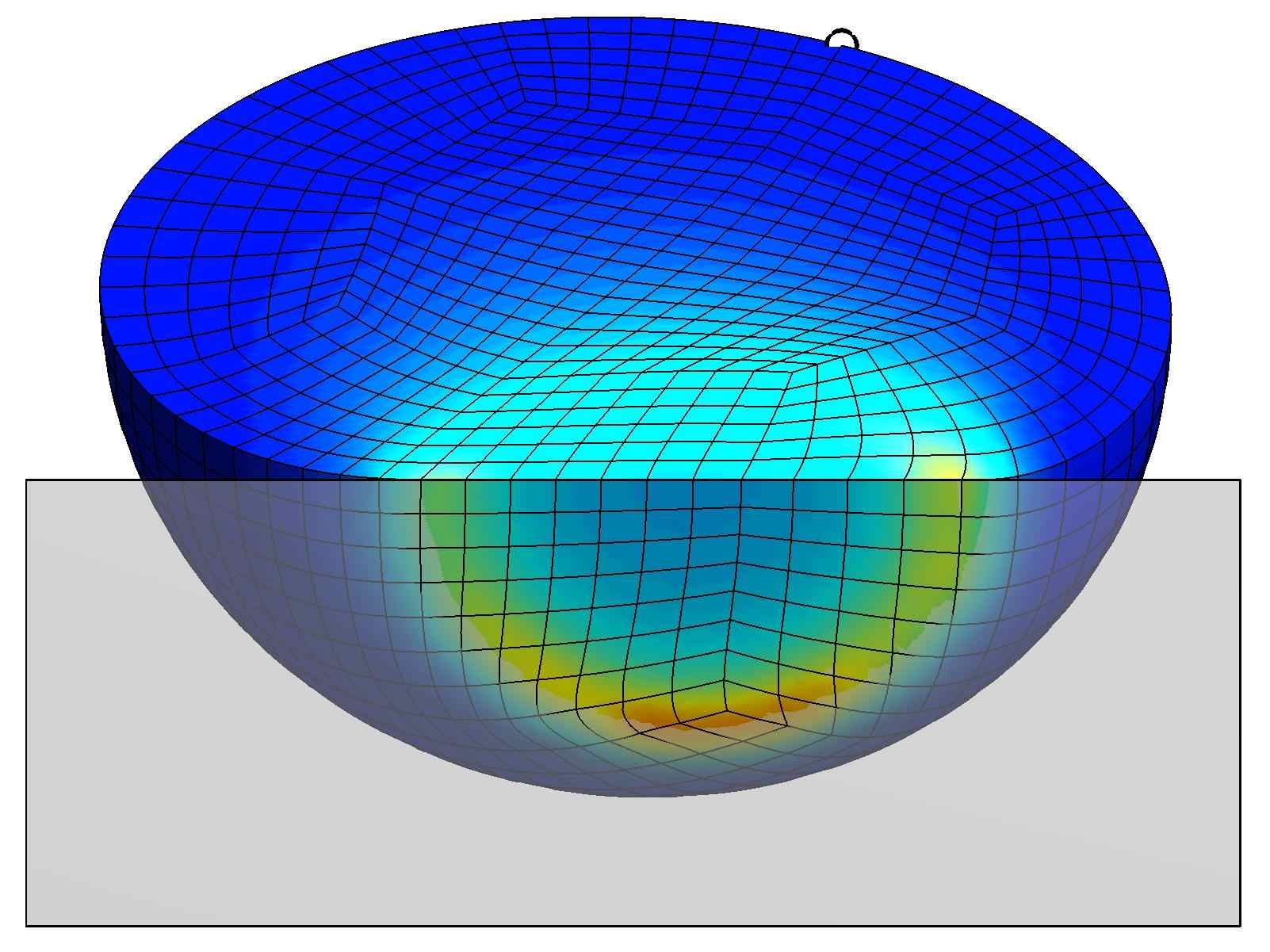}}
\put(4.22,-.2){\includegraphics[height=18.6mm]{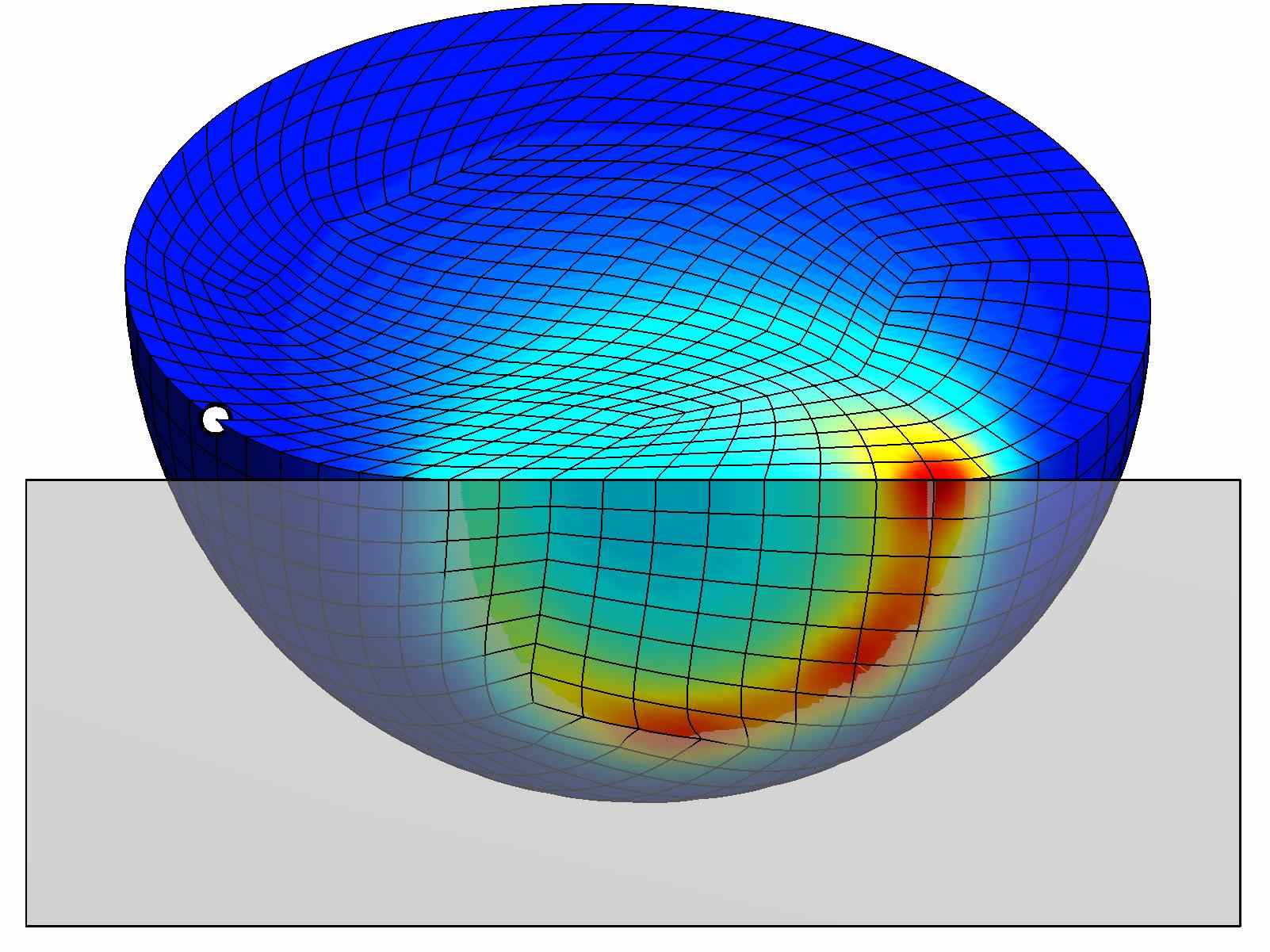}}
\put(-7.6,-.1){a.}
\put(0.55,-.1){b.}
\end{picture}
\caption{Rolling droplet: a.~smoothed vorticity component $2\omega_2$ at $t=50\,T_0$ and $t=200\,T_0$; b.~smoothed dissipation $\sD=\bsig:\bD$ at $t=50\,T_0$ and $t=200\,T_0$; both for $\beta_0=20^\circ$ and $m=8$. The units of $2\omega_2$ are $1/T_0$; the units of $\sD$ are $p_0/T_0$.}  
\label{f:rdrop_WD}
\end{center}
\end{figure}

The droplet setup considers similar parameters as in \citet{dropslide}: 
An initially spherical droplet with radius $R=L_0$ and volume $V=4\pi L_0^3/3$ is considered under gravity loading, such that $\rho g L_0^3 = \gamma L_0$. 
For water at room temperature, with $\rho = 1000\,$kg/m$^3$, $g=9.81\,$m/s$^2$ and $\gamma=72.8\,$mN/m, this corresponds to a droplet with $L_0 = 2.72\,$mm and $V=84.6\,\mu$l. 
The droplet surface has no additional mass, and so $\rho_\mrs = 0$.
For further normalization we choose $g_0 = g$ and $\gamma_0=\gamma$, so that $T_0=16.7\,$ms, $F_0=0.198\,$mN and $p_0=26.7\,$Pa.
A high fluid viscosity is chosen, i.e.~$\eta=11.9\,$Ns/m$^2$, such that the Reynolds number becomes very small.
A suitable definition for the Reynolds number of a rolling droplet is
\eqb{l}
Re = \ds\frac{\rho\,L_\mrc\,v_\mathrm{mean}}{\eta}\,,
\label{e:Re_drop}\eqe
where $L_\mrc$ is the diameter of the contact surface and $v_\mathrm{mean}$ is the mean droplet velocity.
The penalty parameter for sticking according to contact model \eqref{e:fc} is taken as $\epsilon_\mrc=250\,m^2\,p_0/L_0$, where $m$ characterizes the FE resolution according to Tab.~\ref{t:rdrop}.
%------------------------------------------------------------------
\begin{table}[h]
\centering
\begin{tabular}{|r|r|r|r|r|}
  \hline
  $m$ & fluid elements & membrane elements & nodes & dofs \\[0.5mm] \hline 
   & & & & \\[-3.5mm]   
   2 & 128 & 48 & 1,241 & 4,964 \\[1mm] 
   4 & 832 & 192 & 7,407 & 29,628 \\[1mm] 
   8 & 6,656 & 768 & 56,157 & 224,628  \\[1mm] 
   16 & 53,248 & 3,072 & 437,433 & 1,749,732 \\[1mm] 
   \hline
\end{tabular}
\caption{Rolling droplet: Considered FE meshes based on quadratic Lagrange elements.}
\label{t:rdrop}
\end{table}
%------------------------------------------------------------------
Quadratic Lagrange elements are used.
The computational runtime per time step (accounting for residual and tangent matrix assembly, contact computation and Newton-Raphson iteration) is about 1 min.~for $m=4$, 20 mins.~for $m=8$ and 100 mins.~for $m=16$.
\\
Initially the droplet is at rest.
Rolling motion is then induced by inclining the substrate considering the time-varying inclination angle
\eqb{l}
\beta(t) = \ds\frac{\beta_0}{2}\left\{\begin{array}{ll}
\ds1-\cos\frac{\pi t}{t_1} & $for $0\leq t<t_1, \\[2mm]
2 & $for $t_1\leq t\leq t_2, \\[0mm]
\ds1+\cos\frac{\pi (t-t_2)}{t_1} & $for $t_2\leq t\leq t_1+t_2, \\[2mm]
0 & $for $t_1+t_2<t\leq t_3,
\end{array}\right.
\eqe
with $t_1=50\,T_0$, $t_2=200\,T_0$, $t_3=350\,T_0$ and the two cases: 
\\
1. $\beta_0=10^\circ$ with $\Delta t=8\,T_0/m$, and \\
2. $\beta_0=20^\circ$ with $\Delta t=4\,T_0/m$. \\
Fig.~\ref{f:rdrop_vt} shows the finite element results for the mean droplet velocity $v_\mathrm{mean}$ for the two cases.\footnote{The mean droplet velocity $v_\mathrm{mean}$ is determined by computing the volume average of the fluid velocity $\bv$ and then taking its component parallel to the substrate surface.} 
%-------------------------------------------------------------------------------------------------------------------------------
% run codes/cFEAR/Input/iFSI/Droplet/oDropRoll/pDropRoll
%%-------------------------------------------------------------------------------------------------------------------------------
As seen the FE results converge upon mesh refinement.
The figure also shows that steady rolling motion is attained at about $t=150\,T_0$ for $\beta_0=20^\circ$, while it is attained almost instantaneously for $\beta_0=10^\circ$ (i.e.~at $t=t_1$).
The instantaneous response of $v_\mathrm{mean}$ on $\beta$, for low $\beta_0$, can be also seen from the $v_\mathrm{mean}(\beta)$--plot in Fig.~\ref{f:rdrop_vb}. 
%-------------------------------------------------------------------------------------------------------------------------------
% run codes/cFEAR/Input/iFSI/Droplet/oDropRoll/pDropRoll
%-------------------------------------------------------------------------------------------------------------------------------
Both branches (for increasing $\beta$ and decreasing $\beta$, respectively) are almost identical.
For $\beta_0=20^\circ$ on the other hand, the two branches are different.
\\
For further illustration, Fig.~\ref{f:rdrop_v} shows the droplet deformation and velocity field $\norm{\bv}$ during rolling.
%-------------------------------------------------------------------------------------------------------------------------------
% run codes/cFEAR/Input/iFSI/Droplet/oDropRoll/mDropRoll
%-------------------------------------------------------------------------------------------------------------------------------
The deformation is considerable and should not be neglected, as has been done in earlier work \citep{rasool12,rasool13}. 
The figure also shows how the contact surface changes.
Initially the contact surface is circular with a diameter of $L_\mrc=1.36\,L_0$.
During steady rolling the diameter in rolling direction reduces to $L_\mrc=1.04\,L_0$.
Since $v_\mathrm{mean}=0.0268\,L_0/T_0$, the Reynolds number thus becomes $Re=1.04\cdot10^{-3}$ according to \eqref{e:Re_drop}.
Fig.~\ref{f:rdrop_v} clearly shows that the advancing and receding droplet halves are not symmetric during rolling.\\
This can also be seen from the pressure distribution shown in Fig.~\ref{f:rdrop_p}. 
%-------------------------------------------------------------------------------------------------------------------------------
% run codes/cFEAR/Input/iFSI/Droplet/oDropRoll/mDropRoll
%-------------------------------------------------------------------------------------------------------------------------------
The fluid pressure is largest at the advancing front of the contact surface.
Since the contact surface is flat, the fluid pressure is equal to the contact pressure.
Close inspection shows that the pressure is oscillatory in the vicinity of the contact line $\sC$. 
Those oscillations do not converge with mesh refinement, as the velocity field does.
So it seems that the pressure stabilization scheme, described in Sec.~\ref{s:wfF}, is not sufficient to handle the contact boundary of a rolling droplet, even though the static droplet (at $t=0$ and $t=350\,T_0$) poses no problem.
The problem may be related to the discontinuity of the contact pressure: it jumps to zero at the contact boundary.
The way the fluid velocity, fluid pressure and contact pressure are interpolated (quadratic Lagrange interpolation is used here) seem incompatible. 
It seems that this problem has not yet been addressed in the literature.
Further study is required on the topic.
Perhaps $C^1$-continuous interpolation, such as is provided by NURBS, would help.
We note that for $\beta=10^\circ$, pressure oscillations also appear, but they are less pronounced. 
\\
To remove the pressure oscillations, Gaussian smoothing can be used for post-processing. 
Selecting the variance of the Gaussian distribution as $\sig=1/m$, i.e.~on the order of the nodal distance, gives non-oscillatory pressures; see Fig.~\ref{f:rdrop_ps}.
%-------------------------------------------------------------------------------------------------------------------------------
% run codes/cFEAR/Input/iFSI/Droplet/oDropRoll/mDropRoll
%-------------------------------------------------------------------------------------------------------------------------------
The smoothed pressure converges with mesh refinement.
The pressure distribution shows that the advancing contact surface carries most of the droplet weight (component $\cos\beta \times \rho gV$). 
Component $\sin\beta \times \rho gV$ is equilibrated by a tangential sticking force. 
The moment caused by these external forces is equilibrated by the internal moment of the fluid stress. 
\\
The last plot shows the vorticity (i.e.~spin) component $2\omega_2 := \be_2\cdot(\nabla\times\bv)$ (along the axis of rotation $\be_2$) and the dissipation $\sD=\bsig:\bD$ during rolling; see Fig.~\ref{f:rdrop_WD}.
%-------------------------------------------------------------------------------------------------------------------------------
% run codes/cFEAR/Input/iFSI/Droplet/oDropRoll/mDropRoll
%-------------------------------------------------------------------------------------------------------------------------------
Also here smoothing is used.
According to Sec.~\ref{s:ana_spin} the vorticity of a spinning sphere is a constant vector with magnitude 2$\omega$.
In contrast, the vorticity of a rolling droplet is non-constant: 
A maximum is attained at the contact boundary and a minimum occurs on the contact surface. 
Although, away from the contact surface, the vorticity approaches a constant.
The behavior is similar for the dissipation:
Away from the contact surface, the dissipation is zero and thus agrees with the spinning sphere solution.
Non-zero dissipation, associated with shear flow, occurs in the vicinity of the contact surface, with a maximum occurring at the advancing contact front.
For longer rolling droplets, or for higher $\beta$, the shear flow becomes more pronounced, such that an ALE formulation is needed for the mesh.  
On the free surface (which is tracked explicitly within the present scheme) such a formulation needs to be Lagrangian in the normal direction but Eulerian in-plane. 
The formulation of such an ALE scheme is outside the present scope.

\subsection{Flapping flag}\label{s:flag}

\begin{figure}[h]
\begin{center} \unitlength1cm
\begin{picture}(0,5.4)
\put(-4.7,-.2){\includegraphics[height=55mm]{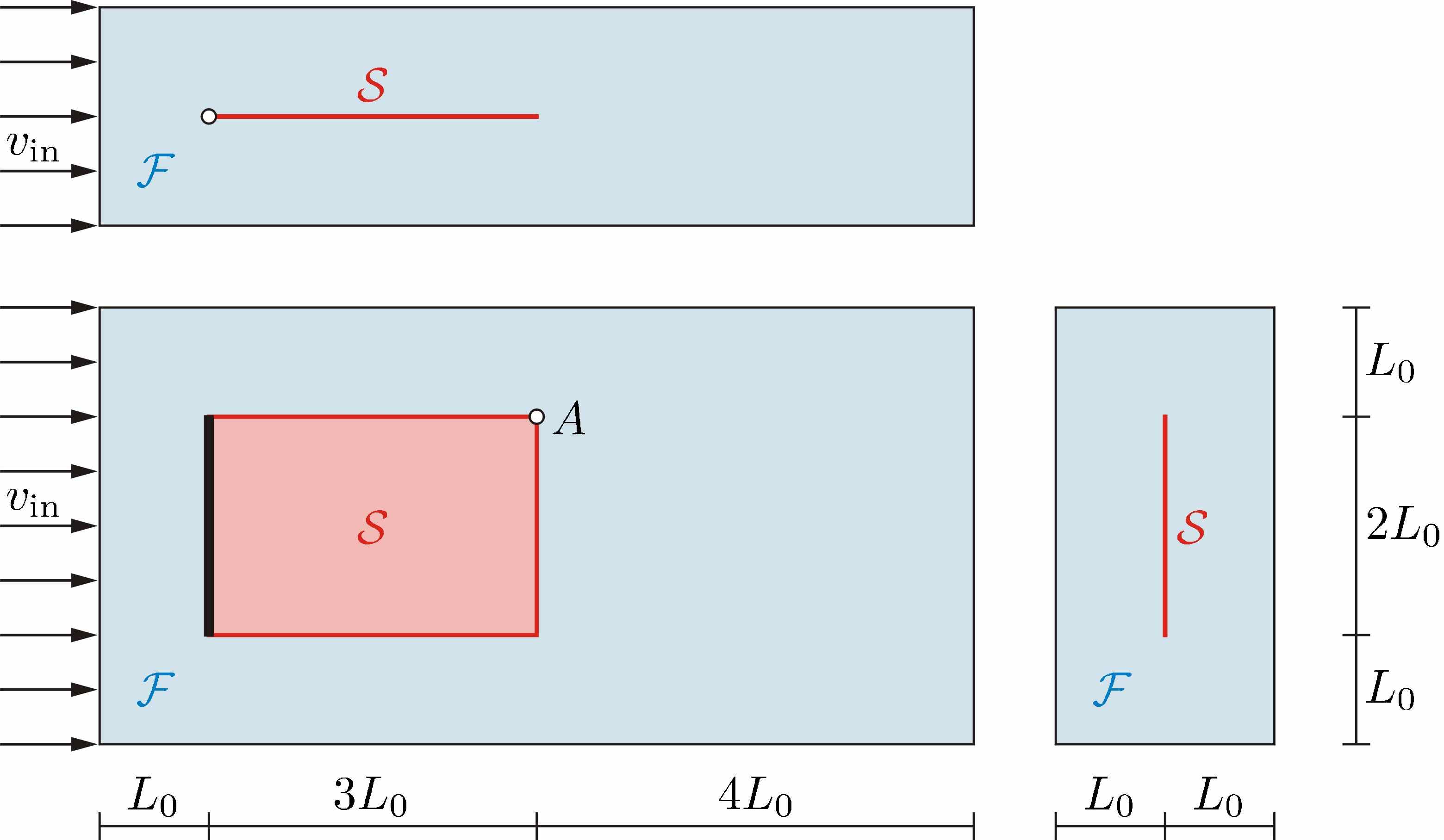}}
\end{picture}
\caption{Flapping flag: Side, top and front view of the problem setup.
The flag is fixed on the left and its lateral displacement and velocity are monitored at point $A$.} 
\label{f:flag_ex}
\end{center}
\end{figure}

\begin{figure}[!ht]
\begin{center} \unitlength1cm
\begin{picture}(0,2.3)
%\put(-7.9,-.2){\includegraphics[height=23mm]{../../codes/cFEAR/Input/iFSI/Flag/oFlag/flagiNDp09/flag_mc2_m8_v1_rhos1_muxp1_cp02_ofsp05_tr5_te50_dtp02/movie_v_dense/flag_2218v.jpg}}
%\put(-4.75,-.2){\includegraphics[height=23mm]{../../codes/cFEAR/Input/iFSI/Flag/oFlag/flagiNDp09/flag_mc2_m8_v1_rhos1_muxp1_cp02_ofsp05_tr5_te50_dtp02/movie_v_dense/flag_2246v.jpg}} 
%\put(-1.6,-.2){\includegraphics[height=23mm]{../../codes/cFEAR/Input/iFSI/Flag/oFlag/flagiNDp09/flag_mc2_m8_v1_rhos1_muxp1_cp02_ofsp05_tr5_te50_dtp02/movie_v_dense/flag_2274v.jpg}} 
%\put(1.55,-.2){\includegraphics[height=23mm]{../../codes/cFEAR/Input/iFSI/Flag/oFlag/flagiNDp09/flag_mc2_m8_v1_rhos1_muxp1_cp02_ofsp05_tr5_te50_dtp02/movie_v_dense/flag_2302v.jpg}}
%\put(4.7,-.2){\includegraphics[height=23mm]{../../codes/cFEAR/Input/iFSI/Flag/oFlag/flagiNDp09/flag_mc2_m8_v1_rhos1_muxp1_cp02_ofsp05_tr5_te50_dtp02/movie_v_dense/flag_2330v.jpg}}
\put(-7.9,-.2){\includegraphics[height=23mm]{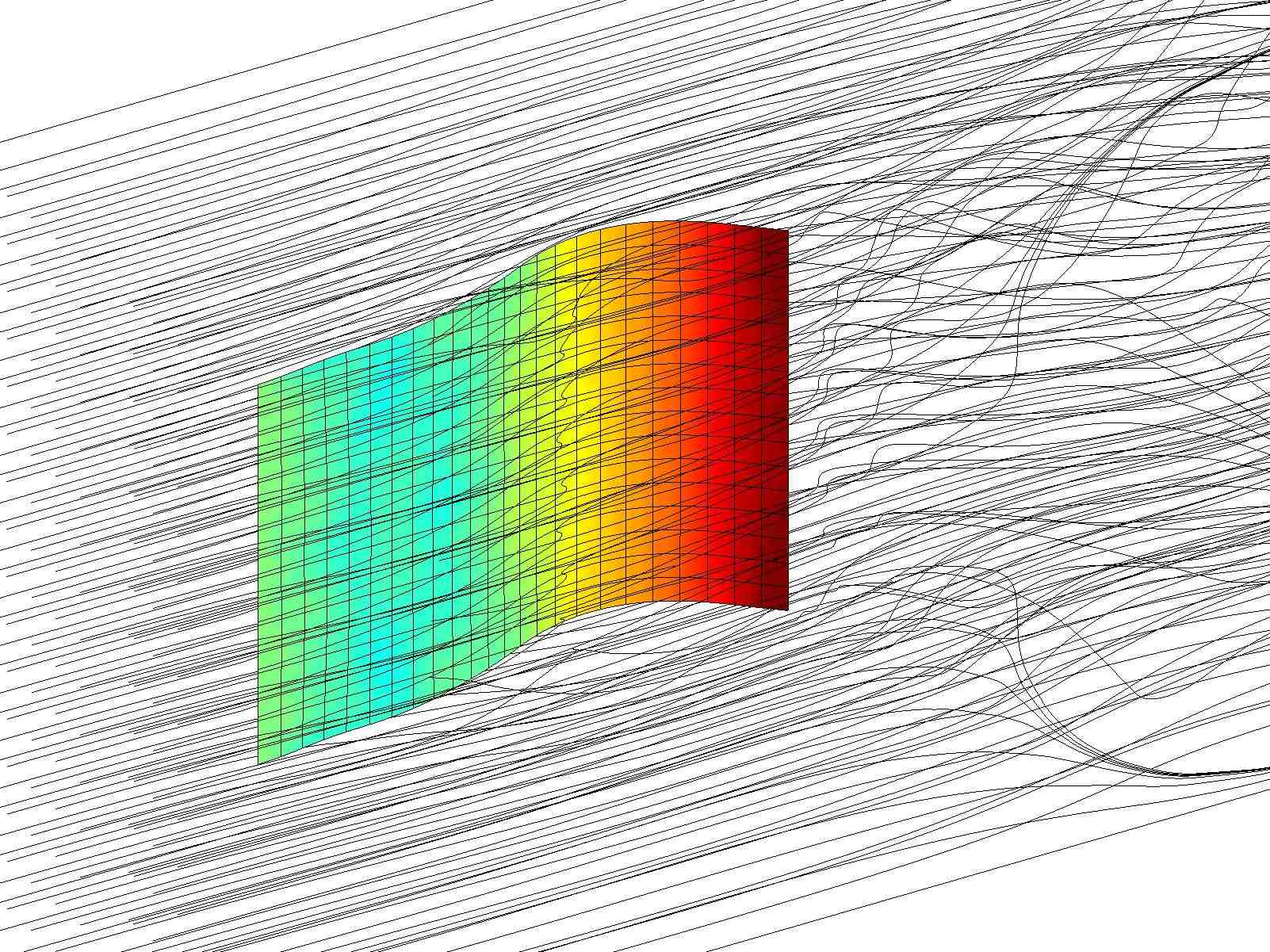}}
\put(-4.75,-.2){\includegraphics[height=23mm]{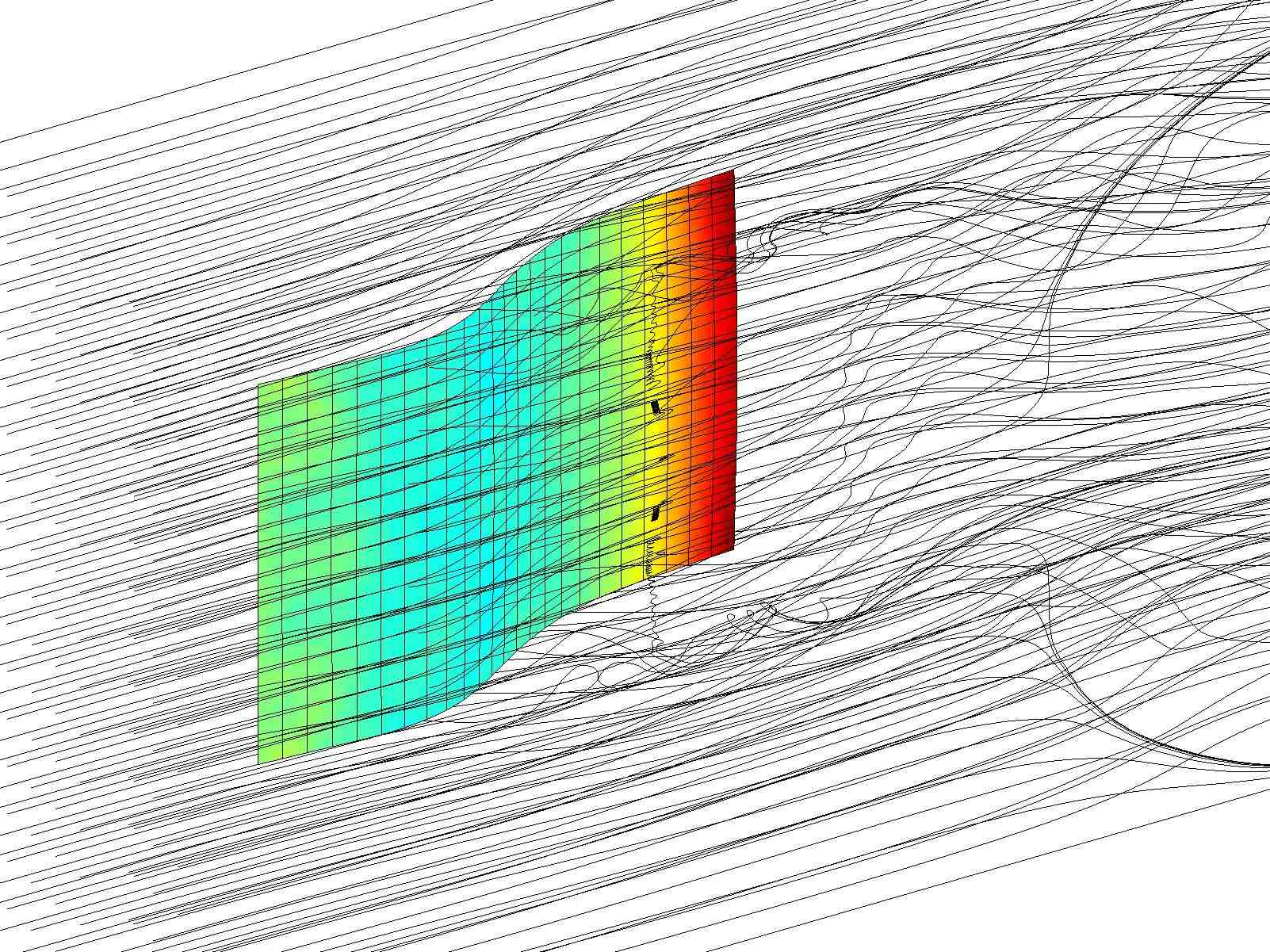}} 
\put(-1.6,-.2){\includegraphics[height=23mm]{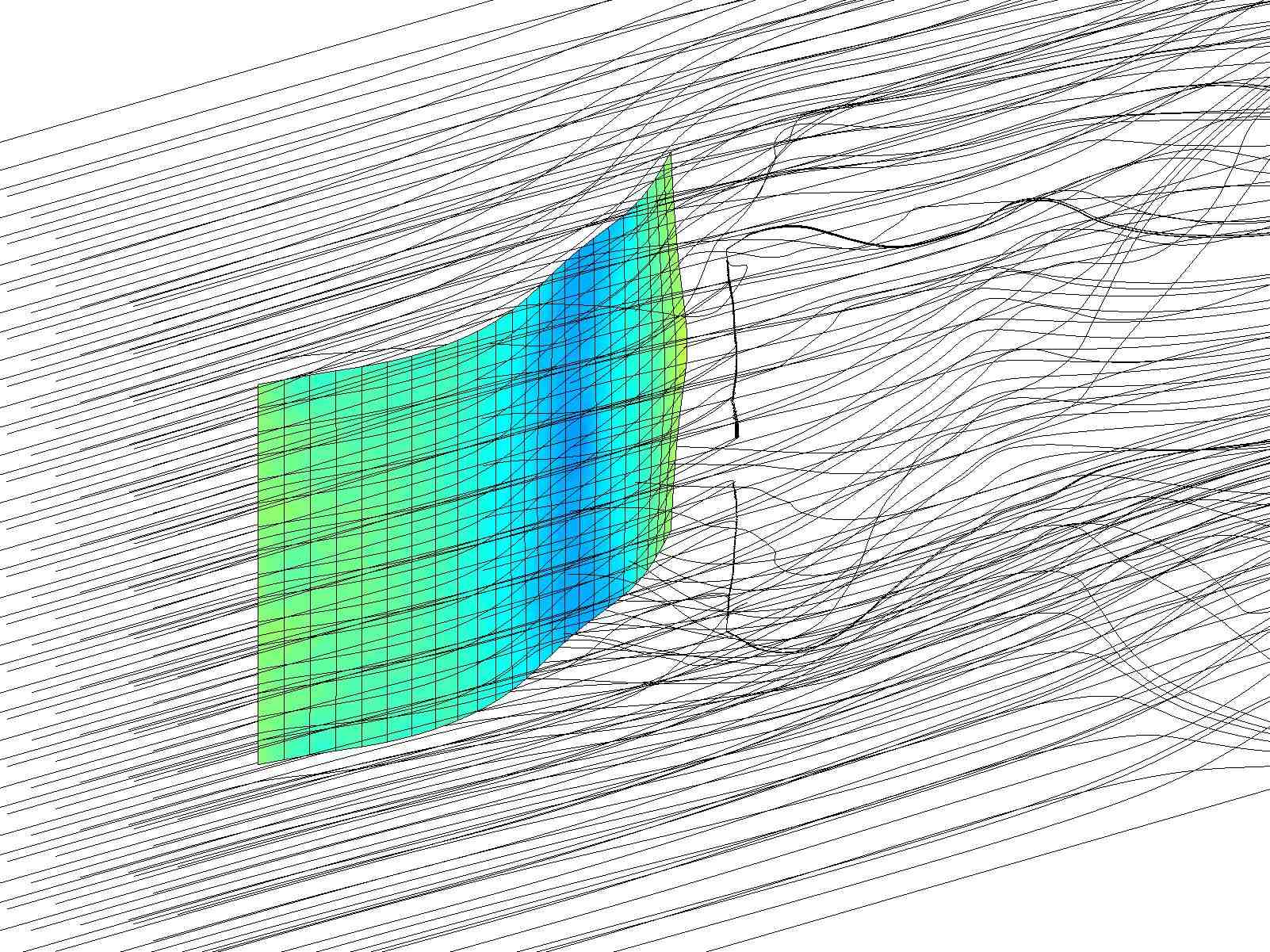}} 
\put(1.55,-.2){\includegraphics[height=23mm]{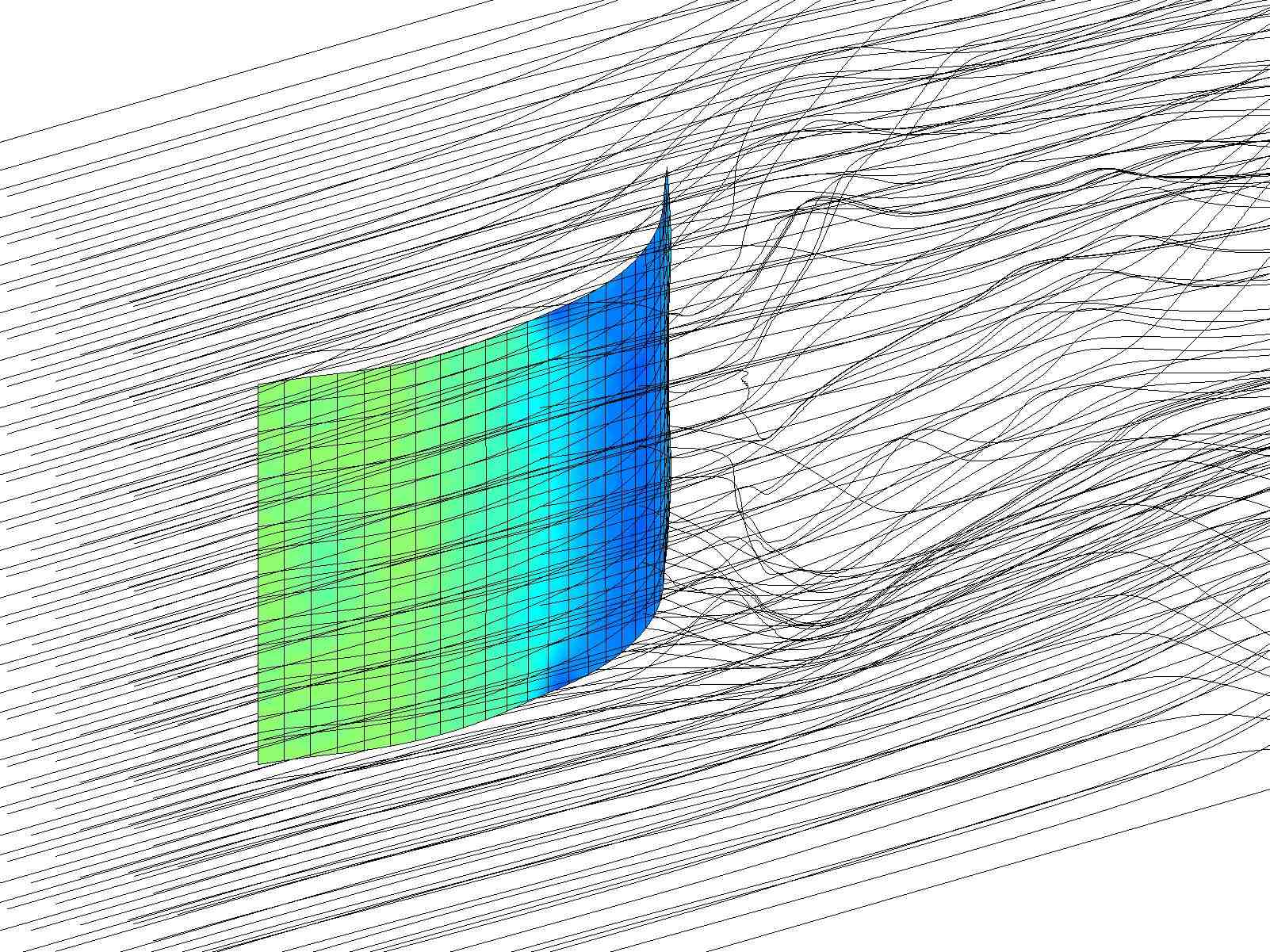}}
\put(4.7,-.2){\includegraphics[height=23mm]{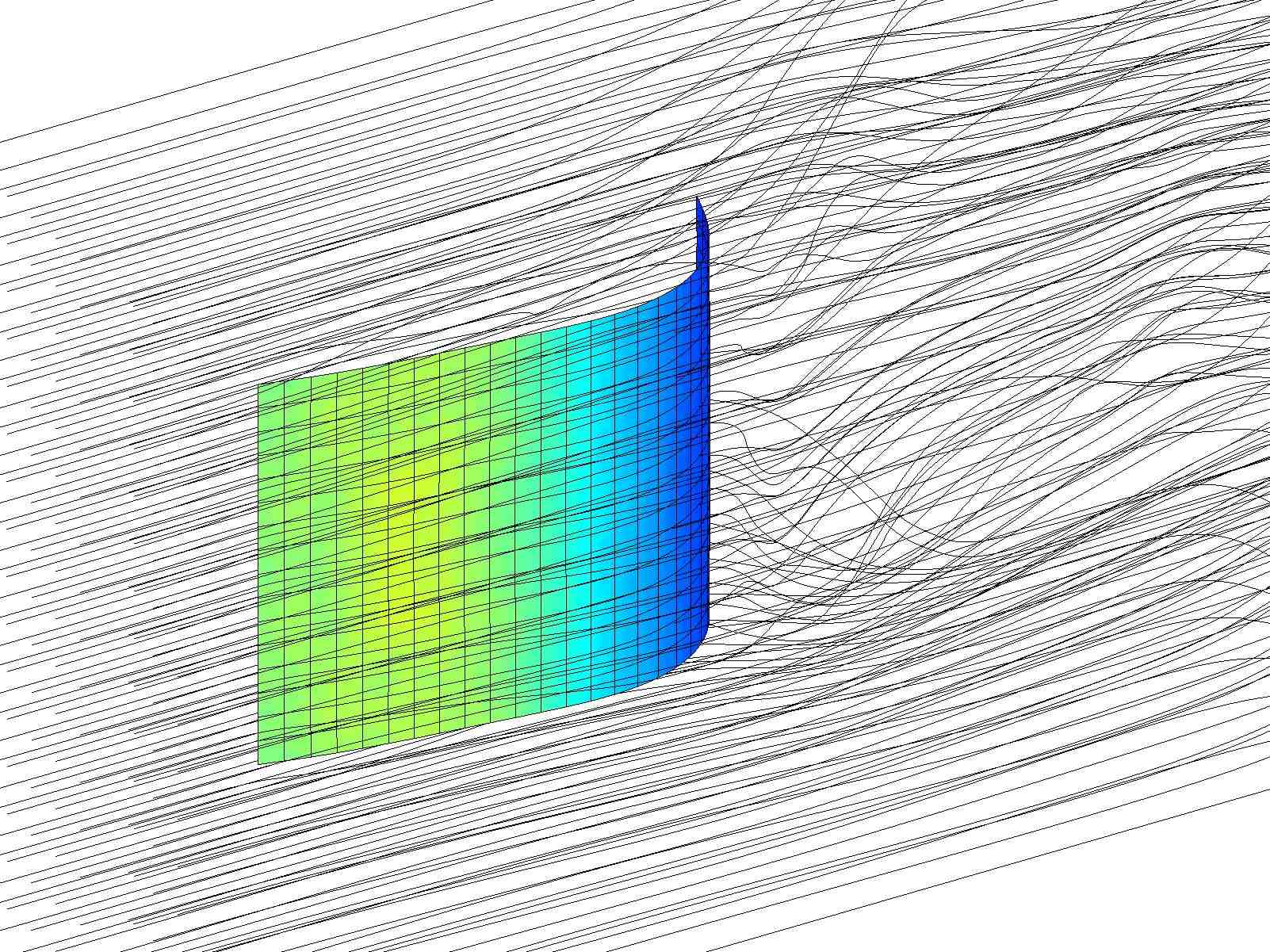}}
\end{picture}
\caption{Flapping flag: Deformation at $t=44.36\,$s, $t=44.92\,$s, $t=45.48\,$s, $t=46.04\,$s and $t=46.60\,$s (left to right) for $m=8$; see also supplementary movie file \texttt{flag\underline{ }v.mpg}.
The coloring shows the lateral velocity component in the range $\{-1,\,1\}v_0$ (from blue to red). 
The streamlines of the flow are also shown.}  
\label{f:flag_x}
\end{center}
\end{figure}

\begin{figure}[!ht]
\begin{center} \unitlength1cm
\begin{picture}(0,3)
%\put(-8.05,1.3){\includegraphics[height=16.3mm]{../../codes/cFEAR/Input/iFSI/Flag/oFlag/flagiNDp09/flag_mc2_m8_v1_rhos1_muxp1_cp02_ofsp05_tr5_te50_dtp02/movie_p_cycle/flag_t2218p.jpg}}
%\put(-2.7,1.3){\includegraphics[height=16.3mm]{../../codes/cFEAR/Input/iFSI/Flag/oFlag/flagiNDp09/flag_mc2_m8_v1_rhos1_muxp1_cp02_ofsp05_tr5_te50_dtp02/movie_p_cycle/flag_t2246p.jpg}}
%\put(2.65,1.3){\includegraphics[height=16.3mm]{../../codes/cFEAR/Input/iFSI/Flag/oFlag/flagiNDp09/flag_mc2_m8_v1_rhos1_muxp1_cp02_ofsp05_tr5_te50_dtp02/movie_p_cycle/flag_t2274p.jpg}} 
%\put(-8.05,-.3){\includegraphics[height=16.3mm]{../../codes/cFEAR/Input/iFSI/Flag/oFlag/flagiNDp09/flag_mc2_m8_v1_rhos1_muxp1_cp02_ofsp05_tr5_te50_dtp02/movie_p_cycle/flag_t2022p.jpg}}  % update figure
%\put(-2.7,-.3){\includegraphics[height=16.3mm]{../../codes/cFEAR/Input/iFSI/Flag/oFlag/flagiNDp09/flag_mc2_m8_v1_rhos1_muxp1_cp02_ofsp05_tr5_te50_dtp02/movie_p_cycle/flag_t2050p.jpg}}  % update figure
%\put(2.65,-.3){\includegraphics[height=16.3mm]{../../codes/cFEAR/Input/iFSI/Flag/oFlag/flagiNDp09/flag_mc2_m8_v1_rhos1_muxp1_cp02_ofsp05_tr5_te50_dtp02/movie_p_cycle/flag_t2078p.jpg}}  % update figure
\put(-8.05,1.3){\includegraphics[height=16.3mm]{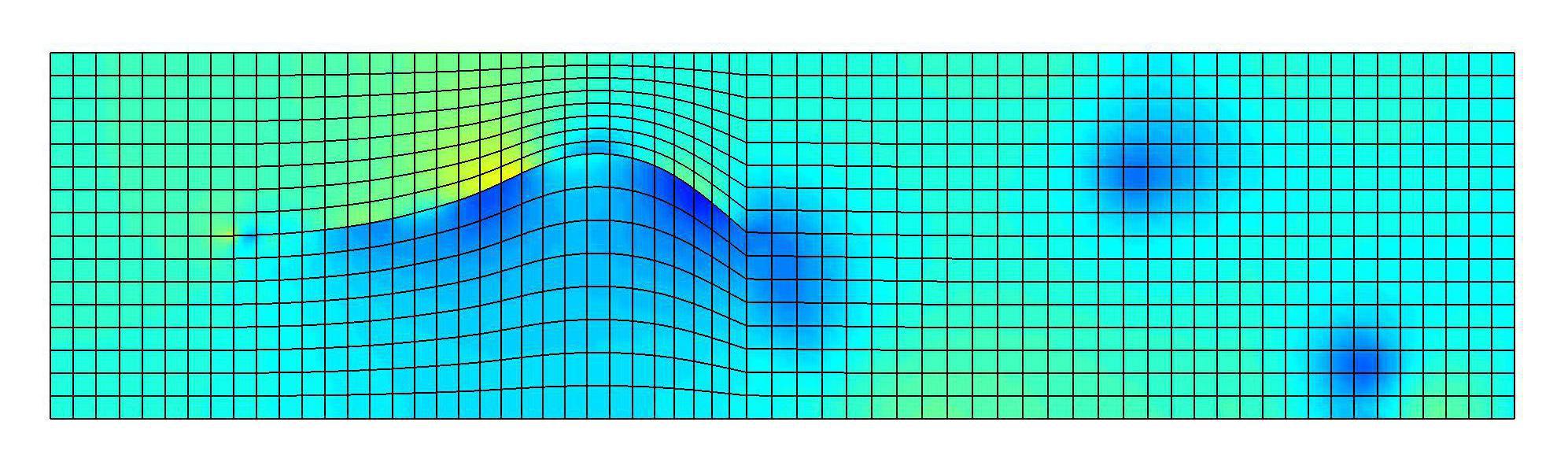}}
\put(-2.7,1.3){\includegraphics[height=16.3mm]{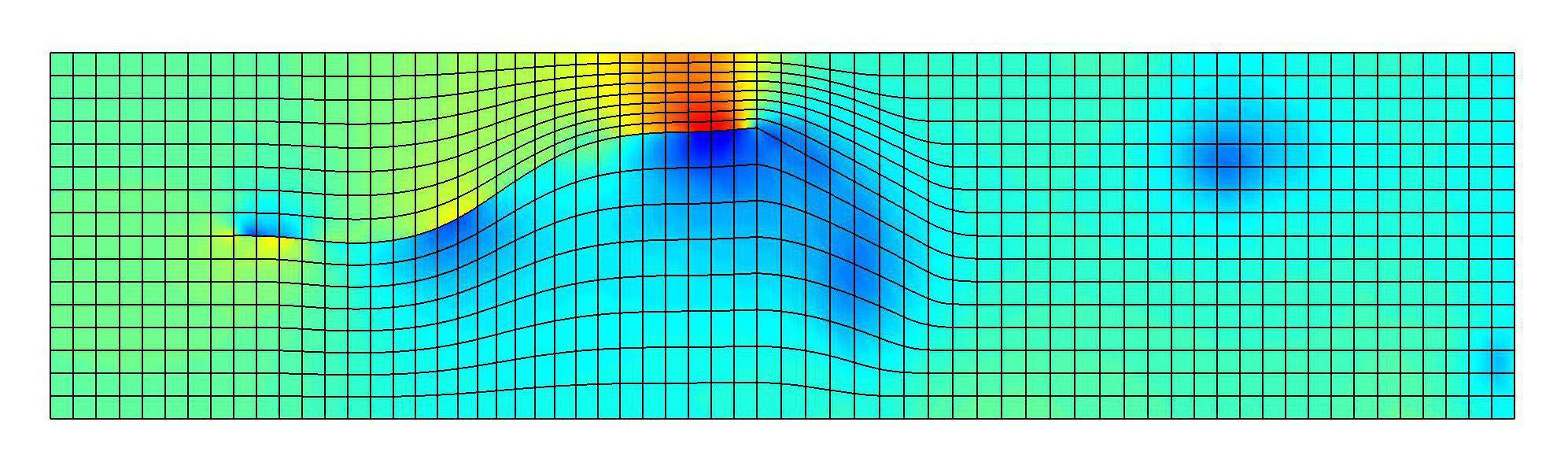}}
\put(2.65,1.3){\includegraphics[height=16.3mm]{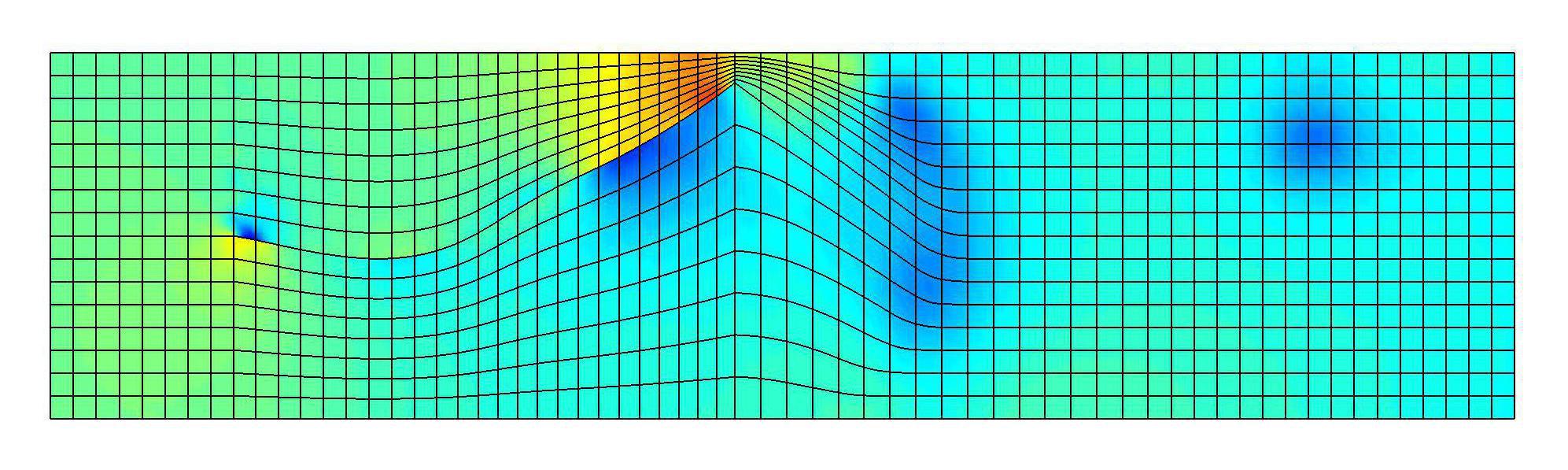}} 
\put(-8.05,-.3){\includegraphics[height=16.3mm]{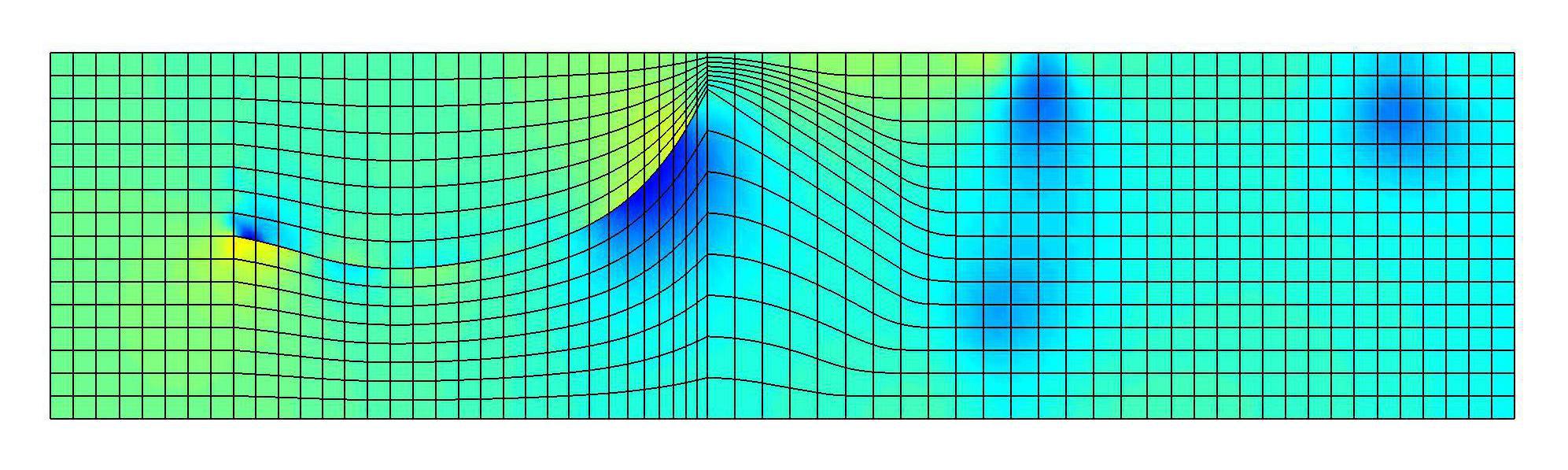}}  % update figure
\put(-2.7,-.3){\includegraphics[height=16.3mm]{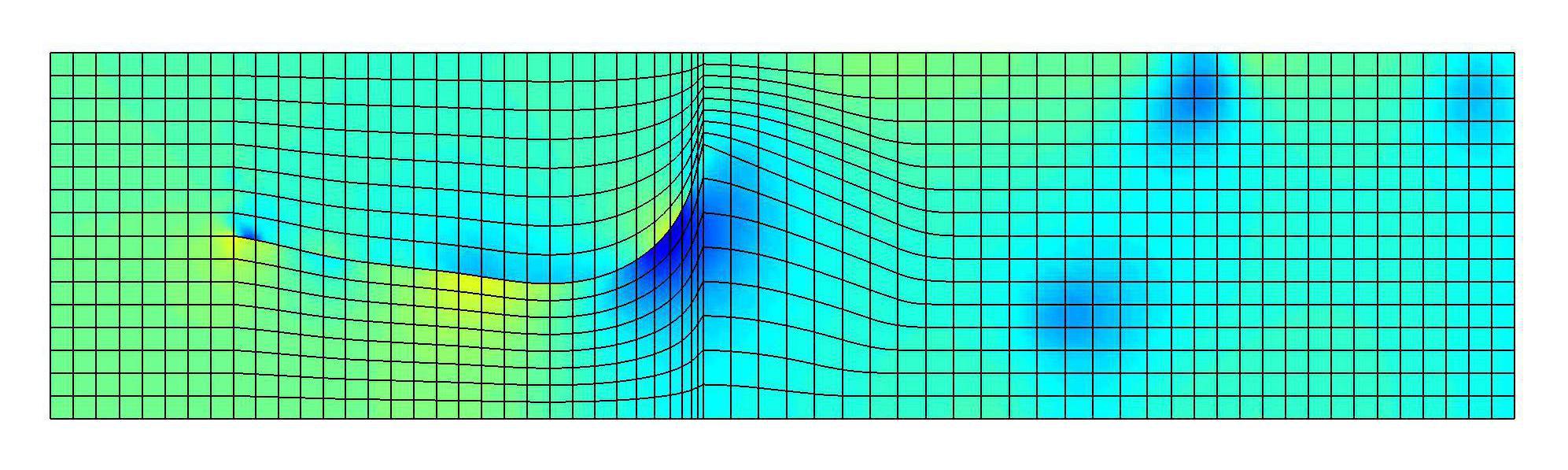}}  % update figure
\put(2.65,-.3){\includegraphics[height=16.3mm]{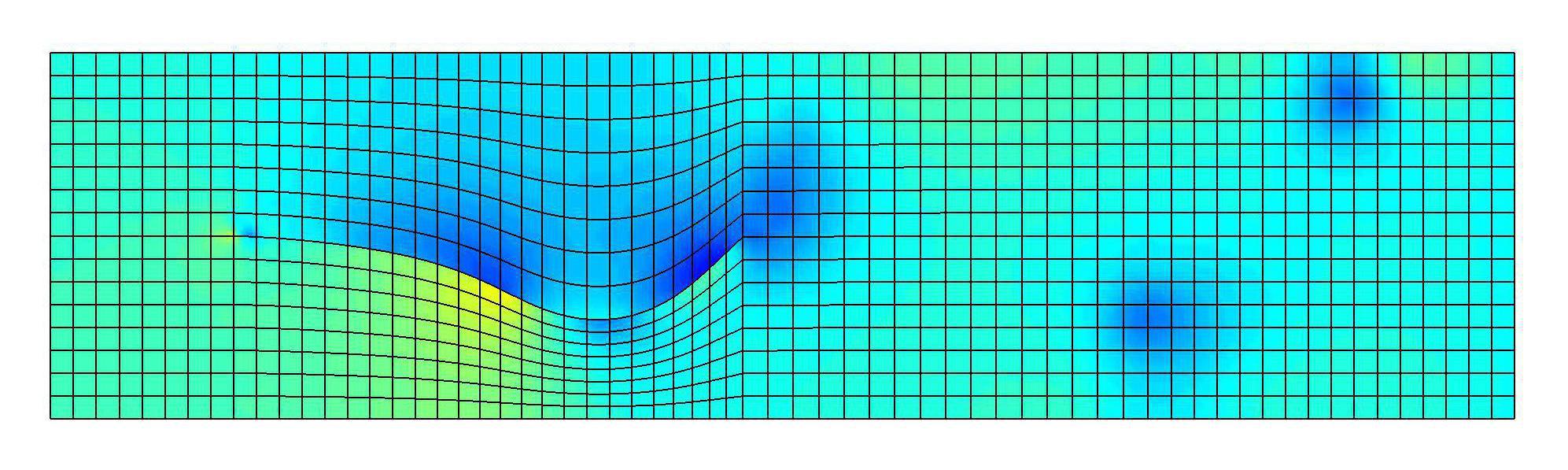}}  % update figure
\end{picture}
\caption{Flapping flag: Fluid pressure in the mid-plane at $t=44.36\,$s, $t=44.92\,$s, $t=45.48\,$s, $t=46.04\,$s, $t=46.60\,$s and $t=47.16\,$s (top left to bottom right) for $m=8$. 
The coloring is in the range $\{-.7,\,1.2\}p_0$ (from blue to red).}  
\label{f:flag_p}
\end{center}
\end{figure}

\begin{figure}[h]
\begin{center} \unitlength1cm
\begin{picture}(0,5.7)
%\put(-7.95,-.1){\includegraphics[height=58mm]{../../codes/cFEAR/Input/iFSI/Flag/oFlag/flag_x-t.pdf}}
%\put(0.15,-.1){\includegraphics[height=58mm]{../../codes/cFEAR/Input/iFSI/Flag/oFlag/flag_v-t.pdf}}
\put(-7.95,-.1){\includegraphics[height=58mm]{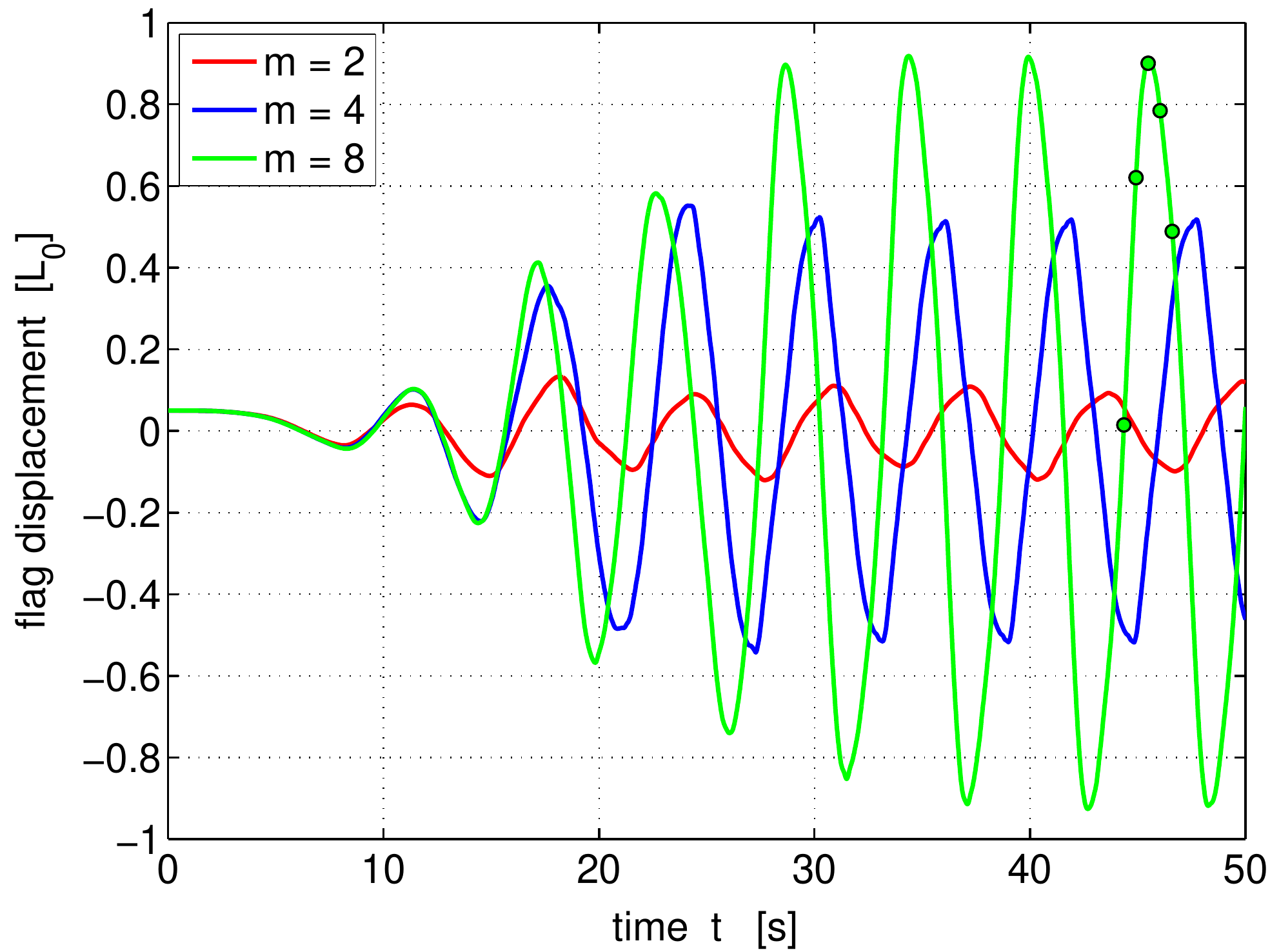}}
\put(0.15,-.1){\includegraphics[height=58mm]{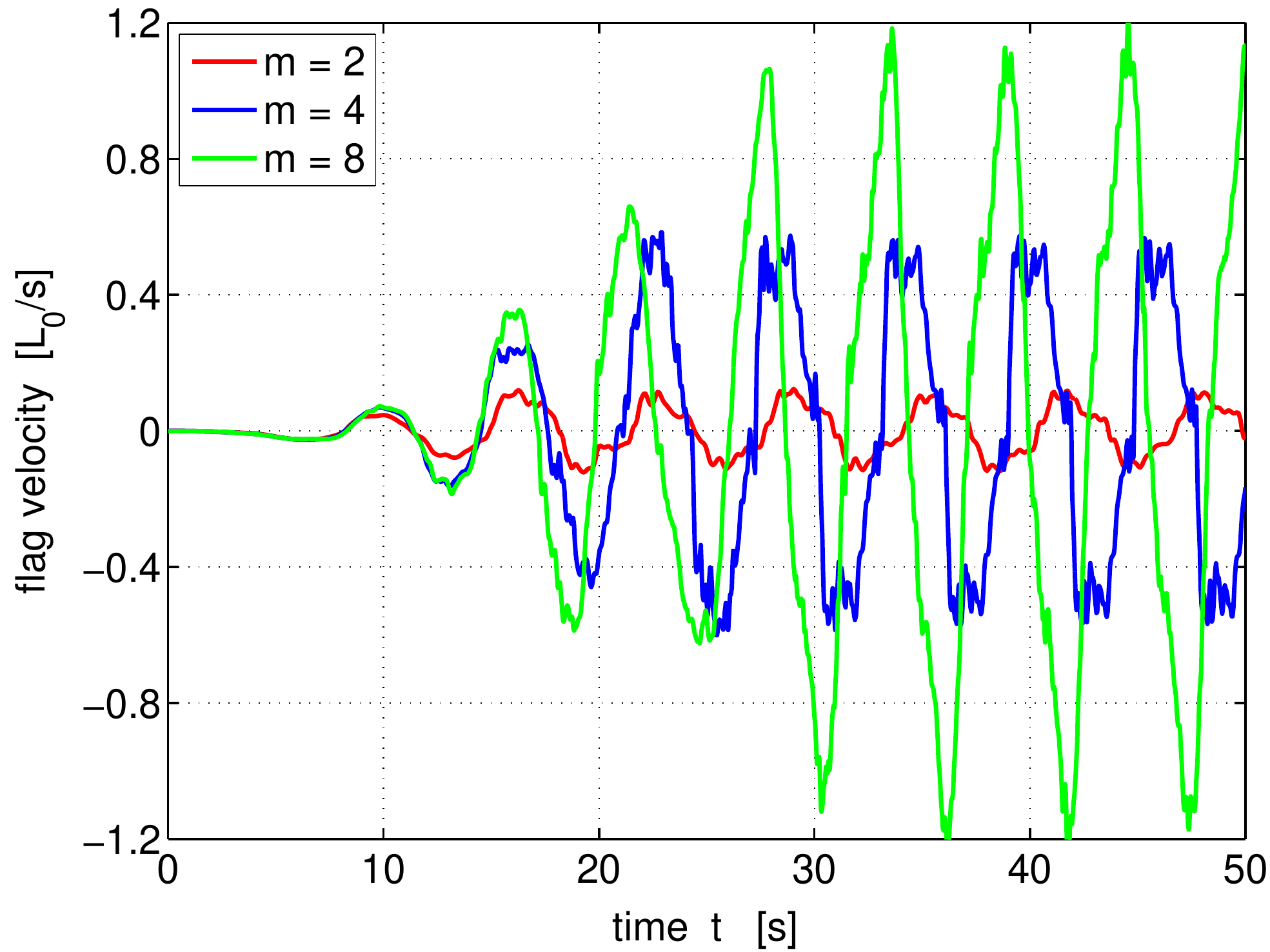}}
\put(-7.85,-.1){a.}
\put(.35,-.1){b.}
\end{picture}
\caption{Flapping flag: Lateral displacement (a) and velocity (b) at point $A$ for various FE discretizations.
Symbol `$\circ$' marks the configurations shown in Fig.~\ref{f:flag_x}.}  
\label{f:flag_xv}
\end{center}
\end{figure}

\begin{figure}[h]
\begin{center} \unitlength1cm
\begin{picture}(0,5.7)
%\put(-7.95,-.1){\includegraphics[height=58mm]{../../codes/cFEAR/Input/iFSI/Flag/oFlag/flag_rho.pdf}}
%\put(0.15,-.1){\includegraphics[height=58mm]{../../codes/cFEAR/Input/iFSI/Flag/oFlag/flag_rho_v-t.pdf}}
\put(-7.95,-.1){\includegraphics[height=58mm]{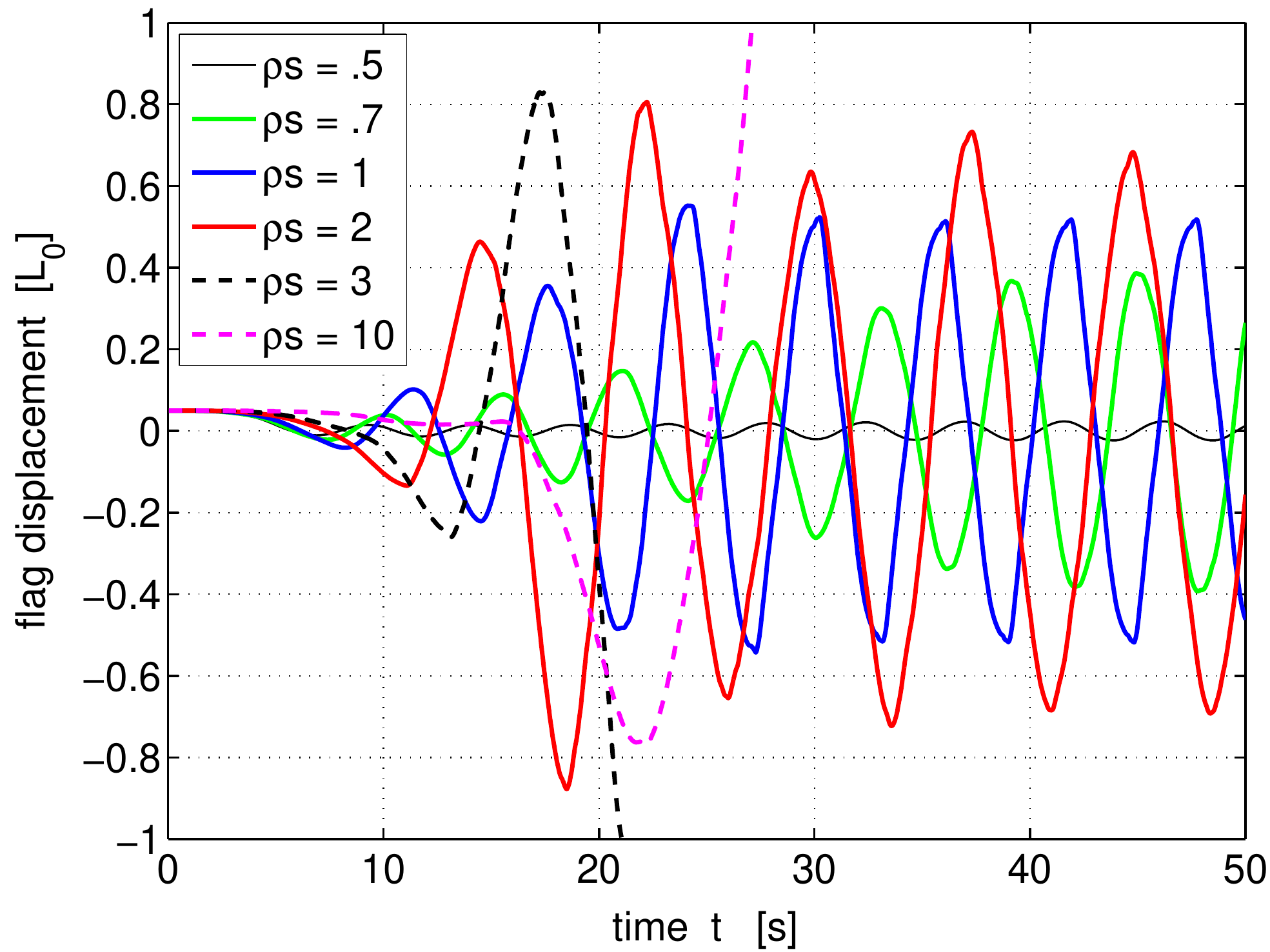}}
\put(0.15,-.1){\includegraphics[height=58mm]{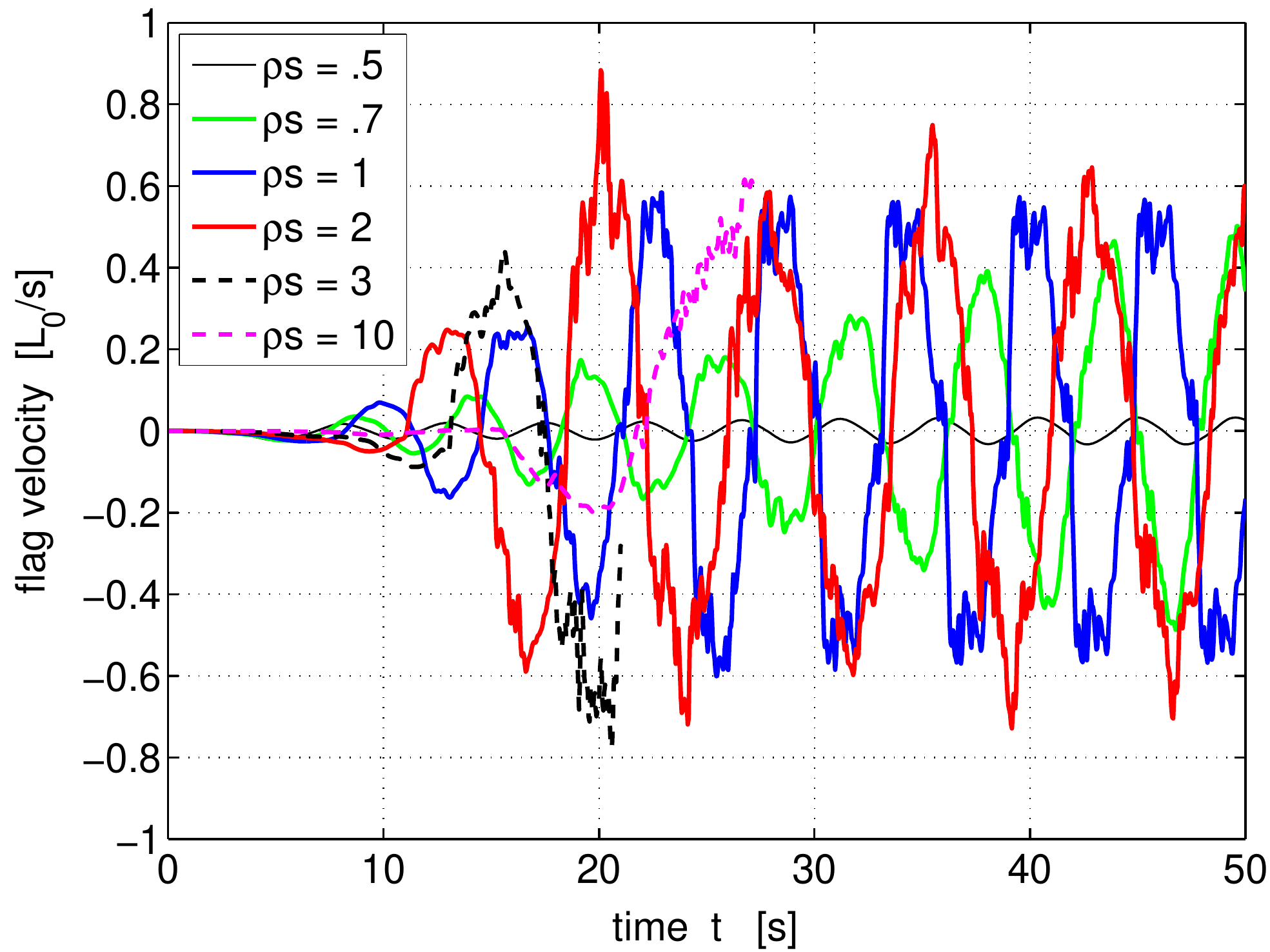}}
\put(-7.85,-.1){a.}
\put(.35,-.1){b.}
\end{picture}
\caption{Flapping flag: Influence of membrane density $\bar\rho_\mrs$ on the flag displacement (a) and velocity (b).
The density ratio affects the frequency and amplitude of vibration as expected.
For $\bar\rho_\mrs=3$ and above, the simulation terminates after the flag penetrates the boundary at $\pm L_0$.}
\label{f:flag_rho}
\end{center}
\end{figure}

The third example simulates the flapping motion of a flag.
The problem setup of this example is shown in Fig.~\ref{f:flag_ex}.
%-------------------------------------------------------------------------------------------------------------------------------
%-------------------------------------------------------------------------------------------------------------------------------
The flag is modeled as a flexible sheet that is supported on the left hand side. 
It is excited by a uniform inflow with velocity $v_\mathrm{in}$.
The length scale $L_0$, the fluid density $\rho_0$ and the time scale $T_0$ are used to normalize the problem.
The remaining parameters are chosen according to Tab.~\ref{t:flag_para}.
%------------------------------------------------------------------
\begin{table}[h]
\centering
\begin{tabular}{|l|l|}
  \hline
  parameter & normalized value \\[0.5mm] \hline 
    & \\[-3.5mm]   
   inflow velocity & $\bar v_\mathrm{in}=1$ \\[0.5mm] 
   density of the fluid & $\bar\rho = 1$ \\[0.5mm] 
   viscosity of the fluid & $\bar\eta = 1.531 \cdot 10^{-3}$ \\[0.5mm] 
   density of the flag & $\bar\rho_\mrs=1$  \\[0.5mm] 
   shear stiffness of the flag & $\bar\mu = 4.167 \cdot 10^3$ \\[0.5mm] 
   bending stiffness of the flag & $\bar c=0.02$ \\[1mm] 
   \hline
\end{tabular}
\caption{Flapping flag: Considered inflow and material parameters.}
\label{t:flag_para}
\end{table}
%------------------------------------------------------------------
Considering $L_0=0.1$m, $T_0=1$s and $\rho_0=1.2\,$kg/m$^3$, the fluid parameters become $\rho=\rho_0$ and $\eta=18.37\,\mu$Ns/m$^2$, which correspond to the values of air at sea level and $20^\circ$C, while the flag parameters become $\rho_\mrs=0.12\,$kg/m$^2$, $\mu = 5\,$N/m and $c=0.24\,\mu$Nm according to Sec.~\ref{s:norm}.\footnote{Following Sec.~\ref{s:norm}, the bending stiffness needs to be normalized by $c_0=F_0\,L_0$, where $F_0=\rho_0\,L_0^4/T_0^2$.}
The Reynolds number of the problem is 
\eqb{l}
Re = \ds\frac{\rho\,L_\mrc\,v_\mathrm{in}}{\eta}\,,
\eqe
where $L_\mrc$ is the chord length of the flag. 
For $L_\mrc=3L_0$ and the considered $\rho$ and $\eta$ follows $Re=1960\,\bar v_\mathrm{in}$.
At this $Re$ and density ratio\footnote{The density ratio $R_1:=\rho_\mrs/(\rho L_\mrc)$, as defined in \citet{shelley11}, is 1/3 here.}, the flag motion can be expected to be chaotic according to the phase diagram of \citet{connell07}.
 \\
The flapping flag example is a good test case since the flag motion and the surrounding flow field can become very complex, as the experimental data reported in \citet{shelley11} show. 
There have been recent 3D simulations that study the problem in detail \citep{hoffman11,banerjee15,gilmanov15,tullio16}. 
In some of those works immersed boundary methods are used instead of ALE.
Such methods are advantageous for very large flag motions that may even involve self-contact. 
In contrast to earlier work, the flag is discretized here with $C^1$-continuous isogeometric shell elements.
Their formulation is the same as the one of Eq.~\eqref{e:f_icfe} with the only exception that $\mf^e_{\sS\mathrm{int}}$ is extended by the internal bending moments according to the formulation of \citet{solidshell} using the Canham bending model.
A shell formulation is used in order to regularize the system with bending stiffness. 
A low stiffness value is used such that the structure remains very flexible.
Below a certain threshold value of $c$, the flapping behavior becomes independent of $c$ as is shown later.
\\
The fluid domain is discretized with $n_{\sF\mathrm{el}}=8m\times2m\times4m$ quadratic 3D NURBS elements, while
the flag is discretized with $n_{\sS\mathrm{el}}=3m\times2m$ quadratic 2D NURBS elements.
The number of nodes and dofs resulting from this discretization\footnote{The number of nodes is $n_\mathrm{no}=(8m+4)(2m+3)(4m+4)$; the number of dofs is $n_\mathrm{dof}=4n_\mathrm{no}+n_{\sS\mathrm{el}}$, due to the double pressure nodes on the flag surface.} are listed in Tab.~\ref{t:flag_mesh}.
%------------------------------------------------------------------
\begin{table}[h]
\centering
\begin{tabular}{|r|r|r|r|r|}
  \hline
  $m$ & fluid elements & membrane elements & nodes & dofs \\[0.5mm] \hline 
   & & & & \\[-3.5mm]   
% m & 64m^3 & 6m^2 & (8m+4)(2m+3)(4m+4) & 4n_no + n_Sel
   2 & 512 & 24 & 1,680 & 6,744 \\[1mm] 
   4 & 4096 & 96 & 7,920 & 31,776 \\[1mm] 
   8 & 32,768 & 384 & 46,512 & 186,432  \\[1mm] 
  %16 & 262,144 & 1,536 & 314,160 & 1,258,176 \\[1mm] 
   \hline
\end{tabular}
\caption{Flapping flag: Considered FE meshes based on quadratic NURBS elements.}
\label{t:flag_mesh}
\end{table}
%------------------------------------------------------------------
On the surface of the flag, double pressure dofs are used to account for pressure jumps as described in Sec.~\ref{s:2xp}. 
The time step is taken as $\Delta t=0.16\,T_0/m$.
The computational runtime per time step is about 3 mins.~for $m=4$ and 25 mins.~for $m=8$.
\\
Fig.~\ref{f:flag_x} shows the flag deformation at selected time steps. 
%-------------------------------------------------------------------------------------------------------------------------------
% run codes/cFEAR/Input/iFSI/Flag/oFlag/mFlagN
%-------------------------------------------------------------------------------------------------------------------------------
Those are snap-shots of the supplementary movie file \texttt{flag\underline{ }v.mpg}.
As expected, the structure performs flag-typical oscillations along its length. 
Close inspection shows that the flag motion also varies in vertical direction.
The pressure field around the flag is shown in Fig.~\ref{f:flag_p}.
%-------------------------------------------------------------------------------------------------------------------------------
% run codes/cFEAR/Input/iFSI/Flag/oFlag/mFlagN
%-------------------------------------------------------------------------------------------------------------------------------
The figure also shows the mesh motion around the flag.
It is based on the interpolation scheme given in App.~\ref{s:flagALE}.
\\
For the chosen parameters, the flapping behavior is still (quite) periodic, as Fig.~\ref{f:flag_xv} shows.
%-------------------------------------------------------------------------------------------------------------------------------
% run codes/cFEAR/Input/iFSI/Flag/oFlag/pFlagFinal
%-------------------------------------------------------------------------------------------------------------------------------
The period of the main oscillation is 5.60\,s.
Apart  from the main oscillations, there are also fine scale oscillations, as Fig.~\ref{f:flag_xv}b shows.
Fig.~\ref{f:flag_xv} also shows that the simulation results converge with mesh refinement. 
For the first 20 seconds, mesh $m=4$ already gives quite good results.
\\
The model parameters of Tab.~\ref{t:flag_para} affect the flapping behavior of the flag.
The influence of $Re$ has been discussed in detail in earlier work, e.g.~see \citet{shelley11}, so the following discussion focuses on the membrane parameters. %Apart from their usual influence on structural vibrations, 
Three aspects are noteworthy:\\
1.~For sufficiently low $c$, the flapping behavior (for given $Re$) remains unchanged, i.e. it becomes independent of $c$. 
%-------------------------------------------------------------------------------------------------------------------------------
\begin{figure}[h]
\begin{center} \unitlength1cm
\begin{picture}(0,5.7)
%\put(-7.95,-.1){\includegraphics[height=58mm]{../../codes/cFEAR/Input/iFSI/Flag/oFlag/flag_c.pdf}}
%\put(0.15,-.1){\includegraphics[height=58mm]{../../codes/cFEAR/Input/iFSI/Flag/oFlag/flag_muv.pdf}}
\put(-7.95,-.1){\includegraphics[height=58mm]{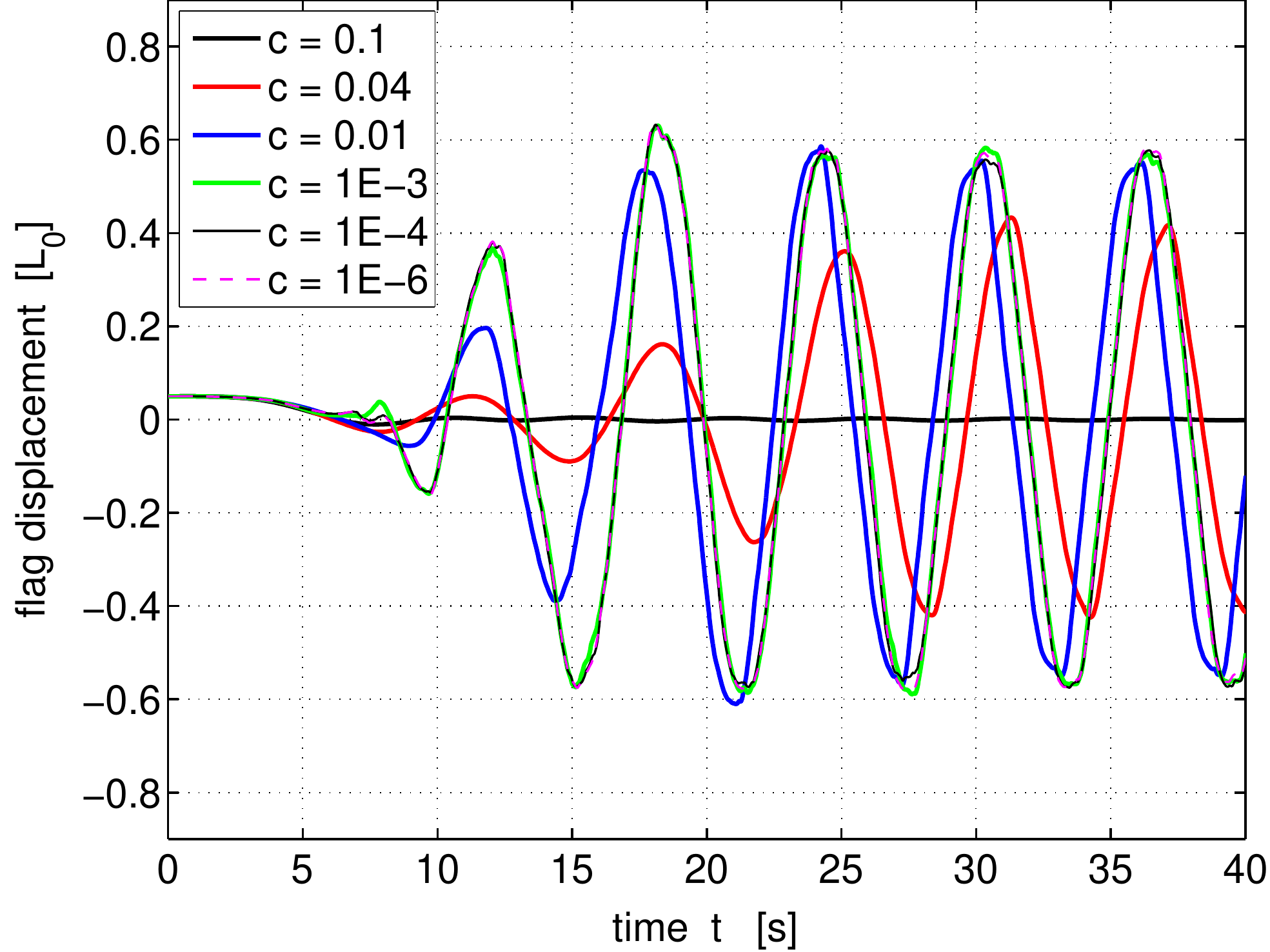}}
\put(0.15,-.1){\includegraphics[height=58mm]{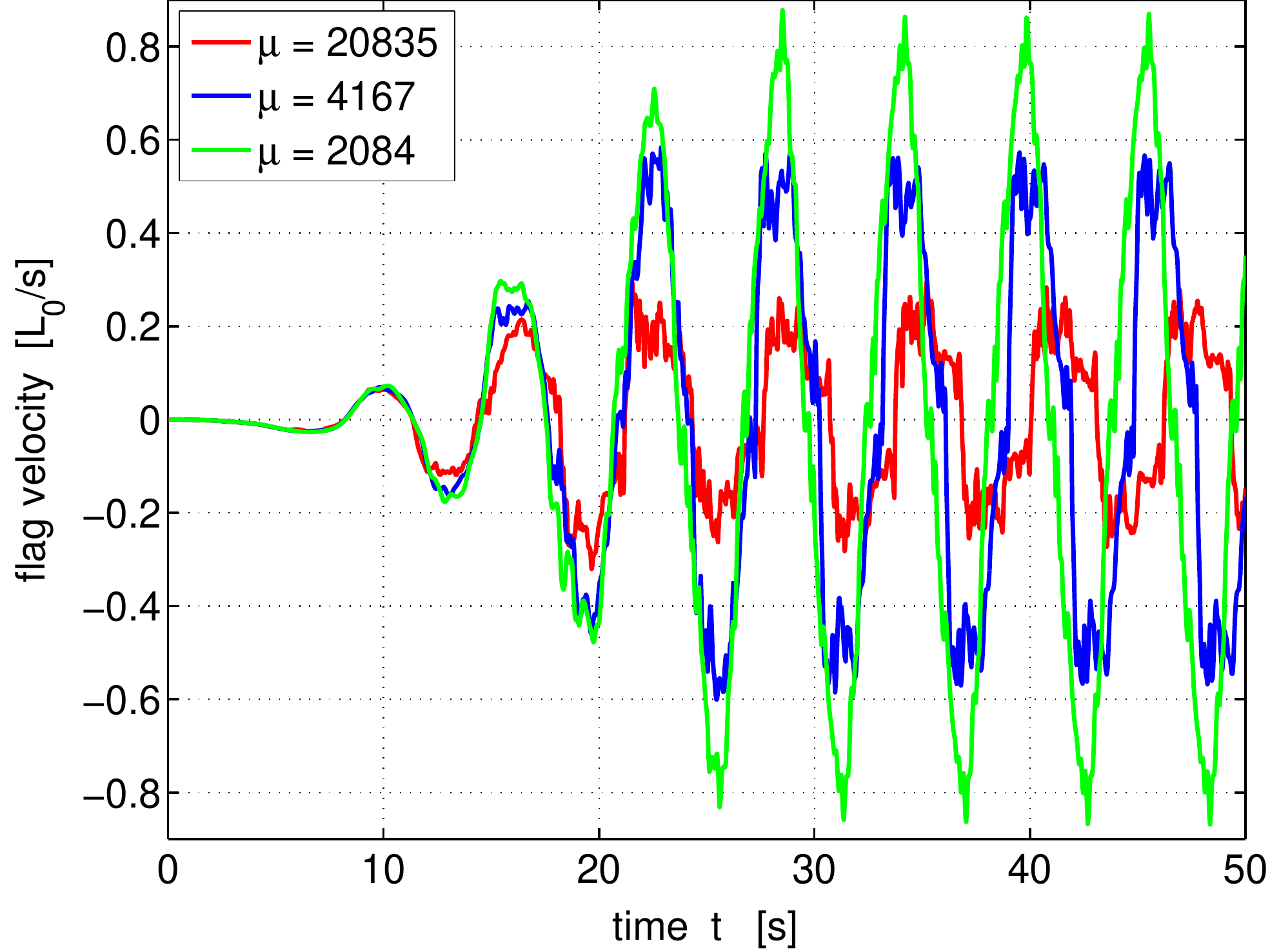}}
\put(-7.85,-.1){a.}
\put(.35,-.1){b.}
\end{picture}
\caption{Flapping flag: Influence of membrane parameters $\bar c$ (a) and $\bar \mu$ (b). 
The influence of $c$ vanishes below a threshold value of $c$. Increasing $\mu$ leads to smaller velocities but increased fine scale oscillations.}  
\label{f:flag_cm}
\end{center}
\end{figure}
% run codes/cFEAR/Input/iFSI/Flag/oFlag/pFlagFinal
%-------------------------------------------------------------------------------------------------------------------------------
According to Fig.~\ref{f:flag_cm}a this occurs below $\bar c\approx10^{-3}$.
Below that $c$, the flag is effectively a membrane without bending stiffness, and $c$ is only helpful for regularizing the numerical solution.\\
2.~Increasing $\mu$ leads to increased fine scale oscillations, as Fig.~\ref{f:flag_cm}b shows. 
Since $\mu$ controls the in-plane stiffness of the flag, those oscillations can be associated with longitudinal vibrations of the flag.\\
3. Increasing the ratio between fluid and membrane density does not degrade the computational robustness of the proposed monolithic scheme:
Fig.~\ref{f:flag_rho} shows the flapping behavior for various density ratios. 
%-------------------------------------------------------------------------------------------------------------------------------
% run codes/cFEAR/Input/iFSI/Flag/oFlag/pFlag18
%-------------------------------------------------------------------------------------------------------------------------------
For $\bar\rho_\mrs = \bar\rho$ ($=1$ here), the nodal FE forces due to fluid and membrane inertia are equal in the limit $h_e\rightarrow0$ (since $\dot\bv\approx$ const.~across the element thickness).
For all the considered density ratios, the Newton-Raphson iteration at each time step converges to a normalized energy residual of $10^{-27.7}$ within an average of six iterations.
The density ratio therefore does not have a negative affect on the computational stability or the conditioning of the system.
%The proposed monolithic scheme therefore does not
This is different to partitioned FSI schemes, which have been shown to suffer from a loss of robustness as the inertia forces of the flow become comparable or larger than those of the structure \citep{letallec01,causin05}.
The reason lies in the strong effect of the fluid on the structure for high fluid densities that is not well captured by weakly coupled partitioned schemes or requires many staggering steps in strongly coupled partitioned schemes.
The extreme case of this effect occurs when $\bar\rho_\mrs=0$, which was considered in the droplet example of Sec.~\ref{s:ex2}.
Also in this case no stability issues were encountered in all simulations.

\section{Conclusion}\label{s:concl}

% review
A unified FSI formulation is presented that is suitable for solid, liquid and mixed membranes.
At free liquid surfaces, sticking contact can be accounted for.
The fluid flow and the structure are discretized with finite elements using a stabilized fluid formulation and a surface-based membrane formulation.
A conforming interface discretization is used between fluid and membrane, which leads to a simple monolithic coupling formulation.
On membrane surfaces surrounded by fluid on both sides, double pressure nodes are required. 
The temporal discretization is based on the generalized-$\alpha$ scheme.
Two analytical and three numerical examples are presented in order to illustrate and verify the proposed formulation.
They consider fluid flow at low and high Reynolds numbers exhibiting strong FSI coupling.
\\
% outlook
The proposed formulation is very general and thus suitable as a basis for further research.
In order to increase efficiency, the formulation can be extended to boundary elements (for low $Re$) or turbulence models (for high $Re$).
Under current study is the use of enriched finite element discretizations \citep{ZEN} that are suitable to efficiently capture boundary layers \citep{ESEflow}.
Another extension of the present formulation is to re-examine the pressure stabilization scheme at contact boundaries. 
This would be especially important in the presence of sharp contact angles.
Such a formulation would then allow for a detailed flow analysis of droplets on rough surfaces.

\bigskip

{\Large\bf Acknowledgements}

The authors are grateful to the German Research Foundation (DFG) for supporting this research under grants GSC 111 and SA1822/3-2.
The authors also wish to thank Maximilian Harmel and Raheel Rasool for proofreading the manuscript.

\appendix

\section{Uniform membrane stretch}\label{s:ana_mem}

For the analytical example of Sec.~\ref{s:ana_infl}, the initial and the current membrane position are described by
\eqb{lll}
\bX(\theta,z) \is R_\mrs\,\be_r + z\,\be_3\,, \\[1mm]  
\bx(\theta,z) \is r_\mrs\,\be_r + z\,\be_3\,.
\eqe
From this follows $\bA_1=R_\mrs\,\be_\theta$, $\ba_1=r_\mrs\be_\theta$ and $\bA_2=\ba_2=\be_3$ with $\be_\theta=-\sin\theta\,\be_1+\cos\theta\,\be_2$. We further find
\eqb{llllll}
[A_{\alpha\beta}] \is 
\begin{bmatrix}
R_\mrs^2 & 0 \\
0 & 1
\end{bmatrix}, & 
[a_{\alpha\beta}] \is 
\begin{bmatrix}
r_\mrs^2 & 0 \\
0 & 1
\end{bmatrix}, \\[5mm]
\big[A^{\alpha\beta}\big] \is 
\begin{bmatrix}
R_\mrs^{-2} & 0 \\
0 & 1
\end{bmatrix} , & 
\big[a^{\alpha\beta}\big] \is 
\begin{bmatrix}
r_\mrs^{-2} & 0 \\
0 & 1
\end{bmatrix},
\eqe
such that $J_\mrs = r_\mrs/R_\mrs=:\lambda$ and
\eqb{l}
\big[\sigma^{\alpha\beta}\big] = \ds\frac{\mu}{\lambda}
\begin{bmatrix}
\ds\frac{1}{R_\mrs^2}
\big(1-\lambda^{-4}\big)
& 0 \\
0 & 1-\lambda^{-2}
\end{bmatrix}.
\eqe
The stress component along $\be_\theta$ is $\sig:=\be_\theta\cdot\big(\sigma^{\alpha\beta}\,\ba_\alpha\otimes\ba_\beta\big)\,\be_\theta$, which yields expression \eqref{e:ana1_sig}.

\section{FE tangent matrices for the time-continuous system}\label{s:FE_k}

\subsection{Fluid element}\label{s:FE_kF}

In order to evaluate the tangent matrix of the finite element force vector $\mf^e_\sF$ defined in \eqref{e:f_eFd}, we require
\eqb{l}
\mL_\mrv := \ds\pa{\big(\mB_\mrv\mv_e)}{\mv_e}\,,
\eqe
which can be written as $\mL_\mrv := [\mL_{\mrv1},\,\mL_{\mrv2},\,...,\,\mL_{\mrv n_e}]$ with
\eqb{l}
\mL_{\mrv I} := N_I\,\bL + B_{\mrv I}\,\bone\,.
\eqe
Therefore
\eqb{l}
\ds\pa{\bff^h_{\!\mathrm{res}}}{\mv_e} = \rho\,\mL_\mrv - \eta\,\mF\,.
\eqe
Based on this, we find the tangent matrices of the fluid forces defined in \eqref{e:f_eF}-\eqref{e:mcd_e}
\eqb{lllll}
\mm^e_\sF \dis \ds\pa{\mf^e_{\sF\mathrm{in}}}{\mv'_e} \is \mm_e\,, \\[4mm]
\mm^e_\mathrm{supg} \dis \ds\pa{\mf^e_\mathrm{supg}}{\mv'_e} \is \ds\int_{\Omega^e}\tau_\mrv\,\rho\,\mB_\mrv^\mrT\,\mN\,\dif v\,, \\[4mm]
\mm^e_\mathrm{pspg} \dis \ds\pa{\mg^e_\mathrm{pspg}}{\mv'_e} \is \ds\int_{\Omega^e}\tau_\mrp\,\rho\,\mG^\mrT\,\mN\,\dif v\,, \\[4mm]
%\eqe
%
%\eqb{lllll}
\mcc^e_\mathrm{con} \dis \ds\pa{\mf^e_\mathrm{con}}{\mv_e} \is \ds\int_{\Omega^e}\rho\,\mN^\mrT\,\mL_\mrv\,\dif v\,, \\[4mm]
\mcc^e_{\sF\mathrm{int}} \dis \ds\pa{\mf^e_{\sF\mathrm{int}}}{\mv_e} \is \mcc_e\,, \\[4mm]
\mcc^e_\mathrm{supg} \dis \ds\pa{\mf^e_\mathrm{supg}}{\mv_e} \is 
\ds\int_{\Omega^e}\tau_\mrv\,\big(\mB_\mrf^\mrT\,\mN + \rho\,\mB_\mrv^\mrT\,\mL_\mrv-\eta\,\mB_\mrv^\mrT\,\mF\big)\,\dif v\,, \\[4mm]
\mcc^e_\mrg \dis \ds\pa{\mg^e_\mrg}{\mv_e} \is \md^\mrT_e\,, \\[4mm]
\mcc^e_\mathrm{pspg} \dis \ds\pa{\mg^e_\mathrm{pspg}}{\mv_e} \is 
\ds\int_{\Omega^e}\tau_\mrp\,\big(\rho\,\mG^\mrT\,\mL_\mrv - \eta\,\mG^\mrT\,\mF\big)\,\dif v\,, \\[4mm]
%\eqe
%
%\eqb{lllll}
\md^e_{\sF\mathrm{int}} \dis \ds\pa{\mf^e_{\sF\mathrm{int}}}{\mpp_e} \is -\,\md_e\,, \\[4mm]
\md^e_\mathrm{supg} \dis \ds\pa{\mf^e_\mathrm{supg}}{\mpp_e} \is \ds\int_{\Omega^e}\tau_\mrv\,\mB_\mrv^\mrT\,\mG\,\dif v\,, \\[4mm]
\md^e_\mathrm{pspg} \dis \ds\pa{\mg^e_\mathrm{pspg}}{\mpp_e} \is \ds\int_{\Omega^e}\tau_\mrp\,\mG^\mrT\,\mG\,\dif v\,.
\label{e:mcd}\eqe
As seen, a major source of complexity are the stabilization terms $\mf^e_\mathrm{supg}$ and $\mg^e_\mathrm{pspg}$.

\subsection{Membrane element}\label{s:FE_kS}

Linearizing the membrane forces in \eqref{e:G} w.r.t.~$\dot\mv_e$ and $\mx_e$ yields the mass matrix
\eqb{l}
\mm^e_\sS := \ds\pa{\mf^e_{\sS\mathrm{in}}}{\dot\mv_e} = \ds\int_{\Omega^e} \rho_\mrs\,\mN^\mrT\mN\,\dif v\,,
\eqe
and the stiffness matrix
\eqb{l}
\mk^e_\sS := \mk^e_{\sS\mathrm{int}} + \mk^e_\mrc\,.
\eqe
The first term of $\mk^e_\sS$ follows from \citet{membrane} as
\eqb{l}
\mk^e_{\sS\mathrm{int}} := \ds\pa{\mf^e_{\sS\mathrm{int}}}{\mx_e} = \mk^e_\mathrm{geo} + \mk^e_\mathrm{mat}\,,
\label{e:kS}\eqe
with
\eqb{lll}
\mk^e_\mathrm{geo} \dis \ds\int_{\Omega_0^e}\tau^{\alpha\beta}\,\mN^\mrT_{,\alpha}\,\mN_{,\beta}\,\dif A\,, \\[4mm]
\mk^e_\mathrm{mat} \dis \ds\int_{\Omega_0^e}c^{\alpha\beta\gamma\delta}\,\mN^\mrT_{,\alpha}\,(\ba_\beta\otimes\ba_\gamma)\,\mN_{,\delta}\,\dif A\,,
\label{e:kmat}\eqe
$\tau^{\alpha\beta}:=J_\mrs\,\sigma^{\alpha\beta}$ and
\eqb{l}
c^{\alpha\beta\gamma\delta} := 2\ds\pa{\tau^{\alpha\beta}}{a_{\gamma\delta}}\,.
\eqe
Here,
\eqb{l}
c^{\alpha\beta\gamma\delta} = \ds\frac{2\mu}{J_\mrs^2}\big(a^{\alpha\beta}\,a^{\gamma\delta}+a^{\alpha\gamma}\,a^{\beta\delta} + a^{\alpha\delta}\,a^{\beta\gamma} \big)
\eqe
for model \eqref{e:sig_sol} and
\eqb{l}
c^{\alpha\beta\gamma\delta} = J_\mrs\gamma\,\big(a^{\alpha\beta}\,a^{\gamma\delta} - a^{\alpha\gamma}\,a^{\beta\delta} - a^{\alpha\delta}\,a^{\beta\gamma} \big)
\label{e:cabcd}\eqe
for model \eqref{e:sig_liq}, see \citet{membrane} and \citet{shelltheo}. 
Inserting these into \eqref{e:kS}, yields the simpler expression
\eqb{l}
\mk^e_\mathrm{mat} := \ds\int_{\Omega_0^e}\frac{2\mu}{J_\mrs^2}\,\mN^\mrT_{,\alpha}\,\big(\ba^\alpha\otimes\ba^\beta+\ba^\beta\otimes\ba^\alpha+a^{\alpha\beta}\bi\big)\,\mN_{,\beta}\,\dif A
\eqe
for model \eqref{e:sig_sol} and
\eqb{l}
\mk^e_\mathrm{mat} := \ds\int_{\Omega_0^e}J_\mrs\,\gamma\,\mN^\mrT_{,\alpha}\,\big(\ba^\alpha\otimes\ba^\beta-\ba^\beta\otimes\ba^\alpha-a^{\alpha\beta}\bi\big)\,\mN_{,\beta}\,\dif A
\eqe
for model \eqref{e:sig_liq}. 
Here $\bi:=\ba_\gamma\otimes\ba^\gamma$ is the identity tensor on surface $\sS$.
With this, $\mk^e_{\sS\mathrm{int}}$ can be further simplified, in particular for model \eqref{e:sig_liq}, see \citet{dropslide}.

The second term of $\mk^e_\sS$ depends on the contact description. 
Here, sticking contact is considered with a rigid substrate using the penalty regularization of Eq.~\eqref{e:fc}.
For this case, we have
\eqb{l}
\mk^e_\mrc := \ds\pa{\mf^e_{\mrc}}{\mx_e} = -\ds\int_{\Omega^e}\mN^\mrT\,\pa{\bff_{\!\mrc}}{\bx}\mN\,\dif a -\int_{\Omega^e}\mN^\mrT\,\big(\bff_{\!\mrc}\otimes\ba^\alpha\big)\,\mN_{,\alpha}\,\dif a\,,
\eqe
with
\eqb{l}
\ds\pa{\bff_{\!\mrc}}{\bx} = \left\{\begin{array}{ll}
-\epsilon\,\bone & $if $\bg\cdot\bn_\mrc < 0\,, \\[1mm]
\mathbf{0} & $else$\,.
\end{array}\right.
\eqe
The front term of $\mk^e_\mrc$ follows directly from Eqs.~\eqref{e:fc} and \eqref{e:bxh}, while the rear term is derived in \citet{spbf}.

\section{FE tangent matrices for the time-discrete system}\label{s:FE_kt}

\subsection{Fluid element}

For a fluid element $\Omega^e\subset\sF^h$, the tangent matrix $\mk^e$ defined by Eq.~\eqref{e:kFE} is given by the $4n_e\times4n_e$ matrix
\eqb{l}
\mk^e = \left[\begin{matrix}
  \ds\pa{\mf^e}{\mv^{n+1}_e} & \ds\pa{\mf^e}{\mpp^{n+1}_e} \\[4mm]
  \ds\pa{\mg^e}{\mv^{n+1}_e} & \ds\pa{\mg^e}{\mpp^{n+1}_e}
\end{matrix}\right],
\eqe
with
\eqb{lll}
\ds\pa{\mf^e}{\mv^{n+1}_e} \is \ds\pa{\mf^e}{\ma^{n+\alpha_\mrm}_e}\,\pa{\ma^{n+\alpha_\mrm}_e}{\ma^{n+1}_e}\,\pa{\ma^{n+1}_e}{\mv^{n+1}_e}
+ \ds\pa{\mf^e}{\mv^{n+\alpha_\mrf}_e}\,\pa{\mv^{n+\alpha_\mrf}_e}{\mv^{n+1}_e}\,, \\[4mm]
\ds\pa{\mg^e}{\mv^{n+1}_e} \is \ds\pa{\mg^e}{\ma^{n+\alpha_\mrm}_e}\,\pa{\ma^{n+\alpha_\mrm}_e}{\ma^{n+1}_e}\,\pa{\ma^{n+1}_e}{\mv^{n+1}_e}
+ \ds\pa{\mg^e}{\mv^{n+\alpha_\mrf}_e}\,\pa{\mv^{n+\alpha_\mrf}_e}{\mv^{n+1}_e}\,.
\label{e:drdv}
\eqe
Based on \eqref{e:globsys}, \eqref{e:mcd}, \eqref{e:Newmark} and \eqref{e:gen-a} we obtain
\eqb{lll}
\ds\pa{\mf^e}{\mv^{n+1}_e} \is \ds\frac{\alpha_\mrm}{\gamma\,\Delta t}\big(\mm^e_\sF+\mm^e_\mathrm{supg}\big)
+ \alpha_\mrf\,\big(\mcc^e_\mathrm{con}+\mcc^e_{\sF\mathrm{int}}+\mcc^e_\mathrm{supg} \big)\,, \\[4mm]
\ds\pa{\mf^e}{\mpp^{n+1}_e} \is \md^e_{\sF\mathrm{int}} + \md^e_\mathrm{supg}\,, \\[4mm]
\ds\pa{\mg^e}{\mv^{n+1}_e} \is \ds\frac{\alpha_\mrm}{\gamma\,\Delta t}\,\mm^e_\mathrm{pspg}
+ \alpha_\mrf\,\big(\mcc^e_\mrg+\mcc^e_\mathrm{pspg} \big)\,, \\[4mm]
\ds\pa{\mg^e}{\mpp^{n+1}_e} \is \md^e_\mathrm{pspg}\,,
\eqe
where the individual building blocks are given in \eqref{e:mcd}.

\subsection{Membrane element}

For a membrane element $\Omega^e\subset\sS^h$, the tangent matrix $\mk^e$ defined by Eq.~\eqref{e:kFE} is given by the $3n_e\times3n_e$ matrix
\eqb{l}
\mk^e = \ds\pa{\mf^e}{\mv^{n+1}_e}
= \pa{\mf^e}{\ma^{n+\alpha_\mrm}_e}\,\pa{\ma^{n+\alpha_\mrm}_e}{\ma^{n+1}_e}\,\pa{\ma^{n+1}_e}{\mv^{n+1}_e} 
+ \pa{\mf^e}{\mx^{n+\alpha_\mrf}_e}\,\pa{\mx^{n+\alpha_\mrf}_e}{\mx^{n+1}_e}\,\pa{\mx^{n+1}_e}{\mv^{n+1}_e}\,.
\eqe
Based on \eqref{e:globsys}, \eqref{e:mcd}, \eqref{e:Newmark} and \eqref{e:gen-a} we find
\eqb{l}
\mk^e = \ds\frac{\alpha_\mrm}{\gamma\,\Delta t}\,\mm^e_\sS + \frac{\alpha_\mrf\,\beta\,\Delta t}{\gamma}\,\mk^e_\sS\,,
\eqe
where $\mm^e_\sS$ and $\mk^e_\sS$ are given in Appendix~\ref{s:FE_kS}.

\section{Mesh motion for the flapping flag example}\label{s:flagALE}

For the flapping flag example in Sec.~\ref{s:flag}, the mesh velocity $\bv_\mrm$ (with Cartesian components $v_{\mrm i}$) at FE node (i.e. control point) $\bx_\mrm$ is defined by the linear interpolation
\eqb{l}
v_{\mrm1}\big(\bx_\mrm,t\big) = v_{\mrs1}\big(\bX_\mrp,t\big)
\left\{\begin{array}{ll}
0 & $for $X_{\mrm1}\leq L_0 $ or $X_{\mrm1}>6L_0 \,, \\[1mm]
1 & $for $L_0 < X_{\mrm1} < 4L_0\,, \\[1mm]
3-X_{\mrm1}/(2L_0) & $for $4L_0 < X_{\mrm1} < 6L_0 \,,
\end{array}\right.
\eqe
for the inflow direction, and
\eqb{l}
v_{\mrm j}\big(\bx_\mrm,t\big) = v_{\mrs j}\big(\bX_\mrp,t\big)
\left\{\begin{array}{ll}
0 & $for $d\geq L_0 \,, \\[1mm]
1-d/L_0 & $for $d < L_0 \,,
\end{array}\right.
\eqe
for the other directions ($j=2,3$).
Here $X_{\mrm i}$ are the components of $\bX_\mrm = \bx_\mrm\big|_{t=0}$, $d(\bX_{\mrm})$ is the distance of $\bX_\mrm$ from the flag surface, and $\bv_\mrs(\bX_\mrp,t)$ is the current flag velocity at the initially nearest membrane gridpoint $\bX_\mrp=\bX_\mrp(\bX_\mrm)$.
Note that $v_{\mrm1}$ is smooth at $X_{\mrm1}=1$ since $v_{\mrs1}$ approaches 0 smoothly as $X_{\mrm1}\rightarrow L_0$.

\bigskip

\bibliographystyle{apalike}
%\bibliographystyle{wileyj}
%\bibliography{../Tex/bibliography}
\bibliography{bibliography,bibTL}

\end{document}